\documentclass[11pt]{article}

\usepackage{verbatim}
\usepackage{amsmath}
\usepackage{amsfonts}
\usepackage{amssymb}
\usepackage{bbm}
\usepackage{latexsym}
\usepackage{graphics}
\usepackage{lscape} 
\usepackage{epsfig}
\usepackage{color}
\usepackage{graphicx}
\usepackage{graphics}
\usepackage{subfig}
\usepackage{mathbbol}
\usepackage{slashbox}
\usepackage{multirow}

\usepackage{cite}

\usepackage{color}

\renewcommand{\d}{\mathrm{d}}

\newcommand{\TR}{\textsc{tr}}

\DeclareMathSymbol{\mg}{\mathrel}{symbols}{"1D}

\newcommand{\ga}{\alpha}
\newcommand{\gb}{\beta}
\renewcommand{\gg}{\gamma}
\newcommand{\gd}{\delta}

\newcommand{\gf}{\phi}
\newcommand{\gvf}{\varphi}

\newcommand{\gm}{\mu}

\newcommand{\gk}{\kappa}
\newcommand{\gl}{\lambda}

\newcommand{\gth}{\theta}

\newcommand{\gs}{\sigma}

\newcommand{\go}{\omega}

\newcommand{\gp}{\pi}

\newcommand{\gL}{\Lambda}

\newcommand{\gPs}{\Psi}

\newcommand{\cF}{{\cal F}}

\newcommand{\cH}{{\cal H}}

\newcommand{\cJ}{{\cal J}}
\newcommand{\cK}{{\cal K}}

\newcommand{\cR}{{\cal R}}

\newcommand{\cV}{{\cal V}}

\newcommand{\Tr}{\mbox{Tr}}
\newcommand{\tr}{\text{tr}}

\newcommand{\Id}{\text{\small 1}\hspace{-3.5pt}\text{1}}

\newcommand{\ra}{\rightarrow}
\renewcommand{\Re}{\text{Re}\ }
\renewcommand{\Im}{\text{Im}\ }

\newcommand{\dsp}{\displaystyle}

\newcommand{\half}{\frac 12 }

\newcommand{\Kh}{K\"{a}hler}
\newcommand{\beq}{\begin{equation}}
\newcommand{\eeq}{\end{equation}}
\newcommand{\barr}{\begin{array}}
\newcommand{\earr}{\end{array}}
\newcommand{\equ}[1]{\begin{gather} #1 \end{gather}}
\newcommand{\equa}[1]{\begin{align} #1 \end{align}}

\newcommand{\arry}[2]{\begin{array}{#1} #2 \end{array}}

\newcommand{\non}{\nonumber}
\newcounter{oldcounter}

\newcommand{\bS}{{\bar S}}
\newcommand{\bT}{{\bar T}}
\newcommand{\bU}{{\bar U}}

\newcommand{\bgth}{{\bar\theta}}

\newcommand{\bgo}{{\bar\omega}}

\newcommand{\Intr}{\mathbb{Z}}
\newcommand{\Cplx}{\mathbb{C}}

\newcommand{\ba}[2]{\[\begin{array}{#2}\label{#1}}
\newcommand{\ea}{\end{array}\]}
\newcommand{\be}{\begin{equation}}
\newcommand{\ee}{\end{equation}}
\newcommand{\bea}{\begin{eqnarray}}
\newcommand{\eea}{\end{eqnarray}}

\newcommand{\U}[1]{\mathrm{U(#1)}}

\newcommand{\rep}[1]{\mathbf{#1}}
\newcommand{\crep}[1]{\overline{\rep{#1}}}

\newcommand{\sm}{{\,\mbox{-}}}

\begin{document}

\thispagestyle{empty}

\begin{flushright}
HD-THEP-09-1 \\
CPHT-RR003.0109\\
LPT-ORSAY-09-04 \\ 
LMU-ASC~03/09
%Version: \today  
\end{flushright}
\vskip 1cm
\begin{center}
{\Large {\bf Heterotic $\boldsymbol{\Intr_\text{6--II}}$ MSSM Orbifolds in Blowup} 
}
\\[0pt]

\bigskip
\bigskip 
{\large
{\bf Stefan Groot Nibbelink$^{a,b,}$\footnote{
{{ {\ {\ {\ E-mail: grootnib@thphys.uni-heidelberg.de}}}}}}},
{\bf Johannes Held$^{a,}$\footnote{
{{ {\ {\ {\ E-mail: johannes@tphys.uni-heidelberg.de}}}}}}},
{\bf Fabian Ruehle$^{a,}$\footnote{
{{ {\ {\ {\ E-mail: fabian@tphys.uni-heidelberg.de}}}}}}},\\
{\bf Michele Trapletti$^{c,}$\footnote{
{{ {\ {\ {\ E-mail: michele.trapletti@cpht.polytechnique.fr}}}}}}} 
{\bf and} 
{\bf Patrick K.S.~Vaudrevange$^{d,}$\footnote{
{{ {\ {\ {\ E-mail: Patrick.Vaudrevange@physik.uni-muenchen.de}}}}}}
\bigskip }\\[0pt]
\vspace{0.23cm}
${}^a$ {\it 
Institut f\"ur Theoretische Physik, Universit\"at Heidelberg, 
Philosophenweg 16 und 19,  D-69120 Heidelberg, Germany 
\\[1ex]  }
${}^b$ {\it
Shanghai Institute for Advanced Study, 
University of Science and Technology of China, 
99 Xiupu Rd, Pudong, Shanghai 201315, P.R.\ China
\\} 
\vspace{0.23cm}
${}^c$ {\it 
Laboratoire de Physique Theorique,
Univ,~Paris-Sud and CNRS, F-91405 Orsay, France 
\\[1ex] 
CPhT, \'Ecole Polytechnique, CNRS,
F-91128 Palaiseau, France 
 \\} 
\vspace{0.23cm}
${}^d$ {\it 
Arnold-Sommerfeld-Center for Theoretical Physics,  
Department f\"ur Physik, Ludwig-Maximilians-Universit\"at M\"unchen, 
Theresienstra\ss e 37, 80333 M\"unchen, Germany 
 \\} }

\bigskip
\end{center}

\subsection*{\centering Abstract}
Heterotic orbifolds provide promising constructions of MSSM--like models in string theory. We investigate the connection of such orbifold models with smooth Calabi-Yau compactifications by examining resolutions of the $T^6/\Intr_\text{6--II}$ orbifold (which are far from unique) with Abelian gauge fluxes. These gauge backgrounds are topologically characterized by weight vectors of twisted states; one per fixed point or fixed line. The VEV's of these states generate the blowup from the orbifold perspective, and they reappear as axions on the blowup. We explain methods to solve the 24 resolution dependent Bianchi identities and present an explicit solution.  Despite that a solution may contain the MSSM particle spectrum, the hypercharge turns out to be anomalous: Since all heterotic MSSM orbifolds analyzed so far have fixed points where only SM charged states appear, its gauge group can only be preserved provided that those singularities are not blown up. Going beyond the comparison of purely topological quantities (e.g. anomalous $\text{U(1)}$ masses) may be hampered by the fact that in the orbifold limit the supergravity approximation to lowest order in $\ga'$ is breaking down.

\newpage 
\setcounter{page}{1}

\section{Introduction}
\label{sc:intro}

One of the central tasks of string phenomenology is to build models which make contact with the observations of the real world. A basic step towards this goal is the construction of models in which gauge interactions and chiral matter are those of a (Minimal) Supersymmetric extension of the Standard Model of Particle Physics (MSSM). In the resulting framework one may hope  to comprehend the nature of supersymmetry breaking, and recover the properties of  the particle masses and couplings as part of the Standard Model. In this approach we implicitly assume that we can disentangle the problem of finding the correct matter spectrum from the issue of breaking four dimensional supersymmetry in string theory.

This basic step of obtaining MSSM--like models from string theory has been faced in the past from many different perspectives with some remarkable successes: Among the others, we would like to mention interesting findings based on purely Conformal Field Theory (CFT) constructions, like the so--called free--fermionic formulation \cite{Faraggi:1989ka}, the Gepner models \cite{Dijkstra:2004ym}, and the rational conformal field theory models \cite{Dijkstra:2004cc}.
Most of the other approaches are geometrical in nature. Among these we would like to remind the reader of the works of \cite{Honecker:2004kb} in the intersecting D--brane context (see also references
therein for models including chiral exotics), those of \cite{Verlinde:2005jr} for what concerns local constructions with D3 branes at singularities in Type IIB string theory, those of \cite{Beasley:2008kw,Beasley:2008dc,Donagi:2008ca,Donagi:2008kj} for similar constructions in a local F--theory language, and those of~\cite{Blumenhagen:2008zz} for globally consistent GUT models from intersecting D7-branes. Finally, there has been recent progress in heterotic model building by \cite{Donagi:1999ez} on smooth (elliptically fibered) Calabi Yau spaces that resulted in interesting constructions \cite{Braun:2005ux,Braun:2005bw,Braun:2005nv,Bouchard:2005ag,Andreas:2007ei}. The results of \cite{Buchmuller:2005jr,Buchmuller:2006ik} on heterotic orbifold model building were further exploited by \cite{Lebedev:2006kn,Lebedev:2007hv}. 

Each construction has peculiar properties and shows a certain amount of complementarity: Models can be global or only local. They may be obtained via elaborate computer scans or in a more
geometric/constructive perspective, and they may or may not  incorporate issues such as moduli stabilization, decoupling of exotics, Yukawa textures, etc. Comparing these diverse approaches can have severe impacts, as one might be able to use the good features of a given construction to overcome the limitations of others. Bringing these different approaches together can be achieved by using the duality properties of string theory (e.g. S--duality linking heterotic strings to type I strings, or T--duality linking IIB with IIA). Often this requires to overcome a language dichotomy by attaining some dictionary between the different terminologies. 

The dichotomy between CFT construction of heterotic strings on orbifolds and the corresponding supergravity compactifications on smooth Calabi--Yau manifolds will be one of the central themes of the current paper. Heterotic orbifolds allow for a systematic computer assisted search that can be very effective: In e.g.~\cite{Buchmuller:2005jr,Buchmuller:2006ik,Lebedev:2006kn,Lebedev:2007hv}, based on the embedding in
string theory of the orbifold-GUT picture (see e.g. \cite{Kobayashi:2004ya}), more than two hundred MSSM--like models have been assembled on the orbifold $T^6/{\mathbb Z}_\text{6--II}$. However, the CFT construction of heterotic orbifold models are only valid at very specific (orbifold) points of the string moduli space. This hinders the introduction of simple moduli stabilization mechanisms such as those due to flux compactifications~\cite{Giddings:2001yu}. Moreover, the generic presence of an anomalous U(1) in orbifold models induces Fayet--Iliopoulos terms driving them out of the orbifold points, which might shed uneasiness on consistency of the orbifold construction. Obtaining good models by compactifying the heterotic supergravity on smooth Calabi--Yau manifolds is a very challenging mathematical problem, and only a handful of such models have been uncovered so far. 
Establishing a more and more detailed glossary between heterotic orbifold and Calabi--Yau compactifications has been one of the essential challenge pursued in the papers~\cite{Honecker:2006qz,Nibbelink:2007rd,Nibbelink:2007pn,Nibbelink:2008tv} for heterotic strings on simple (mostly non--compact) orbifolds and their supergravity counterpart on their explicit blowups and topological  resolutions. Our aim is to extend these results to the heterotic $T^6/{\mathbb Z}_\text{6--II}$ orbifold that has been the spring of the largest set of MSSM--like models constructed from strings to date. 

In this paper we outline how to construct smooth Calabi--Yau manifolds from the orbifold $T^6/{\mathbb Z}_\text{6--II}$, and how to identify the supergravity analog of the $T^6/{\mathbb Z}_\text{6--II}$ heterotic models. These smooth Calabi--Yau's are compiled in steps: The local orbifold singularities are resolved using techniques of toric geometry, and they are subsequently glued together according to the prescriptions presented in~\cite{Lust:2006zh}. During the local resolution process we are able to detect the ``exceptional divisors'':  the four--cycles (compact hyper surfaces) hidden in the orbifold singularities. The local orbifold singularity is blown up once the volumes of the exceptional divisors become non--zero. The compact orbifold in addition has ``inherited cycles'', that are four dimensional sub--tori of $T^6$. Combining the knowledge of the exceptional and inherited cycles we come in the possession of a complete description of the set of two-- and four--cycles/forms of the orbifold resolutions, including their intersection ring (i.e.\ all their intersection numbers). Let us stress that the single $T^6/{\mathbb Z}_\text{6--II}$ orbifold has a very large number of topologically distinct resolutions. Depending on one's perspective this means that out of this orbifold many Calabi--Yaus are constructed, or that the corresponding Calabi--Yau has a large number of phases related by so--called flop transitions. 

The description of cycles is perfectly compatible with the supergravity language, and thus we can consider compactifications of ten dimensional heterotic supergravity on the resolved spaces. By embedding U(1) gauge fluxes on the hidden exceptional cycles we are able to obtain the gauge symmetry breaking and the chiral matter localized on the resolved singularities, that are the supergravity 
counterparts of the action of the orbifold rotation on the gauge degrees of freedom (and Wilson lines), and the twisted states, respectively. In this way we determine the relationship between the CFT data of heterotic $T^6/{\mathbb Z}_\text{6--II}$ orbifold and supergravity and super Yang--Mills on its resolutions. 

Following this procedure we can potentially describe resolutions of every $T^6/{\mathbb Z}_\text{6--II}$ heterotic orbifold model in the supergravity language. To investigate the properties of such resolution models, we apply our approach to a specific MSSM model (``benchmark model 2'' of \cite{Lebedev:2006kn,Lebedev:2007hv}) as a concrete testing case. This example illustrates a number of generic features of such blowups: we can identify a number of generic features of such blowups: We uncover an intimate relation between the specifications of the $\text{U}(1)$ flux background and the twisted states that generate the blowup from the orbifold point of view. The Standard Model hypercharge turns out to be always broken in complete blowup. This is due to the fact that the full blowup requires non--vanishing VEVs for twisted states at all fixed points, and some fixed points only have states charged under the Standard Model, hence at least the hypercharge is always lost. We stress that this does not depend on the specific choice we make for the gauge bundle. We comment in the conclusions about possible phenomenological consequences of this result as well as about how to avoid it.
\\

\noindent 
The paper has been organized as follows:

Section~\ref{sc:hetorbi} briefly reviews the heterotic orbifolds, specifying the details necessary to understand the $T^6/{\mathbb Z}_\text{6--II}$ orbifold of the heterotic E$_{8}\times$E$_{8}$ string. As a particular example of a MSSM--like model the ``benchmark model 2'' of \cite{Lebedev:2006kn,Lebedev:2007hv} is recalled.

Section~\ref{sc:resZsix} explains how to resolve the $T^6/{\Intr}_\text{6--II}$ orbifold
using toric geometry and gluing procedures presented in \cite{Lust:2006zh}. We first describe the three different possible singularities present in the orbifold, namely $\Cplx^2/\Intr_2$, $\Cplx^2/\Intr_3$ and $\Cplx^3/\Intr_\text{6--II}$. The first two singularities are resolved in a unique way. Contrary to this, a $\Cplx^3/\Intr_\text{6--II}$ singularity has five possible distinct resolutions. Since the $T^6/{\mathbb Z}_\text{6--II}$ orbifold contains 12 $\Cplx^3/\Intr_\text{6--II}$ singularities, the number of topologically different resolutions is huge: The most naive estimate would be $5^{12}$; the number of resolutions that lead to distinct models is close to two million. 

Section~\ref{sc:sugrares} considers ten dimensional heterotic supergravity on a generic  resolution of $T^6/\Intr_\text{6--II}$. Following  the procedure of \cite{Nibbelink:2007pn} we introduce U(1) gauge fluxes wrapped on the exceptional divisors. We describe how to single out the gauge fluxes such that they correspond to the embedding of the orbifold rotation and the Wilson lines in the gauge degrees of freedom in the heterotic orbifold theory. 
The Bianchi identity leads to a set of 24 coupled consistency conditions on the fluxes which depend on the local resolutions chosen. Solving them almost seems to be a mission impossible. However, by identifying the localized axions on the blowup with the twisted states of orbifold theory, that generate the blowup via their VEV's, shows that the  $\text{U(1)}$ fluxes are in one--to--one correspondence to the defining gauge lattice momenta of these states. 
The massless chiral spectrum of the model is computed by integrating the ten dimensional gaugino anomaly polynomial and turns out to suffer from a multitude of anomalous U(1)'s, among them the hypercharge. 

Section~\ref{sc:MSSMblow} illustrates our general findings on resolutions of heterotic MSSM--like orbifolds, by specializing to the study of the blowup of 
the MSSM orbifold model ``benchmark model 2''. We outline how solutions to the 24 coupled Bianchi identities can be updated, and illustrate that the line bundle vectors correspond to twisted states. In particular, we illustrate that the hypercharge is broken in full blowup. 
 
Finally, Section~\ref{sc:concl} contains our conclusions, and additional technical details have been collected in the appendices. 

\section{Heterotic $\boldsymbol{T^6/\Intr_\text{6--II}}$ MSSM models}
\label{sc:hetorbi}

%intro

\subsection{Orbifold geometry}
\label{sc:orbigeom}

First we want to give some general properties of orbifolds, as given for example in \cite{Dixon:1985jw,Dixon:1986jc} or \cite{Katsuki:1989bf}. Later we will examine in detail the $T^6/\mathbb{Z}_\text{6--II}$ orbifold on $\text{G}_2 \times \text{SU}(3) \times\text{SO}(4)$, where we use the conventions of \cite{Lebedev:2007hv}.

\subsubsection*{General description of $T^6/\mathbb{Z}_\text{N}$ orbifolds}
A $T^6/\mathbb{Z}_\text{N}$ orbifold is produced by identifying the points of a six--dimensional torus $T^6$ under the action of a discrete symmetry $\mathbb{Z}_\text{N}$. Using complex coordinates $z_i=\frac{1}{\sqrt{2}} \left( x^{2i-1}+i\, x^{2i}\right)$ ($i=1,\ldots,3$), the action of the $\mathbb{Z}_\text{N}$--twist $\theta$ is
\begin{equation}
z ~\mapsto~ \theta\, z  \quad\text{with}\quad \theta^i_j~=~e^{2\pi i \gvf^i}\delta^i_j\;.
\end{equation}
The order $N$ of the symmetry constrains the orbifold twist vector $\gvf$,
\begin{equation}
\theta^N~=~1 \;\Rightarrow\; N \gvf^i~=~0 \text{ mod }1\;.
\label{eq:Nv=0}
\end{equation}
Furthermore, the twist $\theta$ must fulfill the Calabi--Yau condition
\begin{equation}
\sum\limits_i \gvf^i~=~0 \text{ mod }1\;.
\label{eq:sumv=0}
\end{equation}
One can also consider an orbifold as being produced by modding out its space group $\mathbb{S}$ from $\mathbb{R}^6$. $\mathbb{S}$ is defined as a combination of twists and torus shifts $l$. Here $l=m_a e_a$ (summation over $a=1,\ldots,6$), where the $e_a$ define a basis of the torus lattice of $T^6$. The space group yields an equivalence relation, 
\begin{equation}
z ~\sim~ \left(\theta^k,l\right)z \equiv \theta^k z + l\;, 
\end{equation}
on $\mathbb{R}^6$. The elements of $\mathbb{S}$ fulfill the simple multiplication rule $\left(\theta^{k_1},l_1\right)\cdot\left(\theta^{k_2},l_2\right)~=~ \left(\theta^{k_1+k_2},\theta^{k_1}l_2+l_1 \right)$. In this picture, the torus $T^6$ is produced by dividing $\mathbb{R}^6$ by the basis vectors $e_a$, and one can take $\mathbb{R}^6$ as the covering space of the orbifold.

The space group does not act freely, i.e. there are fixed points. A (non-trivial) space group element $\left(\theta^k,l\right)$ specifies a fixed point $f$ up to shifts by the torus vectors: 
\begin{equation}
\label{eq:fixedpoint}
f~=~\left(\theta^k,l\right)f~=~\theta^k f+ l\;, \quad\text{with}\quad l~=~m_a e_a\;, \quad m_a \in \mathbb{Z}\;.
\end{equation}
If one now takes the fundamental domain of the torus as the cover for the orbifold, the fixed points in this domain will have different space group elements with a one--to--one correspondence between them.

If the twist acts trivially in one complex plane, i.e. $\theta^k z_i=z_i$ for one $i$, one obtains a two dimensional fixed subspace. On the cover, such a space looks like a torus and is often referred to as a fixed torus. However, on the orbifold the topology is not necessarily that of a torus, but it can also be a two dimensional orbifold.  Since in any way one complex coordinate is not affected, we also call those subspaces fixed (complex) lines.

\begin{table}[t]
\centering
\begin{tabular}{|c||ccc|ccc|}
\hline & \multicolumn{6}{|c|}{}\\[-2.5ex] 
torus & \multicolumn{6}{|c|}{basis vectors on} 
\\
& & $ \mathbb{R}^2 $ & \multicolumn{1}{c}{$ \mathbb{C} $} & & $ \mathbb{R}^2 $ & $ \mathbb{C}$\\
\hline\hline &&&&&&\\[-2ex]
$T_1^2$ on G$_2$ & $e_1 =$ & $\left(\begin{array}{c} 1\\0 \end{array}\right) $, & $ 1 $ & 
$e_2 =$ & $ \left(\begin{array}{c} - \frac{1}{2} \\ \frac{1}{2\sqrt{3}} \end{array}\right) $, & $ \frac{1}{\sqrt{3}} \, e^{5 \pi i/6}$ \\[3ex]
\hline &&&&&&\\[-2ex]
$T_2^2$ on SU(3)& $e_3 =$ & $ \left(\begin{array}{c} 1\\0 \end{array}\right) $, & $ 1 $ & $
e_4 =$ & $ \left(\begin{array}{c} -\frac{1}{2}\\ \frac{\sqrt{3}}{2} \end{array}\right) $, & $ e^{2\pi i/3}$ \\[3ex]
\hline&&&&&&\\[-2ex]
$T_3^2$ on SO(4) & $e_5 =$ & $ \left(\begin{array}{c} 1\\0 \end{array}\right) $, & $ 1 $ & $
e_6 =$ & $ \left(\begin{array}{c} 0\\1 \end{array}\right) $, & $ i $
\\[2ex]
\hline 
\end{tabular}
\caption{The basis vectors of the root lattice $\text{G}_2 \times \text{SU}(3) \times \text{SO}(4)$, in real and complex notation.}
\label{rootvector}
\end{table}

\subsubsection*{$T^6/\mathbb{Z}_\text{6--II}$ on $\text{G}_2 \times \text{SU}(3) \times \text{SO}(4)$}

\begin{figure}[t]
\centering
\includegraphics[width=0.8\textwidth]{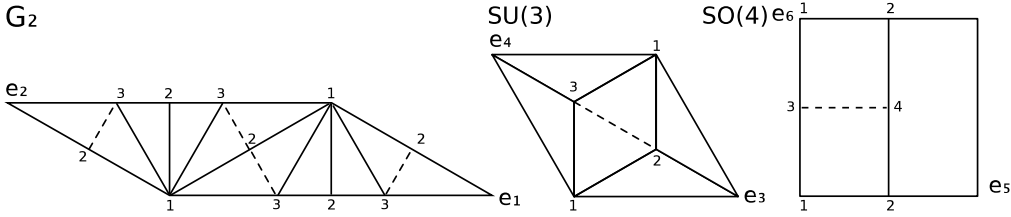}
\caption{The general fixed point structure of the $T^6/\mathbb{Z}_\text{6--II}$ orbifold. For each complex plane, equal numbers denote singularities that are mapped to the same point of the orbifold.}
\label{FunDom}
\end{figure}

We consider the torus $T^6$ obtained by dividing out $\mathbb{R}^6$ by the root lattice of $\text{G}_2 \times \text{SU}(3) \times \text{SO}(4)$. Since the lattice factorizes in three two dimensional parts, the same will be true for the torus. Therefore $T^6$ can be depicted by three parallelograms spanned by the simple root vectors of $\text{G}_2 \times \text{SU}(3) \times \text{SO}(4)$, as given in Table~\ref{rootvector}. The orbifold twist vector for $\mathbb{Z}_\text{6--II}$ is
\begin{equation}
\gvf~=~\frac{1}{6} \left(0,1,2,-3 \right)\;,
\end{equation}
where the $0$--th entry $\gvf^0 = 0$ is included for later use. Therefore, a single twist acts as a counterclockwise rotation of $60^\circ$ and $120^\circ$ on the first and second torus and as a (clockwise) rotation of $180^\circ$ on the third. The general structure of singularities, appearing after modding out the $\mathbb{Z}_\text{6--II}$ action, is shown in Figure~\ref{FunDom}. The numbers denote the locations of the orbifold singularities. Singularities in the covering space (i.e. the torus) that are identified on the orbifold are labeled by the same number.

\begin{figure}[t]
\centering
\includegraphics[width=0.75\textwidth]{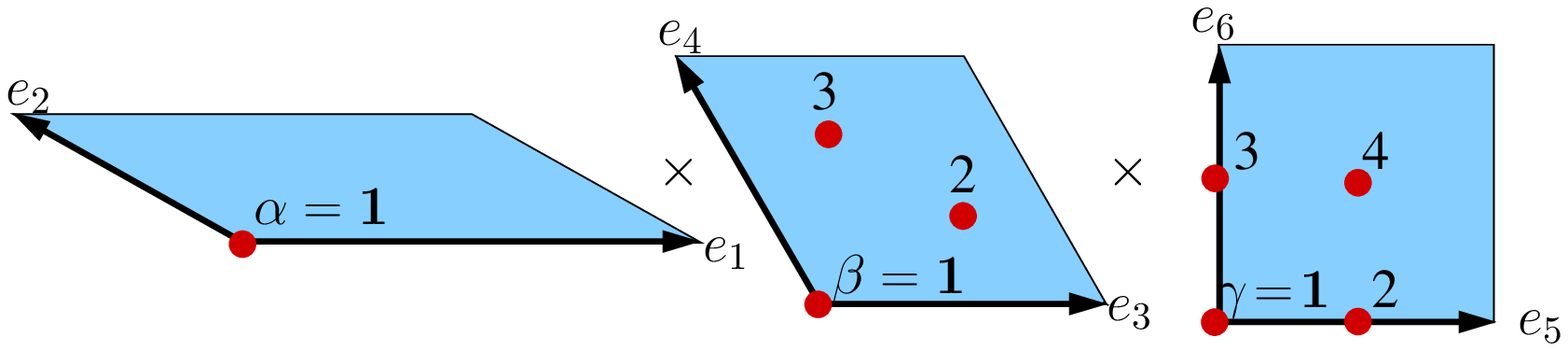}\vspace{0.5cm}
\begin{tabular}{|c||c|c|c|c|}
\hline
\multicolumn{5}{|c|}{torus shifts $l_{\beta\gamma}$ in the $\theta$--sector} \\
\hline
\backslashbox{$\beta$}{$\gamma$} & $1$ & $2$ & $3$ & $4$ \\
\hline\hline 
$1$ & $0$ & $e_5$ & $e_6$ & $e_5+e_6$ \\
$2$ & $e_3$ & $e_3+e_5$ & $e_3+e_6$ & $e_3+e_5+e_6$ \\
$3$ & $e_3+e_4$ & $e_3+e_4+e_5$ & $e_3+e_4+e_6$ & $e_3+e_4+e_5+e_6$\\
\hline
\multicolumn{5}{|c|}{torus shifts $l_{\beta\gamma}$ in the $\theta^5$--sector} \\
\hline
\backslashbox{$\beta$}{$\gamma$} & $1$ & $2$ & $3$ & $4$ \\
\hline\hline 
$1$ & $0$ & $e_5$ & $e_6$ & $e_5+e_6$ \\
$2$ & $e_3+e_4$ & $e_3+e_4+e_5$ & $e_3+e_4+e_6$ & $e_3+e_4+e_5+e_6$ \\
$3$ & $e_4$ & $e_4+e_5$ & $e_4+e_6$ & $e_4+e_5+e_6$ \\
\hline
\end{tabular}
\caption{Upper Figure: the fixed points in the $\theta$-- and $\theta^5$--sector. They are labeled by $\alpha=1$, $\beta=1,2,3$ and $\gamma=1,\ldots,4$. Lower table: the corresponding torus shifts $l_{\beta\gamma}$, see equation~(\ref{eq:fixedpoint}). For example, the space group element associated to the fixed point $\beta=2$ and $\gamma=1$ in the $\theta$--sector reads $(\theta,l_{21})=(\theta,e_3)$.}
\label{theta1}
\end{figure}

\begin{figure}[t]
\centering
\includegraphics[width=0.75\textwidth]{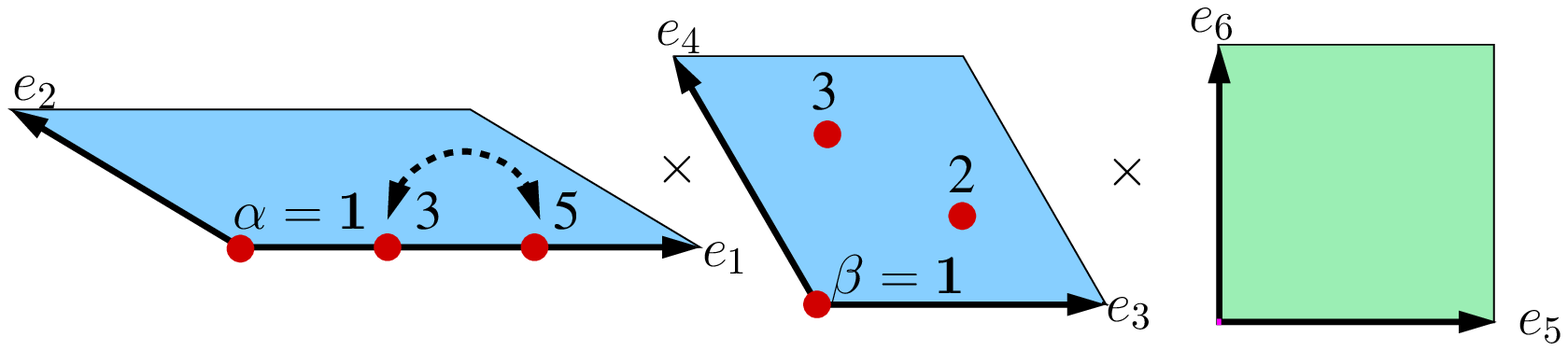}\vspace{0.8cm}
\begin{tabular}{|c||c|c|c|}
\hline
\multicolumn{4}{|c|}{torus shifts $l_{\alpha\beta}$ in the $\theta^2$--sector} \\
\hline
\backslashbox{$\alpha$}{$\beta$} & $1$ & $2$ & $3$ \\
\hline\hline 
$1$ & $0$ & $e_3+e_4$ & $e_4$ \\
$3$ & $-e_2$ & $-e_2+e_3+e_4$ & $-e_2+e_4$ \\
$5$ & $-2e_2$ & $-2e_2+e_3+e_4$ & $-2e_2+e_4$ \\
\hline
\multicolumn{4}{|c|}{torus shifts $l_{\alpha\beta}$ in the $\theta^4$--sector} \\
\hline
\backslashbox{$\alpha$}{$\beta$} & $1$ & $2$ & $3$ \\
\hline\hline 
$1$ & $0$ & $e_3$ & $e_3+e_4$ \\
$3$ & $e_1+e_2$ & $e_1+e_2+e_3$ & $e_1+e_2+e_3+e_4$ \\
$5$ & $2e_1+2e_2$ & $2e_1+2e_2+e_3$ & $2e_1+2e_2+e_3+e_4$ \\
\hline
\end{tabular}
\caption{Upper Figure: the fixed lines in the $\theta^2$-- and $\theta^4$--sector. They are labeled by $\alpha=1,3,5$ and $\beta=1,2,3$, where the points $\alpha=3$ and $\alpha=5$ are identified on the orbifold. Lower table: the corresponding torus shifts $l_{\alpha\beta}$.}
\label{theta2}
\end{figure}

\begin{figure}[t]
\centering
\includegraphics[width=0.75\textwidth]{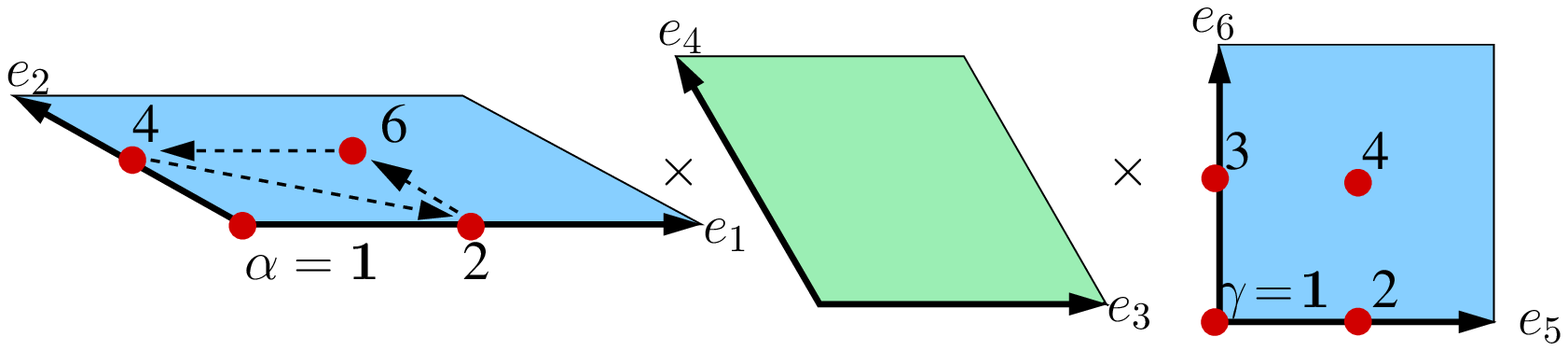}\vspace{0.8cm}
\begin{tabular}{|c||c|c|c|c|}
\hline
\multicolumn{5}{|c|}{torus shifts $l_{\alpha\gamma}$ in the $\theta^3$--sector} \\
\hline
\backslashbox{$\alpha$}{$\gamma$} & $1$ & $2$ & $3$ & $4$ \\
\hline\hline 
$1$ & $0$ & $e_5$ & $e_6$ & $e_5+e_6$ \\
$2$ & $e_1$ & $e_1+e_5$ & $e_1+e_6$ & $e_1+e_5+e_6$ \\
$4$ & $e_2$ & $e_2+e_5$ & $e_2+e_6$ & $e_2+e_5+e_6$ \\
$6$ & $e_1+e_2$ & $e_1+e_2+e_5$ & $e_1+e_2+e_6$ & $e_1+e_2+e_5+e_6$ \\
\hline
\end{tabular}
\caption{Upper Figure: the fixed lines in the $\theta^3$--sector. They are labeled by $\alpha=1,2,4,6$ and $\gamma=1,\ldots,4$, where the points $\alpha=2$, $\alpha=4$ and $\alpha=6$ are identified on the orbifold. Lower table: the corresponding torus shifts $l_{\alpha\gamma}$.}
\label{theta3}
\end{figure}

In order to obtain the detailed fixed point structure we look at every twist $\theta^k$--sector separately. For the twist $\theta$ (and its inverse $\theta^5$) one obtains the full order of the group $\mathbb{Z}_6$. The fixed points are shown in Figure~\ref{theta1}. They are labeled by $\alpha$ in the first torus, by $\beta$ in the second and by $\gamma$ in the third. The lattice shifts needed to bring the points back after a rotation are given in the table of Figure~\ref{theta1}. Since $\alpha=1$ in the first and fifth sector, the fixed points are determined by $\beta$ and $\gamma$. Next we consider the fixed points in the $\theta^2$-- and $\theta^4$--sector with twists $2\gvf=\frac{1}{3}(0,1,2,0)$ and $4\gvf=\frac{1}{3}(0,2,1,0)$, respectively. The order of these twists is $3$ and they act trivially on the third torus. Thus, concentrating solely on the $\theta^2$-- and $\theta^4$--sector, the compactification can be described as a $T^4/\mathbb{Z}_3$ orbifold resulting in a six--dimensional theory. The fixed lines of the $T^4/\mathbb{Z}_3$ orbifold are shown in Figure~\ref{theta2}. By comparing with Figure~\ref{FunDom} we see that the points $\alpha=3$ and $\alpha=5$ correspond to the same point on the orbifold as they are mapped onto each other by further twists. Hence, there are six independent fixed lines, labeled by $\alpha=1,3$ and $\beta=1,2,3$. The corresponding lattice shifts are given in the table of Figure~\ref{theta2}. At last we examine the $\theta^3$--sector. Here, the twist $3\gvf=\frac{1}{2}(0,1,0,-1)$ leaves the second torus invariant and acts with order two. In this case one obtains $T^4/\mathbb{Z}_2$ fixed lines, depicted in Figure~\ref{theta3}. Again one notes by comparing with Figure~\ref{FunDom} that the points $\alpha=2$, $4$ and $6$ are mapped onto each other by further twists and correspond to one point on the orbifold. Hence there are eight independent fixed lines, labeled by $\alpha=1,2$  and $\gamma=1,\ldots,4$. The lattice shifts for this sector are given in the table of Figure~\ref{theta3}.


\subsection{Heterotic orbifold models}
\label{sc:hetmodels}

Next, we review some technical details of the compactification of the heterotic string on orbifolds. The starting point of our discussion is the consideration of boundary conditions for closed strings. On orbifolds, there are new boundary conditions associated to non--trivial elements of the space group, i.e. $g\in\mathbb{S}$ defines a boundary condition $X(\tau,\sigma+2\pi) = g\; X(\tau,\sigma)$ for the six compactified dimensions of the string. If $g$ is not freely--acting (i.e. it has a fixed point), the string is attached to the fixed point and $g$ is called the constructing element of a so--called twisted string. On the other hand, strings with a constructing element $g=\Id$ correspond to the ordinary strings of the ten--dimensional heterotic string theory (being the supergravity and the $\text{E}_8\times\text{E}_8$ gauge multiplets). They are henceforth referred to as untwisted strings.

However, the geometrical action of the space group is not sufficient to define a consistent com\-pac\-ti\-fi\-cation. One needs to accompany the geometrical action of $\mathbb{S}$ by some action in the 16 gauge degrees of freedom, in our case in $\text{E}_8\times\text{E}_8$. In the case of shift embedding, the most general embedding of the space group is
\begin{equation}
\label{localshift}
g~=~\left(\theta^k,m_a e_a\right) \hookrightarrow V_g ~=~ kV + m_a A_a\;.
\end{equation}
That is, whenever a rotation by $\theta^k$ and a translation by $m_a e_a$ is performed in the six compact dimensions of the orbifold, the 16 gauge degrees of freedom are shifted by $V_g = kV + m_a A_a$, summation over $a=1,\ldots,6$. $V$ is called the shift vector and $A_a$ are (up to six) Wilson lines. They are constrained to lie in the $\text{E}_8\times\text{E}_8$ root lattice $\Lambda$ as follows:
\begin{equation}
N\,V ~\in~\Lambda\quad\text{ and }\quad N_a \,A_a ~\in~\Lambda\;,
\end{equation}
no summation over $a$. The order $N_a$ of the Wilson line $A_a$ is determined by the action of the twist in the direction of the Wilson line. In addition, Wilson lines have to be constrained due to further geometrical considerations. In the case of the $\mathbb{Z}_\text{6--II}$ orbifold this results in three independent Wilson lines, $A_3$ (of order 3) and $A_5$, $A_6$ (both of order 2) with the identifications
\begin{equation}
A_1 ~=~ A_2 ~=~ 0\;,\quad A_3 ~=~ A_4 ~=~ W_3\;,\quad A_5 ~=~ W_2\;,\quad\text{and}\quad A_6 ~=~ W_2' \;,
\end{equation}
where $W_3$, $W_2$ and $W_2'$ are introduced for later use.

Additionally, modular invariance of one--loop amplitudes imposes strong conditions on the shifts and Wilson lines. In $\mathbb{Z}_\text{N}$ orbifolds, the order $N$ shift $V$ and the twist $\gvf$ must fulfill~\cite{Dixon:1986jc,Vafa:1986wx}
\begin{equation}\label{eq:znmodularinv}
N \,\left(V^2 - \gvf^2\right)~=~0 \mod 2 \,.
\end{equation}
In the presence of Wilson lines, there are additional conditions
\begin{subequations}\label{eq:newmodularinv}
\begin{eqnarray}
  N_a\,\left(A_a\cdot V\right)      & = & 0 \mod 2\;, \\
  \label{eq:fsmi4}
  N_a\,\left(A_a^2\right)           & = & 0 \mod 2\;, \\
  \label{eq:fsmi5}
  Q_{ab}\,\left(A_a\cdot A_b\right) & = & 0 \mod 2 \quad (a \neq b)\;,
  \label{eq:fsmi6}
\end{eqnarray}
\end{subequations}
where $Q_{ab}\equiv\text{gcd}(N_a,N_b)$ denotes the greatest common divisor of $N_a$ and $N_b$~\cite{Ploger:2007iq}\footnote{In the case of two order 2 Wilson lines in an $\text{SO}(4)$ torus, $Q_{ab} = \text{gcd}(2,2)=2$ can be replaced by $Q_{ab} = 4$.}.

\subsection*{The spectrum}
The coordinates of a string can be split into left-- and right--movers, i.e. $X(\tau,\sigma)=X_L(\tau+\sigma) + X_R(\tau-\sigma)$ on--shell. After quantization, a string is described by a state of the form $|q\rangle_R \otimes \tilde\alpha |p\rangle_L$. Here, $q$ denotes the momentum of the (bosonized) right--mover (describing the space--time properties of the string) and $p$ labels the left--moving momentum of the 16 gauge degrees of freedom (describing the strings representation under gauge transformations). Furthermore, $\tilde\alpha$ denotes possible oscillator excitations. In general, physical states have to satisfy the mass--shell conditions for left-- and right--movers, i.e.
\begin{equation}
\label{eq:massless}
\frac{M_L^2}{8} ~=~ \frac{(p+V_g)^2}{2} + \tilde{N} - 1 + \delta c \quad\text{and}\quad \frac{M_R^2}{8}~=~\frac{(q+\gvf_g)^2}{2} - \frac{1}{2} + \delta c\;,
\end{equation}
and the so--called level--matching condition $M_L^2=M_R^2$. Here, $V_g$ denotes the local shift~\eqref{localshift} corresponding to the constructing element $g=\left(\theta^k,m_a e_a\right)$ of the (twisted) string. Analogously, $\gvf_g = k\gvf$ is called the local twist. Furthermore, $\delta c$ yields a change in the zero--point energy and is given by $\delta c = \frac{1}{2} \sum_{i=1}^3 \omega_i (1 - \omega_i)$, where $\omega_i = (\gvf_g)_i \mod 1$ such that $0 \leq \omega_i < 1$. It is convenient to define the shifted momentum $p_\text{sh} = p+ V_g$, as twisted strings transform according to their weight $p_\text{sh}$ under gauge transformations.

If the local twist $\gvf_g$ is non--trivial, i.e. $\gvf_g^i \neq 0$ for $i=1,2,3$, the compact space is six--dimensional resulting in an effective four dimensional theory. Furthermore, the $0$--th component $q^0$ of the solution $q$ to the right--moving mass--shell condition~(\ref{eq:massless}) defines four dimensional chirality, being $q^0 = \pm\frac{1}{2},0$ in this case. This corresponds to a chiral multiplet of $\mathcal{N}=1$ supersymmetry (and its CPT conjugate). For $\mathbb{Z}_\text{6--II}$, this is the case for the $\theta$ / $\theta^5$--sector, which therefore contains only chiral multiplets of $\mathcal{N}=1$ supersymmetry in four dimensions. On the other hand, if the twist acts trivially in one complex plane, i.e. $\gvf_g^i = 0$ for $i\neq 0$, the compact space is first of all only four dimensional resulting in an effective theory in six dimensions. The massless states are then  hyper multiplets of $\mathcal{N}=1$ supersymmetry in six dimensions.  For $\mathbb{Z}_\text{6--II}$, this is the case for the higher $\theta^k$--sectors, $k\neq 1,5$. However, as we will see in the following, these hyper multiplets are decomposed into chiral multiplets of four dimensional $\mathcal{N}=1$ supersymmetry when forming orbifold invariant states.

\subsection*{Orbifold invariant states}
The general idea is that orbifolded strings have to be compatible with the underlying orbifold space. To ensure this one has to analyze the action of the space group on the string states, i.e. under the action of some element $h\in\mathbb{S}$, the state $|q\rangle_R \otimes \tilde\alpha |p\rangle_L$ with constructing element $g \in\mathbb{S}$ transforms with a phase
\begin{equation}
|q_\text{sh}\rangle_R \otimes \tilde\alpha|p_\text{sh}\rangle_L \; \stackrel{h}{\mapsto}\; \Phi |q_\text{sh}\rangle_R \otimes \tilde\alpha|p_\text{sh}\rangle_L\;.
\end{equation}
The transformation phase $\Phi$ reads in detail
\begin{equation}
\label{eq:transformationphase}
\Phi ~=~ e^{2\pi i\,[p_\text{sh}\cdot V_h - r \cdot \gvf_h ]}\, \Phi_\text{vac}\;, \quad\text{where}\quad\Phi_\text{vac} ~=~ e^{2\pi i\,[-\frac{1}{2}(V_g\cdot V_h - \gvf_g\cdot \gvf_h)]}\;.
\end{equation}
$\Phi_\text{vac}$ is called the vacuum phase; for simplicity we assume that it can be set to $1$ in this Subsection. Furthermore, in order to summarize the transformation properties of $q+\gvf_g$ and of the oscillators we have introduced the so--called R--charge\footnote{These R--charges correspond to discrete R--symmetries of the superpotential in the context of string selection rules for allowed interactions.}
\begin{equation}
r^i ~=~ q^i + \gvf_g^i - \tilde{N}^i + \tilde{N}^{*i} \;.
\end{equation}
$\tilde{N}^i$ and $\tilde{N}^{*i}$, $i=0,\ldots,3$, are integer oscillator numbers, counting the number of left--moving oscillators $\tilde{\alpha}^i$ and $\tilde{\alpha}^{\bar i}$, $i=1,2,3$ and $\bar{i}=\bar{1},\bar{2},\bar{3}$, acting on the ground state $|p\rangle_L$, respectively. In detail, they are given by splitting the eigenvalues of the number operator $\tilde{N}$ according to $\tilde{N} = \omega_i \tilde{N}^i + \bar{\omega}_i \tilde{N}^{*i}$, where $\omega_i = (\gvf_g)_i \mod 1$ and  $\bar{\omega}_i = -(\gvf_g)_i \mod 1$ such that $0 \leq \omega_i, \bar{\omega}_i < 1$.

In general, the transformation phase~(\ref{eq:transformationphase}) has to be trivial in order for a string to be compatible with the orbifold background. In other words, strings with $\Phi\neq 1$ have to be removed from the spectrum. However, for a given string with constructing element $g\in\mathbb{S}$ we do not need to consider the action of all elements $h\in\mathbb{S}$. It is useful to distinguish two cases for $h$:

\subsubsection*{Case 1: $gh=hg$}
In the first case, $g$ and $h$ commute ($gh=hg$). This condition can be interpreted as a string located at the fixed point of $g$ but having still some freedom to move, especially in the direction of $h$ (e.g. when $g$ is from the $\theta^2$--sector of the $\mathbb{Z}_\text{6--II}$ orbifold, it has a fixed torus in the $e_5$, $e_6$ direction. Then, $h=(\Id,e_5),(\Id,e_6)$ corresponds to loops on which the string can move around). In this case the transformation phase~(\ref{eq:transformationphase}) has to be trivial, i.e.
\begin{equation}
\label{eqn:projectout}
p_\text{sh}\cdot V_h - r\cdot \gvf_h  ~\stackrel{!}{=}~0 \mod 1
\end{equation}
In other words, the total vertex operator of the state with boundary condition $g$ has to be single--valued when transported along $h$ if $h$ is an allowed loop, $hg=gh$.

For $\mathbb{Z}_\text{6--II}$, this projection acts for example on the higher $\theta^k$--sectors with $k\neq 1,5$ in two ways: 1) by Wilson lines in the fixed torus and 2) by a projection on $\theta$. We concentrate on the second case. For example, for $\alpha=1$ and $\gamma=1$ in the $\theta^3$--sector, the constructing element $\left(\theta^3,0\right)$ obviously commutes with $\left(\theta,0\right)$, see Figure~\ref{theta3}. This induces the condition $p_\text{sh}\cdot V - r\cdot \gvf  = 0 \mod 1$. In general, this kind of conditions can remove parts of the localized spectrum, or in some cases even the complete massless localized matter of some fixed lines.

\subsubsection*{Example for Case 1: Breaking of $\text{E}_8\times\text{E}_8$}
One further important example of equation~(\ref{eqn:projectout}) is the breaking of the ten dimensional gauge group $\text{E}_8\times\text{E}_8$ by the orbifold compactification. Gauge bosons are untwisted strings (with constructing element $g=\Id$). Hence, all elements $h$ of the space group commute and induce projection conditions. As $r\cdot \gvf_h = q\cdot \gvf_h = 0$ for the gauge bosons, this leads to the following conditions on the roots $p \in \Lambda$ (with $p^2=2$) of the unbroken gauge group
\begin{equation}
p\cdot V  ~\stackrel{!}{=}~0 \mod 1 \quad\text{and}\quad p\cdot A_a  ~\stackrel{!}{=}~0 \mod 1 \quad\text{for }a=1,\ldots,6\;.
\end{equation}

\subsubsection*{Case 2: $gh\neq hg$}
In the second case, $g$ and $h$ do not commute ($gh\neq hg$). Then, $h$ maps the fixed point of $g$ to an equivalent one, which corresponds to the space--group element $hgh^{-1}$. In other words, a string located at $g$ cannot move along the direction of $h$. But still, the state corresponding to $g$ has to be invariant under the action of $h$. Therefore, one has to build linear combinations of states located at equivalent fixed points. These equivalent fixed points are distinguishable only in the covering space of the orbifold (for example, for $\mathbb{Z}_\text{6--II}$, states from the $\theta^2$--sector located at the two fixed points $\alpha = 3,5$ have to be combined, since $\theta^3$ maps the corresponding fixed points to each other, see Figure~\ref{theta2}). These linear combinations can in general involve relative phases $\gamma$, i.e.
\begin{equation}
\sum_{n} \left( e^{-2\pi i\, n\, \gamma}\, |q_\text{sh}\rangle_R \otimes \tilde{\alpha}|p_\text{sh}\rangle_L \otimes |h^n\,g\,h^{-n}\rangle\right) = |q_\text{sh}\rangle_R \otimes \tilde{\alpha} |p_\mathrm{sh}\rangle_L \otimes \big(\sum_{n} e^{-2\pi i\, n\, \gamma}\ |h^n\,g\,h^{-n}\rangle\big)\;,
\label{eqn:physicalstate2}
\end{equation}
where $|g'\rangle = |h^n\,g\,h^{-n}\rangle$ denotes the localization of the state at the fixed point of $g'\in\mathbb{S}$ and $\gamma = \text{integer}/N$. The geometrical part of the linear combination transforms non--trivially under $h$
\begin{equation}
\label{eqn:constr_element_case2}
\sum_{n} e^{-2\pi i n \gamma}\ |h^n\,g\,h^{-n}\rangle\, \stackrel{h}{\mapsto} \ e^{2\pi i\,\gamma} \sum_{n} e^{-2\pi i\, n\, \gamma}\ |h^n\,g\,h^{-n}\rangle\;.
\end{equation}
Now, $h$ has to act as the identity on the linear combination. Consequently, we have to impose the following condition using the equations~(\ref{eq:transformationphase}), (\ref{eqn:physicalstate2}) and~(\ref{eqn:constr_element_case2}) for non--commuting elements:
\begin{equation}
\label{eqn:projectionwithgamma} 
p_\text{sh}\cdot V_h - r\cdot \gvf_h + \gamma~\stackrel{!}{=}~0 \mod 1\;.
\end{equation}
However, given some solution to the mass equations~(\ref{eq:massless}) one can always choose an appropriate $\gamma$ to fulfill this condition. In this sense, equation~(\ref{eqn:projectionwithgamma}) does not remove states from the spectrum and is hence not a projection condition.

\subsection*{Anomalous $\text{U}(1)$}
Using the material discussed so far, one can construct consistent heterotic orbifold models. One way to check their consistency is to analyze whether all gauge anomalies of the massless spectrum vanish. For example, for a $\text{U}(1)$ gauge factor there are several possible anomalies:
\begin{equation}
\begin{array}{lll}
\U{1}-\text{grav}-\text{grav}, & \qquad    & \U{1}-\U{1}-\U{1},\\
\U{1}-G-G,                     &\text{and} & \U{1}-\U{1}'-\U{1}'\;,\phantom{\big(\big)^{I^I}}
\end{array}
\end{equation}
where $G$ denotes a non--Abelian gauge group factor (like $\text{SU}(2)$) and $\U{1}'$ is another $\U{1}$ factor. We denote the 16--dim. vector that generates a $\text{U}(1)$ by $t$ and the associated charge by $Q$. Then, a state with left--moving momentum $p_\text{sh}$ carries a charge $Q=p_\text{sh}\cdot t$. However, it is known that in heterotic compactifications one $\text{U}(1)$ factor can seem to be anomalous, where we denote its generator by $t_\text{anom}$ and its charge by $Q_\text{anom}$. Then, the anomalous $\text{U}(1)$ has to satisfy the following conditions~\cite{Casas:1987us,Kobayashi:1997pb}
\begin{equation}
\label{eq:Anomaly_conditions}
\frac{1}{24}\text{Tr}\,Q_\text{anom} ~=~ \frac{1}{6 |t_\text{anom}|^2}\text{Tr}\,Q_\text{anom}^3 ~=~ \text{Tr}\,\ell Q_\text{anom} ~=~ \frac{1}{2|t|^2}\text{Tr}\,Q^2 Q_\text{anom} ~=~ \frac{1}{2}|t_\text{anom}|^2 \neq 0
\end{equation}
in order to be canceled by the universal Green--Schwarz mechanism, i.e. by a cancelation induced from the anomalous transformation of the axion $\alpha^\text{orb}$. Here, $\ell$ is the Dynkin index\footnote{The Dynkin index $\ell(\boldsymbol{r}^{(f)})$ of some representation $\rep{r}^f$ is defined by $\ell(\rep{r}^{(f)})\,\delta_{ab}=\tr(t_a(\rep{r}^{(f)})\,t_b(\rep{r}^{(f)}))$, using the generator $t_a$ of $G$ in the representation $\rep{r}^f$. The conventions are such that $\ell(\rep{M})=1/2$ for $\text{SU}(M)$ and $\ell(\rep{M})=1$ for $\text{SO}(M)$.} with respect to the non--Abelian gauge group factor $G$. Since all other anomalies vanish this results in an anomaly--free theory.

\begin{table}[t]
\begin{center}
\begin{tabular}{|c||l|l|c|c||l|l|}
\cline{1-3} \cline{5-7}
\# & irrep. & label & & \# & irrep. & label\\
\cline{1-3}\cline{5-7} \\[-2.5ex] \cline{1-3} \cline{5-7}
 3 & $\left(\boldsymbol{3},\boldsymbol{2};\boldsymbol{1},\boldsymbol{1}\right)_{1/6}$             & $q_i$      & &
 3 & $\left(\overline{\boldsymbol{3}},\boldsymbol{1};\boldsymbol{1},\boldsymbol{1}\right)_{-2/3}$ & $\bar u_i$ \\
 7 & $\left(\overline{\boldsymbol{3}},\boldsymbol{1};\boldsymbol{1},\boldsymbol{1}\right)_{1/3}$  & $\bar d_i$ & &
 4 & $\left(\boldsymbol{3},\boldsymbol{1};\boldsymbol{1},\boldsymbol{1}\right)_{-1/3}$            & $d_i$ \\
 8 & $\left(\boldsymbol{1},\boldsymbol{2};\boldsymbol{1},\boldsymbol{1}\right)_{-1/2}$            & $\ell_i$ & &
 5 & $\left(\boldsymbol{1},\boldsymbol{2};\boldsymbol{1},\boldsymbol{1}\right)_{1/2}$             & $\bar \ell_i$ \\
 3 & $\left(\boldsymbol{1},\boldsymbol{1};\boldsymbol{1},\boldsymbol{1}\right)_{1}$               & $\bar e_i$ & &
   &                                                                                              & \\
\cline{1-3}\cline{5-7}
47 & $\left(\boldsymbol{1},\boldsymbol{1};\boldsymbol{1},\boldsymbol{1}\right)_{0}$               & $s^0_i$ & &
26 & $\left(\boldsymbol{1},\boldsymbol{1};\boldsymbol{1},\boldsymbol{2}\right)_{0}$               & $h_i$ \\
20 & $\left(\boldsymbol{1},\boldsymbol{1};\boldsymbol{1},\boldsymbol{1}\right)_{1/2}$             & $s^+_i$ & &
20 & $\left(\boldsymbol{1},\boldsymbol{1};\boldsymbol{1},\boldsymbol{1}\right)_{-1/2}$            & $s^-_i$ \\
 2 & $\left(\boldsymbol{1},\boldsymbol{1};\boldsymbol{1},\boldsymbol{2}\right)_{1/2}$             & $x^+_i$ & &
 2 & $\left(\boldsymbol{1},\boldsymbol{1};\boldsymbol{1},\boldsymbol{2}\right)_{-1/2}$            & $x^-_i$ \\
 4 & $\left(\overline{\boldsymbol{3}},\boldsymbol{1};\boldsymbol{1},\boldsymbol{1}\right)_{-1/6}$ & $\bar \gvf_i$ & &
 4 & $\left(\boldsymbol{3},\boldsymbol{1};\boldsymbol{1},\boldsymbol{1}\right)_{1/6}$             & $\gvf_i$\\
 2 & $\left(\boldsymbol{1},\boldsymbol{2};\boldsymbol{1},\boldsymbol{2}\right)_{0}$               & $y_i$ & &
 9 & $\left(\boldsymbol{1},\boldsymbol{1};\boldsymbol{8},\boldsymbol{1}\right)_{0}$               & $w_i$ \\
 4 & $\left(\boldsymbol{1},\boldsymbol{2};\boldsymbol{1},\boldsymbol{1}\right)_{0}$               & $m_i$ & &
   &                                                                                              & \\
\cline{1-3}\cline{5-7}
\end{tabular}
\end{center}
\caption{The massless spectrum of the benchmark model 2 contains three generations of quarks and leptons plus vector--like exotics. The representations (irrep.) with respect to $\text{SU}(3)\times\text{SU}(2)\times\text{SO}(8)\times\text{SU}(2)$ are shown, where the hypercharge is given as a subscript.
\label{bm2spectrum}}
\end{table}

\subsection{Example: Benchmark model 2}
\label{sc:benchtwo}

The so--called ``benchmark model 2'' \cite{Lebedev:2006kn,Buchmuller:2007zd,Lebedev:2007hv} is defined by the shift $V$ and two non--trivial Wilson lines $W_3$ and $W_2$, i.e.
\begin{subequations}
\begin{eqnarray}
V & =&
\Big(~\frac 13, \sm\frac 12, \sm\frac 12, ~~0^2,~ 0^3\Big)
\Big(~~0,\,\sm\frac 23,\,~0^2,~0^3, 1\Big)~, 
%\left(\tfrac{1}{3},\,-\tfrac{1}{2},\,-\tfrac{1}{2},\,0,\,0,\,0,\,0,\,0\right)\left(\tfrac{1}{2},\,-\tfrac{1}{6},\,-\tfrac{1}{2},\,-\tfrac{1}{2},\,-\tfrac{1}{2},\,-\tfrac{1}{2},\,-\tfrac{1}{2},\,\tfrac{1}{2}\right)\;,
\\
W_2 & =& 
\Big(~\frac 14, \sm\frac 14, \sm\frac 14, \sm\frac 14^2,\,\frac 14^3\Big)
\Big(\,\sm\frac 32,~\frac 12,\,~0^2,~0^3, 0\Big)~, 
%\left(-\tfrac{1}{2},\,-\tfrac{1}{2},\,\tfrac{1}{6},\tfrac{1}{6},\,\tfrac{1}{6},\,\tfrac{1}{6},\tfrac{1}{6},\,\tfrac{1}{6}\right)\left(\tfrac{10}{3},\,0,\,-6,\,-\tfrac{7}{3},\,-\tfrac{4}{3},\,-5,\,-3,\,3\right)\;,
\\
W_3 & =&
\Big(\sm\frac 12,\sm \frac 12, ~~\frac 16, ~\,\frac 16^2,\frac 16^3\Big)
\Big(~~\!\frac 43,~0,\sm \frac 13^2,~\!0^3,0\Big)~. 
%\left(\tfrac{1}{4},\,-\tfrac{1}{4},\,-\tfrac{1}{4},-\tfrac{1}{4},\,-\tfrac{1}{4},\,\tfrac{1}{4},\tfrac{1}{4},\,\tfrac{1}{4}\right)\left(1,\,-1,\,-\tfrac{5}{2},\,-\tfrac{3}{2},\,-\tfrac{1}{2},\,-\tfrac{5}{2},\,-\tfrac{3}{2},\,\tfrac{3}{2}\right)\;,
\end{eqnarray}
\label{eq:DataBM2}
\end{subequations}
and the Wilson line $W_2'$ corresponding to the $e_6$ direction is set to zero, $W_2'=0$ \footnote{The shift and the Wilson lines are given here in a different, but equivalent form compared to~\cite{Lebedev:2006kn}}. These vectors satisfy the modular invariance conditions~(\ref{eq:znmodularinv}), (\ref{eq:newmodularinv}). 
The gauge group of the four dimensional theory is
\begin{equation}
G ~=~ G' \times G'' \;\text{ where }\; G'~=~\text{SU}(3)\times\text{SU}(2)\times\text{U}(1)^5 \;\text{ and }\; G''~=~\text{SO}(8)\times\text{SU}(2)\times\text{U}(1)^3\;.
\end{equation}
$G'$ and $G''$ originate from the first and second $\text{E}_8$, respectively. A $\text{U}(1)_Y$ hypercharge generator can be defined by
\begin{equation}
Y~=~\Big(0,0,0,\frac{1}{2}^2,\sm\frac{1}{3}^3\Big)\Big(0,0,0^2,0^4\Big)\;,
\label{eq:Hypercharge} 
\end{equation}
such that the observable sector $G'$ only contains the Standard Model gauge group times some $\text{U}(1)$ factors, while the hidden sector $G''$ contains further non--Abelian gauge factors. 

The massless matter spectrum is given in Table~\ref{bm2spectrum}. It contains three generations of quarks and leptons plus vector--like exotics. It turns out that one $\text{U}(1)$, generated by
\begin{equation}
t_\text{anom} ~=~ \Big(\sm\frac{7}{3},1,\frac{5}{3},\sm\frac{1}{3}^2,\sm\frac{1}{3}^3\Big)\Big(\sm\frac{2}{3},\frac{2}{3},\frac{2}{3}^{2},0^4\Big)\;,
\label{eq:anomalousCharge}
\end{equation}
is anomalous with $\text{Tr}\,Q_\text{anom} = 416/3$. Obviously, the generator $t_\text{anom}$ mixes hidden and observable sectors. However, the hypercharge is non--anomalous because its generator is orthogonal to the anomalous one, i.e. $Y\cdot t_\text{anom}=0$. Furthermore, as expected, the anomaly fulfills the universality condition~(\ref{eq:Anomaly_conditions}) and consequently can be canceled by the Green--Schwarz mechanism.

Finally, we briefly review the conditions for a supersymmetric vacuum of the benchmark model 2. Due to the anomalous $\text{U}(1)$, the corresponding D--term contains the so--called Fayet--Iliopoulos (FI) term, i.e.
\begin{equation}
D_\text{anom}~\sim~ \sum_\phi Q_\text{anom}^\phi |\phi|^2 + \xi\; \quad\text{with}\quad\xi=\frac{M_s^2 \text{Tr}\,Q_\text{anom}}{192\pi^2}\approx 0.1 M_s^2\;.
\end{equation}
Thus, a supersymmetric vacuum with $D=0$ forces some fields (with negative anomalous $\text{U}(1)$ charge $Q_\text{anom}^\phi <0$) to obtain VEVs. In \cite{Lebedev:2007hv} it is shown that there are non--trivial solutions in which the Standard Model gauge group is left unbroken while all additional $\text{U}(1)$ factors are broken and, furthermore, in which the vector--like exotics get massive and decouple from the low energy effective theory. In these configurations there are some fixed points where more than one twisted state acquires a VEV. In addition, there are also fixed points where no twisted state has a non--trivial VEV, e.g. the fixed point in the $\theta$--sector with $\beta=1$ and $\gamma=2$.

\section{Resolutions of $\boldsymbol{T^6/\Intr_\text{6--II}}$}
\label{sc:resZsix}

%intro 
Since it is crucial for the derivation of the main results of this paper, we want to give a comprehensive review of the techniques needed to resolve compact orbifolds. This is mainly based on \cite{Reffert:2007im,Lust:2006zh,Reffert:2006du,Nibbelink:2007xq, Nibbelink:2007pn}. Mathematical fundamentals can be found in \cite{Nakahara03, Griffiths78,Fulton93}.

Before going into details, we want to outline the general strategy. The main step is to subdivide the problem of resolving a compact orbifold into the easier problem of resolving several non--compact orbifolds. This is done by considering every fixed point separately in the sense that it is ``far away'' from other fixed points and can be locally considered as the fixed point of a non--compact orbifold. Then one can identify the group of this orbifold, which is a subgroup of the group acting in the compact case. This provides all the information needed to resolve the singularities locally.

To obtain the resolution of the compact orbifold, one has to combine the local information in a proper way. This procedure is referred to as ``gluing'' and can be achieved by considering global information coming from the torus $T^6$. The final result of this procedure will be topological informations about the resolved orbifold, which is needed in later computations.

\subsection{Local resolutions}
\label{sc:localres}

First we determine which subgroup of $\mathbb{Z}_\text{6--II}$ acts on which kind of fixed objects. As was stated in Section~\ref{sc:orbigeom} one obtains 12 fixed points under the full action of $\mathbb{Z}_\text{6--II}$ with the labels $\left(\alpha=1,\beta,\gamma\right)$ where $\beta$ runs from $1$ to $3$ and $\gamma$ form $1$ to $4$ (compare also with Figure~\ref{theta1}). Furthermore, there are 6 independent $\mathbb{Z}_\text{3}$ fixed lines out of which 3 are simply fixed lines ($\alpha=1,\beta=1,2,3$) and 3 are the combination of two equivalent fixed lines ($\alpha=3,\beta=1,2,3$; the fixed lines denoted by $\alpha=3$ and $\alpha=5$ in Figure~\ref{theta2} are identified on the orbifold). At last there are 8 independent $\mathbb{Z}_\text{2}$ fixed lines that are subdivided in a similar way: the ones with $\alpha=1$ are just fixed lines and the ones with $\alpha=2$ are a combination of the three equivalent lines that are denoted by $\alpha=2$, $4$, and $6$ in Figure~\ref{theta3}. Therefore we obtain locally three different types of orbifolds that we have to resolve: $\mathbb{C}^3/\mathbb{Z}_\text{6--II}$ for the $\mathbb{Z}_\text{6--II}$ fixed points, $\mathbb{C}^2/\mathbb{Z}_\text{3}$ for the $\mathbb{Z}_\text{3}$ fixed lines and $\mathbb{C}^2/\mathbb{Z}_\text{2}$ for the $\mathbb{Z}_\text{2}$ fixed lines.

How to resolve non--compact orbifolds is a well--known problem in toric geometry. A mathematical introduction to toric geometry is given in \cite{Fulton93}. The orbifold case is covered in \cite{Lust:2006zh,Nibbelink:2007xq, Nibbelink:2007pn}. The main tool in the resolving procedure is the toric diagram of the orbifold, which is constructed in the following way. The orbifold group $\mathbb{Z}_\text{N}$ acts in the d-dimensional complex space $\mathbb{C}^d$ like
\begin{equation}
\theta: \quad \left(z_1,\ldots,z_d\right)~\mapsto~ 
\left(e^{2\pi i \gvf_1} z_1,\ldots, e^{2\pi i \gvf_d} z_d\right).
\end{equation}
We can define $\theta$-invariant monomials $u_j=z_1^{\left(v_1\right)_j} \cdots z_d^{\left(v_d\right)_j}$ ($j=1,\ldots,d$) by fixing a condition on the vectors $v_i$:
\begin{equation}
v_1 \gvf_1 + \ldots +v_d \gvf_d ~=~ 0 \quad\text{mod }1\;.
\label{eq:vectconst}
\end{equation}
From the Calabi--Yau condition (\ref{eq:sumv=0}) one knows that $\gvf_1+\ldots+\gvf_d =0\;\text{mod}\; 1$. Due to this, we can choose the last component of every vector $v_i$ to be equal to $1$, which means that the endpoints of all vectors $v_i$ lie in a plane. The toric diagram of the orbifold is obtained by connecting all those points.

A further statement of toric geometry is that every such vector $v_i$ can be associated with a codimension one hypersurface denoted by $D_i$. These hypersurfaces are called ordinary divisors. Since for each divisor there exists a holomorphic scalar transition function on the orbifold, a holomorphic line bundle can be associated to each divisor, whose first Chern class gives the Poincare dual form of the cycle $D_i$. For a holomorphic line bundle this will be a $\left(1,1\right)$--form. In what follows, the cycle as well as the form is denoted by $D_i$, since the context should make clear which object is meant.

\begin{figure}[t]
\centering
\subfloat[Toric diagram of $\mathbb{C}^2/\mathbb{Z}_\text{2}$. One exceptional divisor is needed for the resolution.]
{\includegraphics[height=2.5cm]{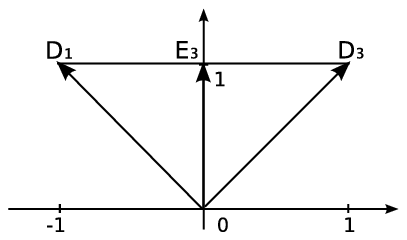}\label{subfig1:toricdiag}}
\hfill
\subfloat[Toric diagram of $\mathbb{C}^2/\mathbb{Z}_\text{3}$. Two exceptional divisors are needed for the resolution.]
{\includegraphics[height=2.5cm]{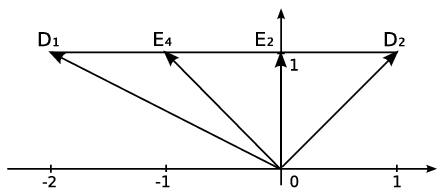}\label{subfig2:toricdiag}}
\hfill
\subfloat[Projection of the toric diagram of $\mathbb{C}^3/\mathbb{Z}_\text{6--II}$. Four exceptional divisors are needed for the resolution.]
{\includegraphics[height=3.5cm]{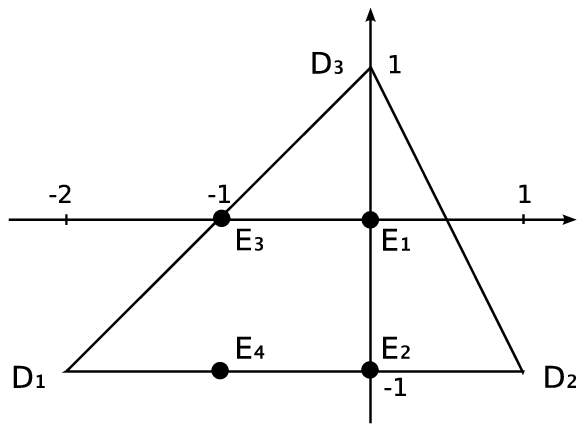}\label{subfig3:toricdiag}}
\caption{The toric diagrams of the orbifolds $\mathbb{C}^2/\mathbb{Z}_\text{2}$, $\mathbb{C}^2/\mathbb{Z}_\text{3}$ and $\mathbb{C}^3/\mathbb{Z}_\text{6--II}$. For the $\mathbb{C}^2/\mathbb{Z}_\text{N}$ orbifolds also the vectors corresponding to divisors are shown.}
\label{toricdiag}
\end{figure}

To resolve the orbifold one introduces a new class of divisors, called exceptional divisors $E_k$. In principle one has to introduce one exceptional divisor for every non--trivial twist $\theta^k\neq \mathbb{1} $. This is the case for $\mathbb{C}^2/\mathbb{Z}_\text{N}$ orbifolds. In the toric diagram (which is a line in this case) the exceptional divisors are placed in such a way that the distances between two divisors are distributed equally. For $\mathbb{C}^3/\mathbb{Z}_\text{N}$ orbifolds a more thorough examination yields the following condition for exceptional divisors, as described in \cite{Aspinwall:1994ev}:\\
If the twist in the $k$--th sector acts like
\begin{equation}
\theta^k: \quad \left(z_1,z_2,z_3\right) ~\mapsto~ \left(e^{2 \pi i g_1} z_1,e^{2 \pi i g_2} z_2, e^{2 \pi i g_3} z_3\right), \quad k~=~1,\ldots,N-1~,
\end{equation}
an exceptional divisor $E_k$ will be placed in the toric diagram at
\begin{equation}
w_k~=~g_1 v_1 + g_2 v_2 + g_3 v_3,\quad \text{if}\quad \sum\limits_{i=1}^3{g_i}~=~1, \text{  and  } 0\leq g_i < 1~.
\label{erest}
\end{equation}
The toric diagrams of the resolved orbifolds $\mathbb{C}^2/\mathbb{Z}_\text{2}$, $\mathbb{C}^2/\mathbb{Z}_\text{3}$ and $\mathbb{C}^3/\mathbb{Z}_\text{6--II}$ are shown in Figure~\ref{toricdiag}. For the $\mathbb{C}^2/\mathbb{Z}_\text{N}$ orbifolds the toric diagram is the line that connects the endpoints of the vectors. There is one exceptional divisor for the $\mathbb{Z}_\text{2}$ orbifold, two for $\mathbb{Z}_\text{3}$ and four for $\mathbb{Z}_\text{6--II}$. The divisors of the $\mathbb{C}^2/\mathbb{Z}_\text{N}$ orbifolds are named in a way convenient for the gluing procedure.

The toric diagram is also encoding equivalences up to cohomology for the divisors. 
Considering for a moment the singular case (i.e. neglecting the exceptional divisors), one can construct invariant monomials from the vectors of the toric diagram:  $u_j=\prod_{i=1}^d{z_i^{\left(v_i\right)_j}}$ is invariant under the action of $\theta$ (where the $i$--th coordinate $z_i$ is associated with the $i$--th vector $v_i$ and corresponding divisor $D_i$). Then it can be shown that the $D_i$'s fulfill the equivalence relation $\sum_{i}{\left(v_i\right)_j D_i} ~\sim~ 0\;$,
where the equivalence becomes an equality if the forms are integrated over a closed boundary. Due to Poincare duality this equivalence up to cohomology of the forms $D_i$ can be turned into an equivalence up to homology of the cycles $D_i$.
This linear equation is modified once the singularity is resolved, since one has to include the exceptional divisors in the invariant monomials. This is done by associating a coordinate $y_r$ to every $E_r$ and introducing a new equivalence relation. Then one can read off the relations between the divisors:
\begin{equation}
u_j~=~\prod\limits_{i,r}{z_i^{\left(v_i\right)_j} y_r^{\left(w_r\right)_j}}\qquad
\Rightarrow\qquad \sum\limits_{i}{\left(v_i\right)_j D_i} + \sum\limits_{r}{\left(w_r\right)_j E_r} ~\sim~ 0~.
\end{equation}
Following this procedure and bringing the relations in such a form that there is only one $D_i$ per relation one obtains from Figure~\ref{toricdiag}:

\begin{equation}
\begin{array}{llccll}
\mathbb{C}^2/\mathbb{Z}_\text{2}: & 2D_1+E_3~\sim~0~, &&& \mathbb{C}^3/\mathbb{Z}_\text{6--II}: & 6D_1+E_1+2E_2+3E_3+4E_4~\sim~0~,\\
                           & 2D_2+E_3~\sim~0~, & &&& 3D_2+E_1+2E_2+E_4~\sim~0~, \\
\mathbb{C}^2/\mathbb{Z}_\text{3}: & 3D_1+E_2+2E_4~\sim~0~, & &&& 2D_3+E_1+E_3~\sim~ 0~.\\
                           & 3D_2+2E_2+E_4~\sim~0~, & &
\end{array}
\label{eq:localleq}
\end{equation}
The main topological information are the intersection numbers of the divisors. Here intersection has a twofold meaning: As long as at least one divisor (or the intersection of two divisors) is compact as a hypersurface, the term can be taken literally. If this is not the case, intersection is not well defined. But via Poincare duality all divisors can be turned into the corresponding forms, so in that case intersection means the integral over all involved divisors considered as forms.

\begin{figure}[t]
\centering
\includegraphics[width=0.7\textwidth]{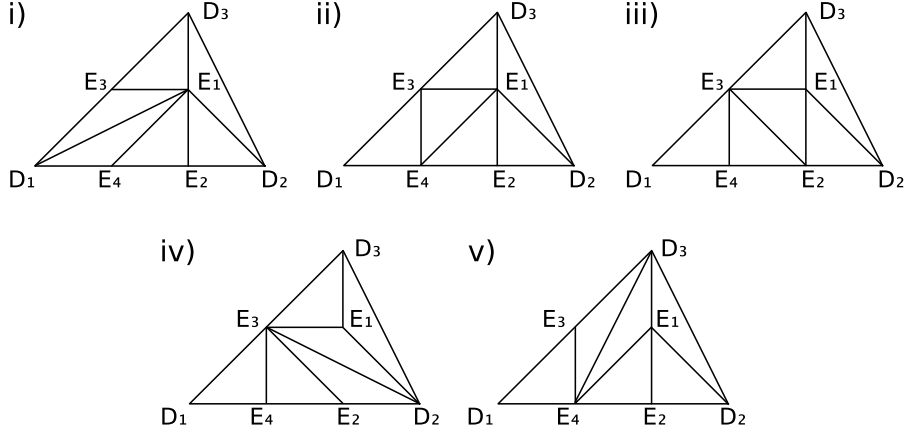}
\caption{The five possible triangulations of the resolved toric diagram of $\mathbb{C}^3/\mathbb{Z}_\text{6--II}$.}
\label{5triang}
\end{figure}

One uses the toric diagram to obtain the intersection numbers. But before one can do so, one has to specify the relative position of all divisors. This is done by triangulating the toric diagram, i.e. by connecting all divisors in the toric diagram with lines in such a way that no lines cross and that no further lines could be added without crossing one another. For $\mathbb{C}^2/\mathbb{Z}_\text{N}$ and some three dimensional orbifolds this is unambiguous. However, in general there are several triangulations possible for higher dimensional orbifolds. Since the toric diagrams of $\mathbb{C}^2/\mathbb{Z}_\text{N}$ orbifolds are just lines, the triangulations of $\mathbb{C}^2/\mathbb{Z}_\text{2}$ and $\mathbb{C}^2/\mathbb{Z}_\text{3}$ are already shown in Figure~\ref{subfig1:toricdiag} and Figure~\ref{subfig2:toricdiag}, respectively. For $\mathbb{C}^3/\mathbb{Z}_\text{6--II}$ one obtains five different triangulations shown in Figure~\ref{5triang}.

The intersection numbers of distinct divisors can be read off from the toric diagram. For two dimensional orbifolds the intersection number of two adjacent divisors is $1$, while  the intersection number of two divisors separated by a third one is $0$. Similarly, for three dimensional orbifolds the intersection of three distinct divisors is $1$ if they lie on the corners of a basic triangle of the triangulation and $0$ if they do not.

The first triangulation of $\mathbb{C}^3/\mathbb{Z}_\text{6--II}$ for example gives as the only non--vanishing intersection numbers with three distinct divisors
\begin{equation}
D_1E_1E_4~=~1 ~,~~ E_1E_2E_4~=~1 ~,~~ D_2E_1E_2~=~1 ~,~~
D_2D_3E_1~=~1 ~,~~ D_3E_1E_3~=~1 ~,~~ D_1E_1E_3~=~1~.
\end{equation}
All other intersection numbers, in particular those containing self--intersections, can be obtained from the intersection of distinct divisors and the linear equivalence relations. For the same example (triangulation i) of $\mathbb{C}^3/\mathbb{Z}_\text{6--II}$) we find: \begin{equation}
\begin{array}{ll} 6D_1+E_1+2E_2+3E_3+4E_4 ~\sim~ 0 & |\cdot D_1E_4 \\ 3D_2+E_1+2E_2+E_4 ~\sim~ 0 & |\cdot D_1E_4 \end{array}
\quad \Rightarrow\quad 
\begin{array}{l} 6D_1^2E_4+1+4D_1E_4^2 ~\sim~ 0 \\ 1+D_1E_4^2 ~\sim~ 0\end{array}
\end{equation}
implying that $D_1^2E_4=1/2$ and $D_1E_4^2=-1$. In a similar way all other self--intersection numbers can be calculated. Therefore we have obtained all the local information that we need and can go on to the gluing procedure.

\subsection{Gluing together the local resolutions}
\label{sc:gluing}

We consider now, how to bring the local information we obtained in the previous Section together in order to characterize the properties of the compact orbifold $T^6/\mathbb{Z}_\text{6--II}$. In our description of this gluing process we follow closely \cite{Lust:2006zh}.

First we determine the total number of divisors of the compact resolution, starting with the ordinary divisors. In the non--compact case one has three ordinary divisors $D_1$, $D_2$ and $D_3$ for each fixed point of a three dimensional orbifold and two for each two dimensional one. From our local information we would expect $12\times3+6\times2+8\times2=64$ ordinary divisors in the compact case. But one has to be careful in order not to overcount. Every ordinary divisor corresponds to one coordinate  of a fixed point. Fixed points which have the same location in one coordinate will thus have the same ordinary divisor for this coordinate. Hence for finding the right number of ordinary divisors one has to count the different locations of fixed points on the tori. As one can see from Figure~\ref{theta1}--\ref{theta3} there are six different locations of fixed points on the first torus ($\alpha=1,\ldots,6$), three different locations on the second torus ($\beta=1,2,3$) and four different locations on the last one ($\gamma=1,\ldots,4$). The corresponding ordinary divisors are denoted by
\begin{equation}
\widetilde{D}_{1,\alpha}~,~~ \alpha~=~1,\ldots,6~; \quad \widetilde{D}_{2,\beta}~,~~ \beta~=1,2,3~; \quad \widetilde{D}_{3,\gamma}~,~~ \gamma~=~1,\ldots,4~.
\end{equation}
But these are divisors on the cover of the orbifold which in particular means that the divisors with $\alpha=3,5$ and $\alpha=2,4,6$ are mapped into each other. In order to obtain invariant objects on the orbifold one has to build invariant combinations out of them. They are given in the first column of Table~\ref{divtable}. After this analysis we conclude that there are ten ordinary divisors for $T^6/\mathbb{Z}_\text{6--II}$.
\begin{table}[t]
\centering
\begin{tabular}{|l||l|}
\hline
ordinary divisors &exceptional divisors\\ 
\hline
\hline
$D_{1,1}~=~\widetilde{D}_{1,1}$ & $E_{1,\beta \gamma}~=~\widetilde{E}_{1,1 \beta \gamma}$ \\
$D_{1,2}~=~\widetilde{D}_{1,2}+\widetilde{D}_{1,4}+\widetilde{D}_{1,6}$ & $E_{2,1 \beta}~=~\widetilde{E}_{2,1 \beta}$ \\
$D_{1,3}~=~\widetilde{D}_{1,3}+\widetilde{D}_{1,5}$ & $E_{3,1 \gamma}~=~\widetilde{E}_{3,1 \gamma}$ \\
$D_{2,\beta}~=~\widetilde{D}_{2,\beta}$ & $E_{4,1 \beta}~=~\widetilde{E}_{4,1 \beta}$ \\
$D_{3,\gamma}~=~\widetilde{D}_{3,\gamma}$ & $E_{2,3 \beta}~=~\widetilde{E}_{2,3 \beta}+\widetilde{E}_{2,5 \beta}$ \\
 & $E_{4,3 \beta}~=~\widetilde{E}_{4,3 \beta}+\widetilde{E}_{4,5 \beta}$\\
 & $E_{3,2 \gamma}~=~\widetilde{E}_{3,2\gamma}+\widetilde{E}_{3,4\gamma}+\widetilde{E}_{3,6\gamma}$\\
\hline
\end{tabular}
\caption{Ordinary and exceptional divisors of $T^6/\mathbb{Z}_\text{6--II}$. $\beta$ runs from $1$ to $3$, $\gamma$ from $1$ to $4$.}
\label{divtable}
\end{table}

Now, we turn to the number of exceptional divisors. Since we know from the previous Section that we get four exceptional divisors for every local $\mathbb{C}^3/\mathbb{Z}_\text{6--II}$ orbifold, two for every $\mathbb{C}^2/\mathbb{Z}_\text{3}$ and one for every $\mathbb{C}^2/\mathbb{Z}_\text{2}$ and since we know which fixed objects belong to which local orbifold, we would expect to get $12\times4+9\times2+16\times 1= 82$ exceptional divisors. But again one is overcounting in this simple estimate. To see what is going wrong one has to consider the twisted sectors separately. From (\ref{erest}) we know that there is one exceptional divisor $E_k$ of $\mathbb{C}^3/\mathbb{Z}_\text{6--II}$ per sector $\theta^k$ ($k=1,\ldots,4$). For the compact case, we denote them in general by $\widetilde{E}_{k,\alpha\beta\gamma}$. Since we have twelve fixed points of $\mathbb{Z}_\text{6--II}$ in the first sector, we obtain twelve divisors $\widetilde{E}_{1,1\,\beta\gamma}$. In the other sectors, only the fixed lines with $\alpha=1$ are fixed under the $\mathbb{Z}_\text{6--II}$ action. Hence we obtain the divisors $\widetilde{E}_{2,1\,\beta}$, $\widetilde{E}_{3,1\,\gamma}$ and $\widetilde{E}_{4,1\,\beta}$ (the missing label $\beta$ or $\gamma$ is due to the fact that labeling in the invariant tori is not possible). Note that for a particular choice of $\beta$ and $\gamma$ one obtains exactly four exceptional divisors for each $\mathbb{C}^3/\mathbb{Z}_\text{6--II}$ singularity, as expected from the local analysis. Next, we consider the exceptional divisors from $\mathbb{C}^2/\mathbb{Z}_\text{3}$. The $\mathbb{Z}_\text{3}$ action is only present in the $\theta^2$ and $\theta^4$ sectors. Since the fixed lines with $\alpha=1$ have already been taken into account there remain only those with $\alpha=3$ or $5$: $\widetilde{E}_{2,3\,\beta}$, $\widetilde{E}_{2,5\,\beta}$, $\widetilde{E}_{4,3\,\beta}$ and $\widetilde{E}_{4,5\,\beta}$. After building invariant linear combinations, this gives for a specific choice of $\alpha$ and $\beta$ the two exceptional divisors of the $\mathbb{C}^2/\mathbb{Z}_\text{3}$ singularity. A similar analysis gives the divisors $\widetilde{E}_{3,2\,\gamma}$, $\widetilde{E}_{3,4\,\gamma}$ and $\widetilde{E}_{3,6\,\gamma}$ as the ones belonging to $\mathbb{C}^2/\mathbb{Z}_\text{2}$. As in the case of the ordinary divisors one has to build combinations of the tilded divisors that are invariant under the orbifold action. These are also shown in Table~\ref{divtable}.

From this examination we see that we have twelve $\mathbb{C}^3/\mathbb{Z}_\text{6--II}$ singularities, giving twelve exceptional divisors from the first sector, three from the second, three from the fourth and four from the third (22 divisors in total). Furthermore, we obtain three $\mathbb{C}^2/\mathbb{Z}_\text{3}$ singularities giving six exceptional divisors and four $\mathbb{C}^2/\mathbb{Z}_\text{2}$ ones giving four divisors. So we see that the total number of exceptional divisors is 32. The identification of the exceptional divisors of fixed lines with the exceptional divisors corresponding to higher twisted sectors of $\mathbb{C}^3/\mathbb{Z}_\text{6--II}$ is the first step of gluing together the non--compact orbifolds. Such an identification takes place each time a fixed point is contained in a fixed torus.

The next step in the gluing procedure is to include explicit information of the six--dimensional torus. On the torus a basis of $\left(1,1\right)$--forms is given by $\d z_i \d\overline{z}_{\overline{j}}$, where a wedge product is understood. Under an orbifold twist this object transforms like $\exp\left[2 \pi i\left(\gvf_i-\gvf_j\right)\right]$. In the $T^6/\mathbb{Z}_\text{6--II}$ case these forms are only invariant and hence well defined on the orbifold if $i=j$. We define the divisors $R_i$ to be the cycles dual to $\d z_i \d\overline{z}_{\overline{i}}$. These divisors are called ``inherited'' divisors because they descend from the torus to the orbifold. Since the forms are well defined on the whole manifold, the $R$'s are also well defined.

On the orbifold there is an equivalence relation between ordinary divisors $D_i$ and inherited divisors $R_i$ (see e.g.\cite{Lust:2006zh}): $R_i ~\sim~ N_i D_{i,\delta}$, where $N_i$ is the order of the group in the $i$--th torus and $\delta$ is the corresponding label for this torus (either $\alpha$, $\beta$ or $\gamma$). From these relations one can obtain the linear equivalence relations on the resolved orbifold by including the relations of the non--compact cases. In order to achieve this, one has to specify one ordinary divisor $D_{i,\delta}$, find all local resolutions involving this divisor, and sum the $E$--part of the associated local equivalence relations. To see how this works in detail we will give the procedure explicitly for $D_{2,1}$ and $D_{1,3}$ and state thereafter all relations for $T^6/\mathbb{Z}_\text{6--II}$.

$D_{2,1}$ belongs locally to the four $\mathbb{C}^3/\mathbb{Z}_\text{6--II}$ orbifolds with $\beta=1$ and also to one $\mathbb{C}^2/\mathbb{Z}_\text{3}$ orbifold (again with $\beta=1$). Since $N_2=3$, one obtains from (\ref{eq:localleq})
\begin{equation}
R_2~\sim~3D_{2,1}+\sum\limits_{\gamma=1}^4{E_{1,1\gamma}}+\sum\limits_{\alpha=1,3}{\left(2E_{2,\alpha\,1}+E_{4,\alpha\,1}\right)}~.
\end{equation}
Note that the sum over $\gamma$ is taken over $E_1$ only since it is the only divisor involved that depends on $\gamma$.

$D_{1,3}$ locally belongs to the three $\mathbb{C}^2/\mathbb{Z}_\text{3}$ orbifolds only. A further subtlety arises here since $D_{1,3}$ is the sum of $\widetilde{D}_{1,3}$ and $\widetilde{D}_{1,5}$ (Table~\ref{divtable}). In such a case one has to divide the group order by the number of elements the divisor is built of. Since $N_1=6$ and $D_{1,3}$ is built out of two elements, one obtains
$R_1~\sim~3D_{1,3}+\sum_\beta (E_{2,3\,\beta}+2E_{4,3\,\beta})$. 
Proceeding in this way it is possible to obtain all linear equivalence relations for the resolution of $T^6/\mathbb{Z}_\text{6--II}$
\begin{align}
\label{eq:leq}
&R_1~\sim~ 6D_{1,1}+\sum\limits_{\beta=1}^3\sum\limits_{\gamma=1}^4{E_{1,\beta\gamma}}+\sum\limits_{\beta=1}^3{\left(2E_{2,1\,\beta}+4E_{4,1\,\beta}\right)}+3\sum\limits_{\gamma=1}^4{E_{3,1\,\gamma}}~, \nonumber \\
&R_1~\sim~2D_{1,2}+\sum\limits_{\gamma=1}^4{E_{3,2\,\gamma}}~,\qquad %\nonumber\\ &
R_1~\sim~3D_{1,3}+\sum\limits_{\beta=1}^3{\left(E_{2,3\,\beta}+2E_{4,3\,\beta}\right)}~,\\
&R_2~\sim~3D_{2,\beta}+\sum\limits_{\gamma=1}^4{E_{1,\beta\gamma}}+\sum\limits_{\alpha=1,3}{\left(2E_{2,\alpha\beta}+E_{4,\alpha\beta}\right)}~, & \beta=1,2,3~,\nonumber\\
&R_3~\sim~2D_{3,\gamma}+\sum\limits_{\beta=1}^3{E_{1,\beta\gamma}}+\sum\limits_{\alpha=1,2}{E_{3,\alpha\gamma}}~, & \gamma=1,2,3,4~. \nonumber
\end{align}
These relations can be seen as the outcome of the gluing procedure since on the one hand we combined several local equivalence relations into one relation and on the other hand they are related to the inherited divisors, which represent the global properties of the torus. Furthermore, if one specifies one fixed point (i.e. $\alpha$, $\beta$, and $\gamma$) and sets all divisors with different labels to zero, one obtains exactly the local equivalence relation associated with that fixed point. This can be seen as a cross check that the gluing procedure respects the properties of the local resolutions. Finally (\ref{eq:leq}) does not depend on the triangulation of the $\mathbb{C}^3/\mathbb{Z}_\text{6--II}$ orbifolds, which will play a role when we consider the intersection numbers of the resolution of $T^6/\mathbb{Z}_\text{6--II}$.

\begin{figure}[t]
\centering
\subfloat[\label{poly1}\footnotesize{The polyhedron for $\mathbb{C}^3/\mathbb{Z}_\text{6--II}$ ($\alpha~=~1$) for triangulation i). Each of the twelve possible polyhedra can have a different triangulation.}]{\includegraphics[width=0.3\textwidth]{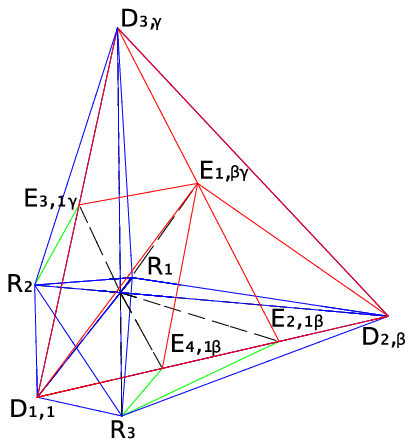}}
\hfill
\subfloat[\label{poly3}\footnotesize{The polyhedron for $\mathbb{C}^2/\mathbb{Z}_\text{3}$ ($\alpha~=~3$). There are three polyhedra of this type. They all have the same triangulation.}]{\includegraphics[width=0.32\textwidth]{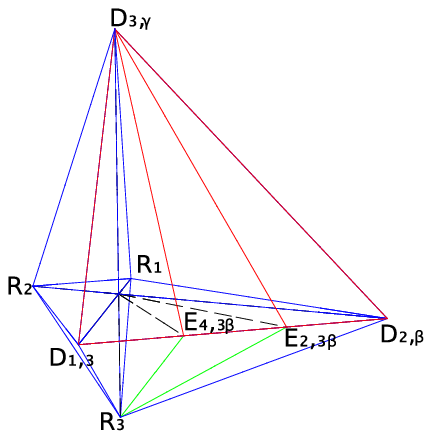}}
\hfill
\subfloat[\label{poly2}\footnotesize{The polyhedron for $\mathbb{C}^2/\mathbb{Z}_\text{2}$ ($\alpha~=~2$). There are four polyhedra of this type. They all have the same triangulation.}]{\includegraphics[width=0.3\textwidth]{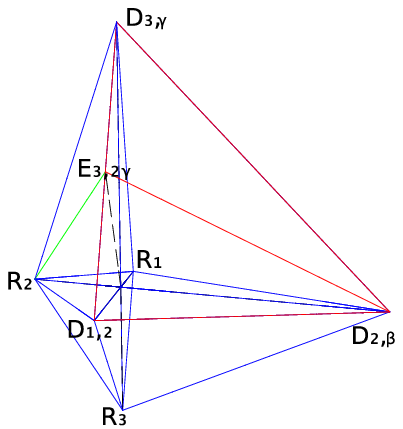}}
\label{polyhedra}
\caption{The auxiliary polyhedra for $T^6/\mathbb{Z}_\text{6--II}$.}
\end{figure}

As in Section~\ref{sc:localres}, after having obtained the linear equivalence relations, we turn to the intersection properties of the compact orbifold. Again, we use information of the local resolutions together with the globally defined inherited divisors $R_i$ to obtain the intersection ring. A very useful method introduced in \cite{Lust:2006zh} is to construct an auxiliary polyhedron for every local non--compact orbifold one has to consider. This is done in accordance with the following rules:
\begin{enumerate}
\item
Take a lattice $N\cong\mathbb{Z}^3$ with basis $f_i=m_ie_i$, $e_i$ being the standard basis vectors and $m_i>0$ such that $m_1m_2m_3=N_1N_2N_3/\left|G\right|$, where $N_i$ is the order of the action of the orbifold group $G$ on the $i$--th coordinate--plane and $\left|G\right|$ is the number of elements of $G$.
\item
Rotate and rescale the toric diagram of $\mathbb{C}^3/G$ in such a way that the divisors $D_i$ correspond to vectors $v_{i+3}=N_if_i$. The position of the $E$'s has to be transformed accordingly.
\item
Add vertices at $v_i=-f_i$ for every inherited divisor $R_i$.
\item
For every strict subgroup $H\subset G$ with action $\mathbb{C}^2/H$ take a second polyhedron which is identical to the original one except that all exceptional divisors which do not appear in $\mathbb{C}^2/H$ are removed. Differently stated, if $z_i$ is invariant, only divisors opposite to $D_i$ are not removed.
\item
Take one polyhedron for each local resolution in such a way that the triangulated toric diagram on which the polyhedron is based is the same as the one used for the resolution, and label $D$'s and $E$'s of the polyhedron accordingly.
\item
For divisors $D_i$ being the sum of $q_i$ tilded divisors, divide the $i$--th component of every vector $v_k$ ($k\geq4$) by $q_i$.
\item
Take a star triangulation of every polyhedron (i.e. every simplex is spanned by $\left<0,v_i,v_j,v_k\right>$) in such a way that the triangulation of the toric diagram is conserved.
\end{enumerate}
For $T^6/\mathbb{Z}_\text{6--II}$, the resulting polyhedra are shown in Figure~\ref{polyhedra}. There are twelve $\mathbb{C}^3/\mathbb{Z}_\text{6--II}$ polyhedra, three $\mathbb{C}^2/\mathbb{Z}_\text{3}$ polyhedra, and four $\mathbb{C}^2/\mathbb{Z}_\text{2}$ polyhedra according to the local resolutions that are part of the resolution of $T^6/\mathbb{Z}_\text{6--II}$.

The polyhedra of the $\mathbb{C}^3/\mathbb{Z}_\text{6--II}$--type can have five different triangulations since locally every fixed point can be resolved with a different triangulation. As the triangulation of every such polyhedron is important for the intersection numbers of $T^6/\mathbb{Z}_\text{6--II}$, these numbers depend on the triangulations chosen for the separate resolutions. Since our later calculations rely strongly on the intersection numbers they also depend on the chosen triangulations.

After having constructed the polyhedra, the intersection numbers of three distinct divisors can be determined by the following rules. If the three divisors do not span a simplex of a polyhedron, the intersection number is zero. In particular every intersection number containing two divisors which are connected by a line running through the polyhedron is zero. Intersection numbers involving two divisors that are separated by a third one also become zero. And most important: all intersections of divisors belonging to different polyhedra are zero. Divisors that span a simplex of the triangulation have the intersection
\begin{equation}
ABC~=~\frac{N}{\left|det\left(v\left(A\right),v\left(B\right),v\left(C\right)\right)\right|}~,
\label{intform}
\end{equation}
with $v\left(Y\right)$ being the vector representing the divisors $Y$ in the polyhedron and $N$ a normalization constant. This constant has to be chosen such that the intersection numbers of three distinct divisors containing no $R$ are the same as in the non--compact case. This ensures that the local intersection properties of the local resolutions remain unchanged. Employing these rules, one obtains all intersections containing three distinct divisors. These intersection numbers are completely determined by the properties of the local resolution. All self--intersections can be calculated from those by multiplying the linear equivalence relations (\ref{eq:leq}) by all combinations of divisors, applying the above rules, and solving the system of linear equations. In this way it is possible to obtain all intersection numbers for all combinations of triangulations.

\subsection{Resolution overview (triangulation independent)}
\label{sc:resoverview}

We give an overview of some properties of the resolved orbifold $T^6/\mathbb{Z}_\text{6--II}$ (which is denoted by $X=\text{Res}(T^6/\mathbb{Z}_\text{6--II})$), that do not depend on the chosen triangulations. As one can see from the linear equivalence relations (\ref{eq:leq}), all ordinary divisors can be expressed completely in terms of inherited and exceptional divisors. Furthermore, these divisors can be shown to be independent. Since they are $\left(1,1\right)$--forms on the resolved orbifold it is possible to view $R$'s and $E$'s as a basis of the cohomology group $H^{1,1}$. Therefore, the number of divisors gives us the dimension $h^{1,1}=35$ of $H^{1,1}$. It is possible to split $H^{1,1}$ into a part coming from the untwisted sector of the orbifold and a part coming from the twisted sector, since $R$'s and $E$'s correspond to untwisted and twisted sectors, respectively.

To find the bases of the other cohomology groups $H^{p,q}$ with $p+q\leq3$ and $p\geq q$ (all others are connected to those by Poincare duality and the symmetry of the Hodge numbers in $p$ and $q$, see e.g. \cite{Nakahara03}) we start by defining $\left(1,0\right)$--forms $\eta_1,\eta_2,\eta_3$, corresponding to $dz_1,dz_2,dz_3$ in the orbifold limit. These forms transform under a $\mathbb{Z}_\text{6--II}$ twist $\theta$ like
\begin{equation}
\theta(\eta_1,\eta_2,\eta_3)~=~(e^{2\pi i/6}\eta_1,e^{2\pi i/3}\eta_2,e^{-2\pi i/2}\eta_3)~=~(e^{\pi i/3}\eta_1,e^{2\pi i/3}\eta_2,-\eta_3),
\end{equation}
i.e. they are not invariant forms. But it is possible to construct invariant forms from them. Namely the holomorphic volume form $\nu=\eta_1\eta_2\eta_3$ and the $\left(2,1\right)$--form $\omega_0=\eta_1\eta_2\overline{\eta}_3$ (a wedge product is understood here and in what follows). Of course, also the forms $\eta_i\overline{\eta}_i$ are invariant. But as noted in Section~\ref{sc:gluing} these forms just correspond to $R_i$'s. Like the $R$'s, $\nu$ and $\omega_0$ correspond to the untwisted sector. Furthermore, there is also the trivial element of $H^{0,0}$.

If one tries to construct other invariant $\left(p,q\right)$--forms one notes that the only possibilities left are $\left(2,1\right)$--forms involving the twisted sector. To see how to construct them one has to remember that in Section~\ref{sc:gluing} we built invariant combinations of tilded divisors (Table~\ref{divtable}), since these tilded divisors are mapped into each other. However, together with the $\eta_i$ one can now construct ten further invariant $\left(2,1\right)$--forms
\begin{gather}
\arry{c}{\dsp 
\omega_{2,\beta}~=~\left(\widetilde{E}_{2,3\,\beta}-\widetilde{E}_{2,5\,\beta}\right)\eta_3~, \qquad 
\omega_{4,\beta}~=~\left(\widetilde{E}_{4,3\,\beta}-\widetilde{E}_{4,5\,\beta}\right)\eta_3~, \\[2ex] 
\omega_{3,\gamma}~=~\left(\widetilde{E}_{3,2\,\gamma}+e^{2\pi i/3}\widetilde{E}_{3,4\,\gamma}+e^{4\pi i/3}\widetilde{E}_{3,6\,\gamma}\right)\eta_2~.
}
\end{gather}
In this way we have constructed maps from $\left(1,0\right)$--forms on the fixed tori to $\left(2,1\right)$--forms on the resolved orbifold. The existence of those maps was used in \cite{Lust:2006zh} to compute $h^{2,1}$ of the twisted sector. The results given there are consistent with ours. The same Hodge number can also be obtained by using orbifold cohomology directly, which was defined in \cite{Chen:2000cy,fantechi-2001}. Furthermore, the forms constructed in such a way correspond to linear combination of states on the orbifold, given in (\ref{eqn:physicalstate2}), if one sets the phase $\gamma$ equal to $-1/2$ for $\omega_{2,\beta}$ and $\omega_{4,\beta}$ and equal to $-1/3$ for $\omega_{3,\gamma}$. Furthermore, one can calculate the inner products of $\left(2,1\right)$-- and $\left(1,2\right)$--forms, we list here the non--vanishing ones
\begin{equation}
\begin{array}{c} \dsp 
\int_X{\omega_{2,\beta}\overline{\omega}_{2,\beta}}~=~E_{2,3\,\beta}^2R_3~=~-4~,\qquad \int_X{\omega_{4,\beta}\overline{\omega}_{4,\beta}}~=~E_{4,3\,\beta}^2R_3~=~-4~,
\\[2ex] \dsp 
\int_X{\omega_{2,\beta}\overline{\omega}_{4,\beta}}~=~E_{2,3\,\beta}E_{4,3\,\beta}R_3~=~2~,\\[2ex] \dsp 
\int_X{\omega_{3,\gamma}\overline{\omega}_{3,\gamma}}~=~E_{3,2\,\gamma}^2R_2~=~-6~, \qquad
%\int_X{\omega_{2,\beta}\overline{\omega}_{3,\gamma}}~=~\int_X{\omega_{4,\beta}\overline{\omega}_{3,\gamma}}~=~0~.
\int_X \go_0  \bgo_0 = - R_1 R_2 R_3 = -6,
\end{array}
\end{equation}
These (2,1)--forms are not orthogonal, but by a change of basis this can be achieved. 

Since all other combinations of $\eta$'s and $\widetilde{E}$'s are not invariant we have found a basis of the cohomology groups $H^{p,q}$ of $X$. The Hodge diamond is
\begin{equation}
\begin{array}{ccccccc}
 & & & 1 & & &\\
 & & 0 & & 0 & &\\
 & 0 & & 3+32 & & 0 &\\
1 & & 1+10 & & 1+10 & & 1\\
 & 0 & & 3+32 & & 0 &\\
 & & 0 & & 0 & &\\
 & & & 1 & & &
\end{array}\nonumber
\end{equation}
The entries are given in the form $a+b$ where $a$ is the contribution of the untwisted sector and $b$ the contributions of the twisted sectors.

From the hodge numbers it is then possible to obtain the Euler number of the manifold, which is 
\begin{equation}
\chi(X)~=~2(1+35-(1+11))=48~.
\end{equation}
The numbers obtained in this way are consistent with the ones given in \cite{Erler:1992ki} (Table~5; case~7) for the orbifold case. We take this as a further successful crosscheck that the resolving process is smooth and therefore topological quantities are not changed.

\subsubsection*{Chern classes}

Further information that can be obtained independently of the triangulation are the Chern classes of the resolved manifold $X$. First of all, all local resolutions are by construction Calabi--Yau manifolds (see the discussion below (\ref{eq:vectconst})). Since our resolution does not change topological quantities, we expect the compact orbifold to stay Calabi--Yau after the resolution. Therefore the first Chern class $c_1(X)$ vanishes. Secondly the third Chern class $c_3(X)$ is the top Chern class for a three dimensional complex manifold. Therefore the integral of $c_3(X)$ over the manifold equals the Euler number. Finally, it is possible to calculate the integral of the second Chern class $c_2(X)$ over a divisor $S$ by making use of the adjunction formula \cite{Griffiths78}
\begin{equation}
\int_{S}{c_2(X)}~=~c_2(X)S~=~\chi(S)-S^3~.
\end{equation}
Therefore, $c_2(X)S$ can be computed, if one knows the topology of $S$ and the intersection number $S^3$. The topology of $S$ depends on the orbifold under consideration and the divisor. It can be found in \cite{Lust:2006zh}; the intersection number can be calculated using the tools from Section~\ref{sc:gluing}.

Although in this way we can obtain all information needed about the Chern classes, it is useful to note that the same results can be obtained if one uses a slightly modified splitting principle to calculate the total Chern class $c(X)$. Since all divisors are associated to complex line bundles a first guess for the total Chern class, motivated by toric geometry in the non--compact case (see e.g.~\cite{Fulton93}), would be $c(X)=\prod\limits_{\text{all divisors}}{(1+S)}$. However, this does not give $c_1(X)=0$ and $\chi(X)=48$ as expected. We use\footnote{The replacement $R_i\rightarrow -R_i$ is due to the fact that one is free to consider instead of the line bundle over $R_i$ the inverse line bundle, which results in an extra minus sign. Squaring the $R_3$--term takes into account that there are more degrees of freedom in the $z_3$--plane, since the two cycles along $e_5$ and $e_6$ are independent.}
\begin{equation}
c(X)~=~\prod\limits_{J=1}^{10}\prod\limits_{r=1}^{32}{(1+D_J)(1+E_r)}(1-R_1)(1-R_2)(1-R_3)^2~.
\label{eq:spprin}
\end{equation}
This gives the expansion for the Chern classes
\begin{align}
\label{2ndChern}
c_1(X)&~=~\sum\limits_{J=1}^{10}{D_J}+\sum\limits_{r=1}^{32}{E_r}-R_1-R_2-2R_3~=~0~, \nonumber\\
c_2(X)&~=~\frac{1}{2!}\sum\limits_{\text{all divisors}}{(c_1(X)-S_i)S_i}~=~ -\frac{1}{2}\sum\limits_{\text{all divisors}}{S_i^2}~, \\
c_3(X)&~=~\frac{1}{3!}\sum\limits_{\text{all divisors}}{(c_1(X)-S_i-S_j)S_iS_j}~=~ -\frac{1}{6}\sum\limits_{\text{all divisors}}{S_i^2S_j+S_iS_j^2}~.\nonumber
\end{align}
If one now replaces all $D$'s via the relations (\ref{eq:leq}) one obtains $c_1(X)=0$ (as indicated), $\chi(X)=48$ and the right values for the integrals over $c_2(X)$. Using this we can express all integrals over Chern classes as linear combinations of intersection numbers.

\subsubsection*{K\"ahler form $J$}

Since we have a basis of $\left(1,1\right)$--forms we can give the K\"ahler form (see for example \cite{Denef:2008wq,Aspinwall:1994ev}) expanded in $R$'s and $E$'s
\begin{equation}
\label{eq:Jexpand}
J~=~\sum\limits_{i=1}^3{a_iR_i}-\sum\limits_{r=1}^{32}{b_rE_r}~,
\end{equation}
where we have introduced a shorthand for sums involving all exceptional divisors by giving them a multi--index $r$ running from $1$ to $32$.  For $r=1,\ldots,12$ the sum runs over $E_{1,\beta\gamma}$ ($E_1=E_{1,1\,1},E_2=E_{1,1\,2},\ldots$). $r=13,\ldots,r=18$ corresponds to $E_{2,\alpha\beta}$, $r=19,\ldots,r=26$ to $E_{3,\alpha\gamma}$, and $r=27,\ldots,32$ to $E_{4,\alpha\beta}$. The coefficients $a_i$, $b_r$ have to be chosen such that the volumes of any compact curve, any divisor, and the manifold $X$ are all positive. This means that the following integrals have to be positive
\begin{equation}
\text{Vol}(C)~=~\int_{C}{J}~, \qquad 
\text{Vol}(S)~=~\frac{1}{2!}\int_{S}{J^2}~,  \qquad 
\text{Vol}(X)~=~\frac{1}{3!}\int_X{J^3}~.
\label{eq:Kint}
\end{equation}
The restrictions on $a_i,b_r$ by the positivity of the volumes are only valid if the considered manifold does not develop singularities and the  geometry stays ``classical'' in this sense. It has been shown in \cite{Aspinwall:1993xz} that applying a so--called ``algebraic'' measure positive volumes and areas on one Calabi--Yau manifold can become negative on Calabi--Yaus connected to the former by blow down or blowup. In particular in the orbifold limit all $b_r$ become $-\infty$.

\subsection{Triangulation dependence of resolutions}
\label{sc:triangdep}

In Section~\ref{sc:gluing} we have given a method to compute all intersection numbers for a given triangulation. Here we want to examine the intersection numbers with regard to the triangulation dependence they show. Since the linear equivalence relations are equal in all cases, it is possible to extract intersection numbers from intersections containing only $R$'s and $E$'s. Hence we only have to consider intersection numbers of inherited and exceptional divisors\footnote{If the linear equivalence relations would not be the same for every triangulation, it would still be possible to express all intersections in terms of intersections just involving $R$'s and $E$'s. But in this case those numbers would no longer be comparable.}. Secondly, since we started our calculation of intersection numbers with the construction of the auxiliary polyhedra, we can check in which points this construction is equal for different triangulations and hence conclude where the similarities in the intersection properties lie. All the dependence on the triangulations comes from the $\mathbb{C}^3/\mathbb{Z}_\text{6--II}$ polyhedra, since they are constructed from triangulation dependent toric diagrams. Still there are intersection numbers which are independent of the triangulation, namely the ones containing at least one inherited divisor $R_i$. This comes from the fact that they can only be connected by lines with those exceptional divisors that sit on the boundary of the toric diagram. Therefore they do not ``see'' the triangulation, which is an effect of the interior of the toric diagram. So the only intersection numbers that are truly triangulation dependent are those consisting of $E$'s only.

This raises the question of how strong the dependence is, or differently stated: Do the intersection numbers depend on just one triangulation of a certain fixed point, or is there information transferred, connecting several fixed points? To clarify this question, let us first consider intersections involving a certain $E_{1,\beta\gamma}$. Since this divisor lies locally on the position of the fixed point $(\alpha=1,\beta,\gamma)$, its intersections are completely determined by the triangulation chosen for that fixed point. The same is true whenever $\beta$ and $\gamma$ are specified in an intersection number (e.g. in $E_{2,1\,\beta}^2E_{3,1\,\gamma}$).

Therefore the only intersection numbers depending on more than one triangulation are those containing divisors that specify only $\beta$ or $\gamma$ ($E_{a,1\,\beta}E_{b,1\,\beta}^2$, with $a,b\in\lbrace2,4\rbrace$ and $E_{3,1\,\gamma}^3$). The intersection number containing only $\beta$ depend on the triangulation of all fixed points with this $\beta$. Analogously the ones specifying only $\gamma$ depend on all fixed points with that $\gamma$. This brings some structure in the triangulation dependence of the intersection numbers.

Furthermore, we want to give a description of the compact curves lying in the resolved manifold. There are some curves occurring in all triangulations, while the existence of others is triangulation dependent. Since our manifold is compact, the intersection of two divisors (if it exists) is a compact curve (since the divisors are hypersurfaces of complex dimension two, the intersection of two gives a hypersurface of complex dimension one, i.e. a curve). It is possible to read off from the auxiliary polyhedra which divisors can intersect and which cannot, namely all divisors that are connected by a line in a given triangulation intersect. Therefore, one can identify the lines of an auxiliary polyhedron with compact curves of the manifold. After this consideration it is obvious that the only triangulation dependent compact curves are those represented by lines of the toric diagram of $\mathbb{C}^3/\mathbb{Z}_\text{6--II}$. All other curves are triangulation independent. The curves existing in all triangulations are given in Table~\ref{indcc}; Table~\ref{depcc} gives the curves existing only for certain triangulations.
\begin{table}[t]
\centering
\begin{tabular}{llllllll}
$R_1R_2$ & $R_1R_3$ & $R_2R_3$ & $R_1D_{2,\beta}$ & $R_1D_{3,\gamma}$ & $R_2D_{1,1}$ & $R_2D_{3,\gamma}$ & $R_3D_{1,1}$\\
$R_3D_{2,\beta}$ & $R_2E_{3,1\,\gamma}$ & $R_3E_{2,1\,\beta}$ & $R_3E_{4,1\,\beta}$ & $D_{2,\beta}D_{3,\gamma}$ & $D_{2,\beta}E_{2,1\,\beta}$ & $E_{2,1\,\beta}E_{4,1\,\beta}$ & $D_{1,1}E_{4,1\,\beta}$ \\
$D_{1,1}E_{3,1\,\gamma}$ & $D_{3,\gamma}E_{3,1\,\gamma}$ & $D_{1,3}E_{4,3\,\beta}$ & $E_{2,3\,\beta}E_{4,3\,\beta}$ & $D_{2,\beta}E_{2,3\,\beta}$ & $D_{2,\beta}D_{3,\gamma}$ & $D_{1,3}D_{3,\gamma}$ & $R_3E_{4,3\,\beta}$ \\
$R_3E_{2,3\,\beta}$ & $D_{3,\gamma}E_{4,3\,\beta}$ & $D_{3,\gamma}E_{2,3\,\beta}$ & $D_{1,3}D_{2,\beta}$ & $D_{2,\beta}D_{3,\gamma}$ & $D_{3,\gamma}E_{3,2\,\gamma}$ & $D_{1,3}E_{3,2\,\gamma}$ & $R_2E_{3,2\,\gamma}$\\
$D_{2,\beta}E_{3,2\,\gamma}$ & & & & & & &\\
\end{tabular}
\caption{The compact curves of Res$\left(T^6/\mathbb{Z}_\text{6--II}\right)$ existing in all triangulations.}
\label{indcc}
\end{table}
\begin{table}[t]
\centering
\begin{tabular}{|c||llllll|}
\hline
triangulation & \multicolumn{6}{|c}{additional compact curves}\\
\hline
\hline
i)& $D_{1,1}E_{1,\beta\gamma}$ & $E_{1,\beta\gamma}E_{4,1\,\beta}$ & $E_{1,\beta\gamma}E_{2,1\,\beta}$ & $D_{2,\beta}E_{1,\beta\gamma}$ & $D_{3,\gamma}E_{1,\beta\gamma}$ & $E_{1,\beta\gamma}E_{3,1\,\gamma}$\\
ii)& $E_{3,1\,\gamma}E_{4,1\,\beta}$ & $E_{1,\beta\gamma}E_{4,1\,\beta}$ & $E_{1,\beta\gamma}E_{2,1\,\beta}$ & $D_{2,\beta}E_{1,\beta\gamma}$ & $D_{3,\gamma}E_{1,\beta\gamma}$ & $E_{1,\beta\gamma}E_{3,1\,\gamma}$\\
iii)& $E_{3,1\,\gamma}E_{4,1\,\beta}$ & $E_{2,1\,\beta}E_{3,1\,\gamma}$ & $E_{1,\beta\gamma}E_{2,1\,\beta}$ & $D_{2,\beta}E_{1,\beta\gamma}$ & $D_{3,\gamma}E_{1,\beta\gamma}$ & $E_{1,\beta\gamma}E_{3,1\,\gamma}$\\
iv)& $E_{3,1\,\gamma}E_{4,1\,\beta}$ & $E_{2,1\,\beta}E_{3,1\,\gamma}$ & $D_{2,\beta}E_{3,1\,\gamma}$ & $D_{2,\beta}E_{1,\beta\gamma}$ & $D_{3,\gamma}E_{1,\beta\gamma}$ & $E_{1,\beta\gamma}E_{3,1\,\gamma}$\\
v)& $E_{3,1\,\gamma}E_{4,1\,\beta}$ & $E_{1,\beta\gamma}E_{4,1\,\beta}$ & $E_{1,\beta\gamma}E_{2,1\,\beta}$ & $D_{2,\beta}E_{1,\beta\gamma}$ & $D_{3,\gamma}E_{1,\beta\gamma}$ & $D_{3,\gamma}E_{4,1\,\beta}$\\
\hline
\end{tabular}
\caption{The compact curves of Res$\left(T^6/\mathbb{Z}_\text{6--II}\right)$ that exist only for a certain triangulation.}
\label{depcc}
\end{table}

\subsection{Examples of $\boldsymbol{T^6/\Intr_\text{6--II}}$ resolutions}
\label{sc:examplres}

\begin{table}[t]
\centering
\begin{tabular}{p{3.5cm}p{3.5cm}p{3.5cm}p{3.5cm}}
$ R_1R_2R_3~=~6~, $&$ R_{2} E_{3,1\,\gamma}^2~=~-2~, $&$ R_{2} E_{3,2\,\gamma}^2~=~-6~, $&$ R_{3} E_{2,1\,\beta}^2~=~-2~, $\\
$ R_{3} E_{2,3\,\beta}^2~=~-4~, $&$ R_{3} E_{4,1\,\beta}^2~=~-2~, $&$ R_{3} E_{4,3\,\beta}^2~=~-4~, $&$ R_{3} E_{2,1\,\beta} E_{4,1\,\beta}~=~1~, $\\
$ R_{3} E_{2,3\,\beta} E_{4,3\,\beta}~=~2~.$
\end{tabular}
\caption{The triangulation independent intersections of Res$\left(T^6/\mathbb{Z}_\text{6--II}\right)$. Intersection numbers not listed involving $E$'s with $\alpha\neq1$ or $R$'s are zero.}
\label{indint}
\end{table}
\begin{table}[t]
\centering
\begin{tabular}{p{3.5cm}p{3.5cm}p{3.5cm}p{3.5cm}}
$E_{1,\beta\gamma}^3~=~6~, $&$ E_{2,1\,\beta}^3~=~8~, $&$ E_{3,1\,\gamma}^3~=~8~, $&$E_{4,1\,\beta}^3~=~8~, $\\
$ E_{1,\beta\gamma}E_{2,1\,\beta}^2~=~-2~, $&$ E_{1,\beta\gamma}E_{3,1\,\gamma}^2~=~-2~, $&$E_{1,\beta\gamma}E_{4,1\,\beta}^2~=~-2~, $&$ E_{1,\beta\gamma}E_{2,1\,\beta}E_{4,1\,\beta}~=~1~, $\\
$ E_{2,1\,\beta}^2E_{4,1\,\beta}~=~-2~.$
\end{tabular}
\caption{The intersection numbers for the case that all fixed points have triangulation i). Only divisors with $\alpha=1$ are involved; all other intersections are zero.}
\label{res1int}
\end{table}

Here we give some  illuminations of the results of  the previous Subsections. The triangulation independent intersection numbers are given in Table~\ref{indint}. All intersections involving $E$'s with $\alpha\neq1$ and $R$'s that are not listed are zero. All other intersection numbers depend on the triangulation. The remaining non--zero intersection numbers for the case that all $\mathbb{C}^3/\mathbb{Z}_\text{6--II}$ fixed points are resolved according to triangulation i) are listed in Table~\ref{res1int}. 

Using this set of intersection numbers one can calculate some further interesting quantities. First of all we want to give the results for the second Chern class integrated over divisors. Using the expansion of the total Chern class to second order (\ref{2ndChern}) and the information from Table~\ref{indint}, one obtains for the second Chern class
\begin{align}
c_2(X)=&-\sum\limits_{\beta,\gamma}{\left[\frac{25}{36}E_{1,\beta\gamma}^2+\frac{5}{18}E_{1,\beta\gamma}E_{2,1\,\beta}+\frac{1}{3}E_{1,\beta\gamma}E_{3,1\,\gamma}+\frac{2}{9}E_{1,\beta\gamma}E_{4,1\,\beta}+\frac{1}{6}E_{,1\,\beta}E_{3,1\,\gamma}+\frac{1}{3}E_{3,1\,\gamma}E_{4,1\,\beta}\right]}\nonumber\\
&-\sum\limits_{\beta}{\left[\frac{7}{9}(E_{2,1\,\beta}^2+E_{2,3\,\beta}^2+E_{4,1\,\beta}^2+E_{4,3\,\beta}^2)+\frac{4}{9}(E_{2,1\,\beta}E_{4,1\,\beta}+E_{2,3\,\beta}E_{4,3\,\beta})\right]} 
\\ 
&-\sum\limits_{\gamma}{\left[\frac{3}{4}(E_{3,1\,\gamma}^2+E_{3,2\,\gamma}^2)\right]}~. \nonumber
\end{align}
This can now be easily integrated using $\int_{S}c_2(X)=c_2(X)S$ to give
\begin{equation}
\text{\begin{tabular}{l@{\hskip.5cm}l@{\hskip.5cm}l@{\hskip.5cm}l}
$c_2(X) E_{1,\beta\gamma}~=~0~,$ & $c_2(X) E_{2,1\,\beta}~=~-4~,$ & $c_2(X) E_{3,1\,\gamma}~=~-4~,$ & $c_2(X) E_{4,1\,\beta}~=~-4~,$\\
$c_2(X) E_{2,3\,\beta}~=~0~,$ & $c_2(X) E_{3,2\,\gamma}~=~0~,$ & $c_2(X) E_{4,3\,\beta}~=~0~,$ & \\
$c_2(X) R_1~=~0~,$ & $c_2(X) R_2~=~24~,$ & $c_2(X) R_3~=~24~.$ &
\end{tabular}}
\label{c2int}
\end{equation}
The first line of (\ref{c2int}) is triangulation dependent, whereas the other results hold for all triangulations.

Finally, we derive the restrictions on the expansion coefficients $a_i,b_r$ of the K\"ahler form $J$ defined in (\ref{eq:Jexpand}) by using the integrals of the K\"ahler form given in (\ref{eq:Kint}). Taking the integral over all curves in any triangulation, we get as a result that all $a_i$ and $b_r$ are larger than zero for all triangulations. Furthermore, only if an exceptional divisor $E$ gets a volume larger than zero, the fixed point corresponding to this divisor gets a finite size. Therefore the corresponding integral has to be larger than zero. On the other hand since the $R$'s are associated to the cycles of the torus, their volume should be larger than zero in any case, unless one wants to shrink one complex dimension of the torus to zero. The results of the integrals (\ref{eq:Kint}) are listed in appendix~\ref{sc:detailsres}.

\subsection{Summary of the resolution procedure}
\label{sc:ressummary}

We want to summarize the results obtained in the previous Subsections. Using local resolutions of fixed points and fixed lines and the globally defined divisors $R$, which are inherited from the torus, we were able to construct resolutions of the $T^6/\mathbb{Z}_\text{6--II}$ orbifold. These resolutions are described by the linear equivalence relations (\ref{eq:leq}), which are independent of the triangulations chosen, and the intersection ring, which is highly triangulation dependent. The knowledge of the intersection numbers is essential for our later computations since it allows us to calculate integrals of quantities that can be expanded in terms of divisors, such as the Chern classes, the gauge field strength and the K\"ahler form. Since the intersection numbers do depend on the chosen triangulation, in general every calculation that we perform later is triangulation dependent.

This raises the question about how many different possibilities to resolve the orbifold there are. A rough estimate would be $5^{12}$ since there are five triangulations possible at each of the twelve fixed points. But since there are permutation symmetries between the fixed points, this number gets reduced to $1.797.090$. This can be interpreted as a large number of distinct Calabi--Yau manifolds or as phases of the same manifold produced by flop transitions. A detailed description of how to obtain the number of different triangulations will be given in appendix~\ref{sc:counting}.

\section{Heterotic supergravity on resolutions}
\label{sc:sugrares}

%intro 

In the previous Section we reviewed how one can determine the properties of resolutions of compact orbifolds, the $T^6/\Intr_\text{6--II}$ in particular.  We now use these topological characterizations of the resulting Calabi--Yau spaces, to describe compactifications of ten dimensional heterotic E$_{8}\times$E$_{8}$ supergravity to four dimensions. After we have described the gauge backgrounds considered in this paper, we study consequences for the effective four dimensional theory. 

\subsection{Abelian gauge flux}
\label{sc:gaugeflux}

As the construction of stable vector bundles on Calabi--Yaus, like the orbifold resolutions described previously, is an extremely difficult task, we focus our attention here on Abelian gauge backgrounds  only. 
Such gauge backgrounds need to fulfill various conditions: 
First of all the gauge flux needs to be properly quantized: The gauge flux integrated over any compact curve has to be equal to an E$_8\times $E$_8$ lattice vector. Secondly, since the main objective of this paper is to compare compactifications on resolutions with those on heterotic orbifolds, we need to indicate how to identify the orbifold gauge shift and Wilson lines with these fluxes. Thirdly, the gauge background has to be chosen such that stringent consistency requirements imposed by the Bianchi identity are fulfilled. Finally, apart from these strict topological conditions, the gauge background has to be a solution to the Hermitian Yang--Mills equation. In the following we investigate the consequences of the topological conditions in detail, postponing the ``metric'' requirements of the Hermitian Yang--Mills equation to Subsection~\ref{sc:sugraDfour}.

We consider Abelian gauge backgrounds, therefore we can choose a Cartan basis in the E$_8\times $E$_8$ gauge group, with generators $H_I$, in which we expand the field strength two--form $\cF$.  (If we want to distinguish the Cartan algebra generators of the first and second E$_{8}$, we denote them by $H'_{I}$ and $H''_{I}$, respectively. Similarly we write $\cF = \cF'+\cF''$, where $\cF'$ lies in the first E$_{8}$ and $\cF''$ in the second.) Since the Hermitian Yang--Mills equation requires the gauge flux to be a $(1,1)$--form, we can expand it in terms of divisors. In Subsection~\ref{sc:resoverview} we saw that resolutions of $T^6/\Intr_\text{6--II}$ have three inherited divisors $R_i$ and 32 exceptional divisors $E_r$, hence in principle we can expand the gauge flux in all of them. In the completely blow down  limit we should recover the situation of the heterotic orbifold theory back. On the orbifold we have only allowed for gauge shifts and Wilson lines that correspond to non--trivial boundary conditions around orbifold fixed points and fixed lines, but not to magnetized tori. As this means that the gauge field strength vanishes everywhere on the orbifold except for the singularities, we assume that the gauge flux is supported at the exceptional divisors only: 
\equ{
\frac{\cF}{2\gp}   ~=~ E_r\, V_r{}^I \, H_I ~=~ 
E_r \big(V_r^{\prime I} \, H'_I + V_r^{\prime\prime I} \, H''_I \big)~, 
\label{gaugeFluxExpansion}
}
since they lie inside the singularities in the orbifold limit. The set of 32 vectors $V_r$ encodes how the gauge flux is embedded into the E$_8\times $E$_8$ gauge group. The bundle vectors in the first or the second E$_{8}$ are denoted by $V'_{r}$ and $V''_{r}$, respectively, hence collectively $V_r= (V'_r; V''_r)$. 

These vectors are severely restricted by the requirement that the gauge flux $\cF$ can be identified with the gauge shift $V$ and the $\Intr_3$ and $\Intr_2$ Wilson lines $W_3$ and $W_2$, respectively. (We consider only two Wilson line models for simplicity.) In order to identify the heterotic orbifold data with the characterization of the bundle one first considers the fixed points and fixed lines with the associated local orbifold gauge shifts individually, as defined in Subsection~\ref{sc:hetmodels}. As was used repeatedly in the previous Section, such singularities separately have non--compact resolutions. As was observed in~\cite{Nibbelink:2007pn} the identification between local gauge shift and the local Abelian bundle flux is obtained on the resolutions by integrating over an appropriately chosen non--compact curve built out of ordinary divisors. 

Here we extend this methodology to the different singularities of compact orbifolds by integrating over similar curves of ordinary divisors. For the $\Intr_\text{6--II}$ singularity this procedure can only be applied to the curve of the divisors $D_{2,\gb}D_{3,\gg}$, as this curve is not interrupted by exceptional divisors in the projected toric diagram given in Figure~\ref{5triang}. The identification therefore reads
\equ{
V_{(\gth,l_{\gb\gg})} ~\equiv~ \int_{D_{2,\gb}D_{3,\gg}} 
\frac{\cF}{2\gp}\Big|_{1\gb\gg} ~=~ V_{1,\gb\gg}~,
}
where the gauge flux has been restricted to fixed points $(\ga=1,\gb\gg)$. The local orbifold shift vector $V_{(\gth,l_{\gb\gg})}$ is characterized by its space group element $(\gth,l_{\gb\gg})$, where the lattice shifts $l_{\gb\gg}$ are given in the table below Figure~\ref{theta1}. Since the orbifold gauge shift and Wilson lines themselves are only determined up to lattice vectors, the matching can also only be performed up to them, as indicated by ``$\equiv$''. In Subsection~\ref{sc:triangdep} we emphasized that the local properties of the $\Intr_\text{6--II}$ singularities are triangulation dependent. This ambiguity does not affect the identification here, because it relies on the intersection $D_{2} D_{3}E_{1}$ only which is triangulation independent. 

The other bundle vectors are supported on exceptional divisors of complex codimension two singularities, hence the matching has to be performed in two complex dimensions. For the $\Intr_3$ singularities  this then amounts to computing the integrals
\equ{
V_{(\gth^2,l_{\ga\gb})} ~\equiv~ \int_{D_{2,\gb}} \frac{\cF}{2\gp}\Big|_{\ga\gb} ~=~ V_{2,\ga\gb}~,   \qquad  
V_{(\gth^4,l_{\ga\gb})} ~\equiv~ \int_{D_{1,3}} \frac{\cF}{2\gp}\Big|_{\ga\gb} ~=~ V_{4,\ga\gb}~, 
}
according to Figure~\ref{subfig2:toricdiag}. Here the lattice shifts $l_{\ga\gb}$ are defined in the table below Figure~\ref{theta2}. The gauge flux has been restricted to the fixed line $\ga\gb$ by setting all other exceptional divisors in $\cF$ to zero. Since the orbifold action for the second and fourth twisted sector is opposite, the identification on the orbifold requires that 
$V_{(\gth^4,l_{\ga\gb})} \equiv - V_{(\gth^2,l_{\ga\gb})}$. The same relations holds for the line bundle vectors $V_{2,\ga\gb}$ and $V_{4,\ga\gb}$. Finally, for the $\Intr_2$ fixed lines the identification reads
\equ{
V_{(\gth^3,l_{\ga\gg})} ~\equiv~ \int_{D_{3,\gg}} \frac{\cF}{2\gp}\Big|_{\ga\gg} ~=~ V_{3,{\ga\gg}}~,
}
see Figure~\ref{subfig1:toricdiag}, with $l_{\ga\gg}$ summarized in the table below Figure~\ref{theta3}. This analysis identifies for all 32 distinct fixed points and fixed lines the bundle vectors $V_r$ with the local gauge shift vectors $V_g$, given in~\eqref{localshift}, up to addition of lattice vectors.

This identification is written out in terms of the gauge shift and the Wilson lines in the following relations: On the exceptional divisors 
$E_{1,\gb\gg}$ inside the $\Intr_\text{6--II}$ fixed points we have
\equ{
\arry{l}{
V_{1,11} ~\equiv~ V_{1,13} ~\equiv~ V~, \\[0ex] %1  
V_{1,12} ~\equiv~ V_{1,14} ~\equiv~ V + W_2~, \\[0ex] %2
V_{1,21} ~\equiv~ V_{1,23} ~\equiv~ V + W_3~, \\[0ex] %5
V_{1,22} ~\equiv~ V_{1,24} ~\equiv~ V + W_2 + W_3~, \\[0ex] %6
V_{1,31} ~\equiv~ V_{1,33} ~\equiv~ V + 2 W_3~, \\[0ex] %9 
V_{1,32} ~\equiv~ V_{1,34} ~\equiv~ V + W_2 + 2 W_3~, %10
}
\label{idV1}
}
for $\gb = 1,2,3$ and $\gg = 1,\ldots, 4$. 
On exceptional divisors $E_{2,\ga\gb}$ and $E_{4,\ga\gb}$  inside the 
$\Intr_3$ singularities one obtains
\equ{
\arry{l}{
V_{2,11} ~\equiv~ V_{2,31} ~\equiv~ 2 V~,\\[0ex] %13
V_{2,12} ~\equiv~ V_{2,32} ~\equiv~ 2 V + 2 W_3~, \\[0ex] %14
V_{2,13} ~\equiv~ V_{2,33} ~\equiv~ 2 V + W_3~, %15
}
\quad 
\arry{l}{ 
V_{4,11} ~\equiv~ V_{4,31} ~\equiv~ -2 V~, \\[0ex] %27
V_{4,12} ~\equiv~ V_{4,32} ~\equiv~ -2 V - 2 W_3~, \\[0ex] %28
V_{4,13} ~\equiv~ V_{4,33} ~\equiv~ -2 V - W_3~,  %29
}
\label{idV24}
}
for $\ga = 1,3$ and $\gb = 1,2,3$. 
Finally, inside the $\Intr_2$ fixed lines on the exceptional divisors $E_{3,\ga\gg}$ the identification reads
\equ{ 
\arry{l}{
V_{3,11} ~\equiv~ V_{3,13} ~\equiv~ V_{3,21} ~\equiv~ V_{3,23} ~\equiv~ 3 V~, \\[0ex] %19
V_{3,12} ~\equiv~ V_{3,14} ~\equiv~ V_{3,22} ~\equiv~ V_{3,24} ~\equiv~ 3 V + W_2~, \\[0ex] %20
}
\label{idV3}
}
for $\ga = 1,2$ and $\gg = 1,\ldots, 4$. Once the bundle vectors $V_{r}$ have been defined in this way, the quantization conditions on all compact curves inside the resolution of $T^{6}/\Intr_\text{6--II}$ are  automatically fulfilled. As solving these quantization requirements is generically a difficult exercise, the matching with the orbifold gauge shift and Wilson lines is advantageous. 

The central consistency requirement of heterotic Calabi--Yau compactification is the Bianchi identity 
\equ{
\d H ~=~ \frac {\ga'}4 \Big(\TR\, \cR^{2} \,-\,\TR\, \cF^{2} \Big)~.
}
Here the trace $\TR$ is normalized as the trace in the fundamental representation of SO--groups. In the following we also encounter the trace $\Tr$ in the adjoint of an E$_{8}$ group, and traces $\tr$ in the fundamental of SU--groups. Since the gauge background $\cF$ is Abelian, these different definitions of the traces are related to each other 
\equ{
\Tr \cF^{2} ~=~ 30\, \TR\, \cF^{2} ~=~ 60\, \tr \cF^{2}~;
}
for higher powers similar identities exist (for gauge field strengths in the adjoint of E$_{8}\times$E$_{8}$ only the trace $\Tr$ is defined, and these identities are then interpreted as formal definitions). Since the left--hand--side of the Bianchi identity is exact, it vanishes when integrated over any of the 35 independent compact divisors 
\equ{
\int _{S} \Big\{ \tr \cF^{2} \,-\, \tr \cR^{2} \Big\} ~=~ 0~, 
\label{allBianchis}
}
with $S = R_{i}$ and $E_{r}$. 

This results in only 24 Bianchi consistency conditions: 11 equations are trivially satisfied.  They correspond to integrals over $R_{1}, E_{3,2\gg}$ and $E_{2,3\gb}, E_{4,3\gb}$, respectively. This can be understood by considering which intersections are needed when integrating over one of these divisors. In detail, the integral of the Bianchi identity over $R_{1}$ only gives non--vanishing contributions when terms proportional to $R_{2}R_{3}$ are present. Since by definition $\cF^{2}$ does not contain this combination and neither does the second Chern class $c_{2}(X) = -\tr \cR^{2}/8\gp^2$, given in~\eqref{2ndChern}, the integral over $R_{1}$ vanishes identically. In the same spirit we note that the only non--vanishing intersection involving $E_{3,2\gg}$ is $R_{2}E_{3,2\gg}^{2}$ (see Table~\ref{indint}). As $R_{2}E_{3,2\gg}$ is neither contained in $\cF^{2}$ and $c_{2}(X)$, also the conditions obtained by integrating over $E_{3,2\gg}$ are identically zero. Using similar arguments, also the integrals of the Bianchi identity over $E_{2,3\gb}$ and $E_{4,3\gb}$ vanish identically. Counting shows that there are in total $1+4+2\times 3=11$ trivial equations. 

Out of the 24 non--trivial Bianchi identities two are universal, while the others depend on the local triangulation of the $\Intr_\text{6--II}$ resolutions. The two universal Bianchi conditions, 
\equ{
\sum_\gg V_{3,1\gg}^2 \,+\, 3\sum_\gg V_{3,2\gg}^2 ~=~ 24~, 
\qquad 
\sum_\gb (V_{2,1\gb};V_{4,1\gb}) 
\,+\, 2 \sum_\gb (V_{2,3\gb};V_{4,3\gb})  ~=~ 24~, 
\label{RBianchis}
}
with $(v;w) = v^{2}+w^{2}-v\cdot w$, are obtained by integrating over $R_{2}$ and $R_{3}$, respectively. Note that neither condition involves the bundle vectors $V_{1,\gb\gg}$ and they are the same as the Bianchi identities on K3, that has gravitational instanton number $24$. This can be understood by noting that $R_{2}$ and $R_{3}$ have the topologies of the resolutions of K3 orbifolds $T^{4}/\Intr_{3}$ and $T^{4}/\Intr_{2}$, respectively~\cite{Aspinwall:1996mn,Nahm:1999ps,Wendland:2000ry}. The Bianchi consistency conditions obtained by integrating over $E_{1,\gb\gg}$ only depend on the local triangulation of the resolution of the $\Intr_\text{6--II}$ fixed point $(\ga=1,\gb\gg)$. The resulting five possible forms of the local Bianchi identity are listed in Table~\ref{tb:E1Bianchis}. The other Bianchi consistency requirements depend on the triangulations of different $\Intr_\text{6--II}$ resolutions simultaneously, therefore it becomes rather involved to indicate all the possible expressions for them. In the latter part of this paper we will only give them for very specific choices of triangulations. 

\begin{table}
\[
\arry{|c||l|}{
\hline  &\\[-2ex] 
\text{Res} & \text{Bianchi identity on } E_{1,\gb\gg}
\\[2ex] \hline\hline  &\\[-2ex] 
i) & \dsp 
3\, V_{1,\gb\gg}^{2} ~=~ (V_{2,1\gb};V_{4,1\gb}) \,+\, V_{3,1\gg}^{2}
\\[2ex] \hline & \\[-2ex]  
ii) & \dsp 
7\, V_{1,\gb\gg}^{2} ~=~ 4 \,+\, 2 \,(V_{2,1\gb};V_{4,1\gb}) 
\,+\, (V_{3,1\gg};V_{4,1\gb}) \,-\, 2\, V_{4,1\gb}^{2} 
+2\, V_{1,\gb\gg}\cdot (V_{3,1\gg}+V_{4,1\gb})
\\[2ex] \hline & \\[-2ex]  
iii) & \dsp 
8\, V_{1,\gb\gg}^{2} ~=~ 8 \,-\, 2\, V_{2,1\gb}\cdot V_{4,1\gb} 
\,+\,2\, V_{1,\gb\gg}\cdot (V_{2,1\gb}+V_{4,1\gb})
\\[2ex] \hline & \\[-2ex]  
iv) & \dsp 
9\, V_{1,\gb\gg}^{2} ~=~ 12  \,-\, V_{3,1\gg}^{2} 
\,+\, 6\, V_{1,\gb\gg}\cdot V_{3,1\gg}
\\[2ex]\hline & \\[-2ex]  
v) & \dsp 
4\, V_{1,\gb\gg}^{2} ~=~ 4 \,+\, (V_{2,1\gb};V_{4,1\gb}) 
\,-\, V_{4,1\gb}^{2} 
\,+\, 2\, V_{1,\gb\gg}\cdot V_{4,1\gb}
\\[2ex] \hline 
}
\]
\caption{\label{tb:E1Bianchis}
The Bianchi identity on the exceptional divisor $E_{1,\gb\gg}$ at a $\Intr_\text{6--II}$ resolution of fixed point $(\ga=1,\gb\gg)$ depends on which triangulation has been employed. }
\end{table}

This completes our description of the conditions on the Abelian bundle vectors $V_{r}$ to obtain a well defined resolution model. Before continuing investigating the resulting physics, let us emphasize a few important issues: The matching of the bundle vectors $V_{r}$ with the orbifold gauge shift and Wilson lines is universal, whereas a large portion of the Bianchi identities depend crucially on the local $\Intr_\text{6--II}$ triangulations chosen. For a fixed choice of local triangulations, the Bianchi identities already constitute a complicated system of 24 quadratic equations in 32 vectors $V_{r}$, each of which has 16 components. Given that they are all determined up to addition of E$_{8}\times$E$_{8}$ lattice vectors, finding a solution means to solve 24 Diophantine equations with 512 unknowns, which is a formidable task. Moreover, the different triangulations of the $\Intr_\text{6--II}$ resolutions lead to a large number of (almost two million) compact Calabi--Yau manifolds, and for each of them we get such a system of equations. Therefore solving the system of 24 Bianchi identities is a very difficult task in general. We will solve this system for a
specific case in Section~\ref{sc:MSSMblow}.

\subsection{Four dimensional spectrum and anomaly analysis}
\label{sc:anom}

Given a resolution and a compatible set of 32 line bundle vectors the spectrum of the resulting model can be computed. To do this we start from the anomaly polynomial of the gaugino in ten dimensions, and integrate over the resolution. In this way we obtain the multiplicity operator 
\equ{
N ~=~ \int_X \Big\{ 
\frac 16\, \Big(\frac{\cF}{2\gp}\Big)^{3}
\,-\, \frac 1{24}\, \tr \Big(\frac{\cR}{2\gp}\Big)^{2}\, \frac{\cF}{2\gp}
\Big\}~. 
\label{multiplicityOperator}
}
By acting with this operator on the 496 states of the E$_{8}\times$E$_{8}$ gaugino, one can determine the number of times each of these states appears on the resolution. Since this operator is defined as the integral over the whole compact resolution $X$, its expression depends on the local triangulations. 

The chiral spectrum computed using this multiplicity operator is free of non--Abelian anomalies because the Bianchi identities are fulfilled on all compact divisors~\cite{Witten:1984dg}. However, Abelian and mixed anomalies do in general arise for Abelian gauge backgrounds~\cite{Blumenhagen:2005pm,Blumenhagen:2005ga,Weigand:2005ng,Honecker:2006dt}, which are canceled via four and six dimensional variants of the Green--Schwarz mechanism~\cite{Green:1984sg,Green:1984bx,Dine:1987xk,Atick:1987gy,gsw_2}. Using the trace identities of E$_{8}$ 
\equ{
\Tr T^{4} ~=~ \frac 1{100}\, \Big( \Tr T^{2}\Big)^{2}~, 
\qquad 
\Tr T^{6} ~=~ \frac 1{7200}\, \Big( \Tr T^{2}\Big)^{3}~, 
}
the four dimensional anomaly polynomial can be written as~\cite{Blumenhagen:2005ga} 
\equa{
2\gp\, I_{6} ~=~ \frac 1{(2\gp )^{5}} \int_X \Big\{
& \frac 16\, \Big( \tr [\cF' F'] \Big)^{2} 
\,+\,  \frac 14\, \Big( \tr {\cF'}^{2} - \frac 12\, \tr \cR^{2} \Big) 
\tr {F'}^{2}
\\[2ex] 
&\,-\, \frac 1{16}\, \Big( \tr {\cF'}^{2} - \frac 5{12}\, \tr \cR^{2} \Big) \TR\, R^{2}
\Big\} \, \tr[\cF' F'] ~+~ ('\ra '')~. 
\non
}
Here $F', F''$ and $R$ denote the four dimensional gauge field strengths for both E$_{8}$ factors and curvature, respectively. 

This formula tells us that the pure U(1), the mixed U(1)--gravitational and the mixed U(1)--non--Abelian anomalies cannot be all absent at the same time. This holds in particular for the hypercharge $Y$: The four dimensional gauge field strength in the observable sector $F' = Y F_{Y} + \ldots$ contains the hypercharge U(1) gauge field $F_{Y}$; the dots denote the SU(2)$\times$SU(3) of the SM and other non--Abelian and  U(1) factors. In order that all anomalies involving the hypercharge U(1) are canceled, it is necessary that 
\equ{
\tr [\cF' Y] ~=~ V'_{r}\cdot Y\, E_{r}~
}
vanishes, i.e.\ that the hypercharge is perpendicular to all Abelian bundle vectors $V_{r}$. For the blowup models under investigation this is impossible, because one of the Wilson lines is responsible for breaking a certain GUT group down to the Standard Model: Since the bundle vectors are constructed from linear combinations of the gauge shift and Wilson lines, up to lattice vectors, some of  the inner products $Y\cdot V_{r}$ are non--zero. This is consistent with the general statement that U(1)'s of type i according to the classification defined in~\cite{Distler:1987ee,Dine:1987bq}, i.e.\ those that lie inside the structure group of the bundle, are broken. Generically there are then pure hypercharge, mixed U(1)--hypercharge, mixed gravitational--hypercharge and non--Abelian--hypercharge anomalies. Under certain circumstances it is possible that one of the latter two is absent. This analysis therefore indicates that at first sight all U(1) symmetries, including the hypercharge, are anomalous. The multitude of anomalous U(1)'s do not render the compactification inconsistent because the Green--Schwarz mechanism is at work to cancel these mixed anomalies. 

\subsection{Axions and twisted states}
\label{sc:Axions} 

This motivates us to consider the Green--Schwarz mechanism in four dimensions. This will lead us to investigate properties of axions and their reinterpretation as twisted states with VEV's that generate the blowup from the orbifold perspective. The starting point  is the bosonic part of the heterotic supergravity action in ten dimensions, given by 
\equ{
S_{het} ~=~ \frac 1{2\gk_{10}^{2}}
 \int \d^{10}x \sqrt{-\det g}\, e^{-2\gf}
\Big\{
R \,+\, 4\, | \d \gf|^{2} \,-\, \half\, |H_3 |^{2} 
\,-\, \frac{\ga'}4\, \TR\, |F|^{2}
\Big\}~, 
\label{hetsugraaction}
}
where in the conventions of~\cite{pol_2} we have $2 \gk_{10}^{2} = (2\gp)^{7} \ga^{\prime 4}$
and $F = \d A + A^2$ is the E$_{8}\times$E$_{8}$ gauge field strength of the gauge potential $A$ and 
\equ{
H_3 ~=~ \d B_{2} \,+\, \frac{\ga'}4 \, X_{3}~, 
\qquad  \d X_{3} ~=~ X_{4} ~=~
\TR\, R^{2} - \TR\, {F}^{2}~. 
}
The Green--Schwarz mechanism in ten dimensions relies on the fact that the anomaly polynomial factorizes 
$I_{12} = X_{4} X_{8}$, where 
\equa{
X_{8} ~=~ \frac 14\, \Big( \TR\, {F'}^{2}\Big)^{2} \,+\, &
 \frac 14\, \Big( \TR\, {F''}^{2}\Big)^{2} \,-\, 
  \frac 14\,  \TR\, {F'}^{2}\, \TR\, {F''}^{2}
\\[2ex] 
& \,-\, \frac 18
\Big( \TR\, {F'}^{2} \,+\, \TR\, {F''}^{2} \Big) \TR\, R^{2} 
\,+\, \frac 18 \TR\, R^{4}\,+\, \frac 1{32}\, \Big( \TR\, R^{2}\Big)^{2}~, 
\non
}
so that the anomalies in ten dimensions can be canceled using the Green--Schwarz interaction term 
\equ{
S_{GS} ~=~ \frac 1{48(2\gp)^{5}\ga'} \int B_{2} X_{8}~.  
}

Compactifying to four dimensions on a resolution of $T^{6}/\Intr_\text{6--II}$, one expands the two--form $B_{2}$ in terms of the 35 harmonic (1,1)--forms corresponding to the inherited and exceptional divisors 
\equ{
B_{2} ~=~ b_{2} \,+\, 2\gp \ga'\, ( \ga_{i}\, R_{i} \,+\, \gb_{r}\, E_{r})~. 
\label{Bexpand}
}
Here $b_{2}$ is the two--form in four dimensions and the $\ga_{i}$ and $\gb_{r}$ are scalars. The normalization of these scalars in~\eqref{Bexpand} has been chosen such that, under Abelian gauge transformations $\gd A^{I} = \d \gl^{I}$ with gauge parameter $\gl^{I}$, the scalars $\gb_{r}$ transform as axions 
\equ{
\gd \gb_{r} ~=~ V_{r}^{I}\, \gl^{I}~, 
\qquad 
\gd \ga_{i} ~=~0~, 
}
while the scalars $\ga_{i}$ are inert. This can be seen by realizing that $H_{3}$ contains 
\(
H_{3} \supset 2\gp \ga' \d_{4}\ga_{i}R_{i} 
+2\gp\ga'(\d_{4}\gb_{r} - V_{r}^{I} A^{I})E_{r}
\)
with $\d_{4}$ the exterior derivative in four dimensions, where we have used the freedom to choose $X_3 \supset -2\, \TR\, A\cF$ (see e.g. \cite{Kaloper:1999yr}). 

Because we chose the compactification to preserve supersymmetry, all states have to fall in supersymmetric multiplets. The scalars in the expansion of the $B_{2}$, the scalars $\ga_{i}$ and the axions $\gb_{r}$ form the scalar components of the multiplets $T_{i}$ and $U_{r}$. These components are defined by the expansion of the dimensionless complexified \Kh\ form 
\equ{
i \Big\{ 
\frac {B_2-b_2}{2\gp \ga'} \,+\, i \, J
\Big\}
~=~ 
T_i| \, R_i \,+\, U_r|\, E_r
}
in terms of the ordinary and exceptional divisors. Explicitly 
their lowest components are given by 
\equ{
T_{i}| ~=~ -a_{i} \,+\, i\, \ga_{i}~, 
\qquad 
U_{r}| ~=~ b_{r} \,+\, i\, \gb_{r}~,  
\label{multipletScalarAxionsEquations}
}
where the $|$ indicates setting all Grassmann coordinates $\gth, \bgth$ to zero. The real parts are the components of the \Kh\ form~\eqref{eq:Jexpand} in terms of the inherited divisors $R_{i}$ and exceptional divisors $E_{r}$. 

The axion states $\gb_{r}$ can be interpreted as twisted states from the heterotic orbifold point of view: As was discussed in Subsection~\ref{sc:hetmodels} the $T^{6}/\Intr_\text{6--II}$ orbifold has four twisted sectors: The first twisted sector contains genuine four dimensional states, while the second, fourth and third twisted sectors defines fields in six dimensions. Similarly, the exceptional divisors $E_{1,\gb\gg}$ correspond to the codimension six singularities, and therefore the scalars $\gb_{1,\gb\gg}$ live in four dimensions. The states $\gb_{r}$ corresponding to the other exceptional divisors, i.e.\ $E_{2,\ga\gb}$, $E_{4,\ga\gb}$ and $E_{3,\ga\gg}$,  all define six dimensional states, because they live on exceptional divisors of codimension four singularities. However, for this interpretation to work in all fine prints, also all the charges w.r.t.\ the 16 Cartan generators have to match. Since the twisted states transform linearly under gauge transformations, while axions transform with shifts, this identification does not directly work. 

A second place where there seems to be a mismatch between heterotic orbifold models and their blowup candidates is the following: Heterotic orbifold models have a single universal axion, which is consistent with the observation~\cite{Kobayashi:1997pb} that such models have at most a single anomalous U(1). On Calabi--Yaus there can be multiple anomalous U(1)'s and many axions supported on their divisors~\cite{Blumenhagen:2005pm,Blumenhagen:2005ga}. These two statements seem to be in contradiction when one considers models on orbifolds in blowup. This paradox is resolved by realizing that the blowup is generated by Higgsing, i.e.\ switching on VEVs for twisted states, and results in localized model dependent axions~\cite{GrootNibbelink:2007ew,Nibbelink:2008tv}. 

In the present case this mechanism sorts out these problems as well. As we just noted, the states $\gb_{r}$ are localized on the exceptional divisors $E_{r}$. The twisted states can be thought of being localized precisely on the exceptional divisors in blowup. If we consider the superfield redefinition 
\equ{
\gPs_{r} ~=~ M_s\, e^{2\gp\, U_{r}} 
~=~ M_s\, e^{2\gp(b_r+i \gb_r)}~, 
\qquad 
\gd \gPs_{r} ~=~ e^{2\gp i \,V_{r}^{I} \gL^{I}}\, \gPs_{r}~, 
\label{idTwU}
}
where $M_{s} = 2/\sqrt{\ga'} = 4\gp/\ell_{s}$ is the string scale, we see that $\gPs_{r}$ transforms linearly under gauge transformations: 
In fact, this shows that $\gPs_{r}$ is a definite twisted state from the orbifold perspective: The identifications of the Abelian bundle vectors $V_{r}$ and the orbifold gauge shift and Wilson lines, see~\eqref{idV1}--\eqref{idV3}, is the same as the local orbifold shifts, $V_{g}$, up to lattice vectors.  The twisted states are identified by their shifted momenta $p_\text{sh} = p+ V_{g}$, see Subsection~\ref{sc:hetmodels}. Putting these two ingredients together implies that each bundle vector $V_{r}$ defines a shifted momentum and therefore each superfield $\gPs_{r}$ corresponds to a definite twisted state.

\begin{figure}[t]
\begin{center} 
\includegraphics[width=90mm]{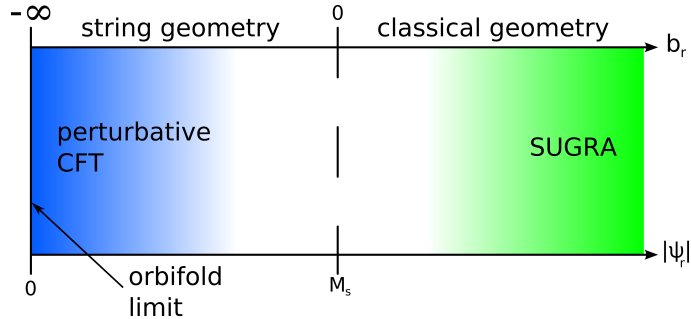}
\end{center}
\caption{
The supergravity approximation uses ``classical'' geometry with positive volumes of exceptional cycles, i.e.\ $b_r > 0$. Contrary, the CFT orbifold description is valid, when the VEV's of twisted states $\gPs_r$ are  much smaller than the string scale $M_s$. Because there $b_r < 0$ this is the regime of ``string'' geometry. 
\label{fig:CFTvsSUGRA}}
\end{figure}

Taking this identification seriously has some striking consequences: From the orbifold perspective the blow down would correspond to having vanishing VEV for the twisted state $\gPs_r$. According to~\eqref{idTwU} this limit is obtained by taking $b_r \ra -\infty$: The areas of some curves actually become negative! This astonishing result, telling us that we enter the ``string'' geometry regime, was  discussed in~\cite{Aspinwall:1993xz}. There it was argued that the volumes of exceptional divisors defined by the ``algebraic'' measure tend to $-\infty$ in the blow down limit. We see that the expectation value of the twisted states precisely corresponds to this measure. The situation is schematically depicted in Figure~\ref{fig:CFTvsSUGRA}. 
In~\cite{Aspinwall:1993xz} another measure, called the ``$\gs$--model'' measure, was defined. Using this measure the volume of exceptional divisors approaches zero in the orbifold limit as one intuitively would expect. But to construct this measure explicitly is much more involved, and not pursued in this work. 

The identification~\eqref{idTwU} also has phenomenological consequences in particular for the hypercharge symmetry of heterotic orbifold models in full blowup. A survey of the spectra of these models reveals that all heterotic MSSM orbifolds have at least one fixed point with all twisted states there charged under the Standard Model. Since going to full blowup corresponds to giving VEV's to at least one twisted state per fixed point or fixed line, always some of its gauge symmetries get broken. Moreover, if one wants to derive the resulting model from a supergravity approach, according to~\eqref{idTwU} all these VEV's need to be large: at least of the string scale. One could choose these VEV's carefully such that the SU(2)$\times$SU(3) remains unbroken, but then necessarily the hypercharge is lost. This is in accordance with the anomaly analysis of the previous Subsection~\ref{sc:anom}, where we found that the hypercharge U(1) generically suffers from pure and mixed anomalies. We conclude that when all fixed points have been blown up the MSSM is necessarily broken in all of the heterotic MSSM orbifolds considered in \cite{Lebedev:2006kn,Lebedev:2007hv}.

\subsection{Effective \Kh\ potential in four dimensions}
\label{sc:sugraDfour}

In the previous Subsection we identified the four and six dimensional axions that are crucial in the anomaly cancelation. Because of supersymmetry the structure of the four dimensional low energy action up to second order in the derivatives can be encoded by three functions: the gauge kinetic function, the superpotential and the \Kh\ potential. For our purposes the \Kh\ potential is the main object of interest. 

In the previous Subsection we already identified the chiral multiplets $T_{i}$ and $U_{r}$ that arose from the expansion of the anti--symmetric tensor field $B_{2}$ and the \Kh\ form $J$. Let $G$ denote the Hermitian metric on the internal six dimensional resolution $X$, and $*_4$ is the Hodge--dualization in four dimensions. The integrals over $X$ are distinguished from those in four dimensions by a subscript $X$ under the integral for the former. The volume of the resolution $X$ is obtained as 
\equ{ 
\text{Vol}(X) ~=~ \int_X \d^{6}z\, \det G
~=~ l_s^6 \int_X \frac 16 J^3~, 
}
where in the last equal sign we have used that the \Kh\ form $J$ is defined to be dimensionless and  $\ell_{s}^{2} = (2\gp)^{2}\ga'$ sets the string length. From the kinetic term and the theta--term of the gauge field in four dimensions,
\equ{
\int \d^{4}x\d^{4}\gth \, S\, \TR\, W^{\ga}W_{\ga} \,+\, \text{h.c.}
~\supset~
-\frac 14 \int \Re S\, \TR\, F *_{4} F 
\,-\, \frac 14\int \Im S\, \TR\, F^{2}~, 
}
we can identify the dilaton multiplet $S$ with scalar component
\equ{
S| ~=~ \frac 1{2\gp} \Big( 
\frac{\text{Vol}(X)} {e^{2\gf}\ell_{s}^{6}}
\,-\, i\, \gb_{0}
\Big)~, 
}
as follows: This multiplet contains the universal axion $\gb_{0}$, that is obtained from dualizing the anti--symmetric tensor $b_{2}$ via
\equ{
 \text{Vol}(X)\,  *_{4}\, \d_{4} b_{2} ~=~ e^{2\gf} \ell_{s}^{6}~\d_{4} \gb_{0}~, 
}
where the dilaton takes its constant VEV. To determine the transformation of this axion, start from parts of the action of $b_{2}$ given by
\equa{
S(b_{2}) ~\supset~ - &
\frac 1{4\gk_{10}^{2}} \int \d_{4} b_{2} * \d_{4}b_{2} \, 
\int_X \d^{6}z\, e^{-2\gf}\det G
\\[2ex] &
\,-\, \frac 1{8\ga'(2\gp)^{2}} \int \d_{4}b_{2} \, A'_{I}\, 
\frac 1{(2\gp)^{2}} \int_X \Big( \tr \cF^{\prime 2} -\frac 12 \tr \cR^{2} \Big) 
E_{r} V_{r}^{\prime I} \,+\, ('\ra '').
\non 
}
Here we have used that 
\(
X_{2,6} = 6 \big( \tr \cF^{\prime 2} - \frac 12 \tr \cR^2 \big) 
\tr (\cF' F') + ('\ra ''),
\)
is the expansion of $X_8$ to first order in four dimensional gauge fields. This can be rewritten in terms of the axion $\gb_{0}$ as 
\equ{
S(\gb_{0}) ~\supset~ \frac {1}{4\gp \ga'} \int 
\frac{e^{2\gf}\ell_{s}^{6}}{\text{Vol}(X)} 
\Big\{ \d_{4}\gb_{0} *_{4} \d_{4}\gb_{0}
\,+\, 2\, q_{I}\, \d_{4} \gb_{0} *_{4}  A^{I} 
\Big\}
  \, -\,  \frac 1{8\gp} 
  \int \gb_{0} \Big( \TR\, R^{2} - \TR\, F^{2}\Big)~. 
}
This axion and therefore the superfield $S$ transforms under gauge transformations as 
\(
\gd \gb_{0} = -q_{I} \gl^{I} = - q'_{I}\, \gl^{I} - q''_{I} \gl^{''I} 
\)
with 
\equ{
q'_{I} ~=~ \frac 1{16\gp} \int_X \frac 1{(2\gp)^{2}} 
\Big( \tr {\cF'}^{2} - \half \tr \cR^{2}\Big) 
E_{r}\, V_{r}^{\prime I}~,
}
and similarly for the second E$_8$ i.e.\ $'\ra ''$.  

When there are no anomalous U(1)'s present, like Calabi--Yau compactifications using the standard embedding, the moduli \Kh\ potential is given by~\cite{Strominger:1985ks,Candelas:1990pi} 
\equ{
\cK ~=~ - \ln \cH \,-\, \ln \int_X \frac 16 \cJ^{3}~, 
\label{Kahler}
}
with $\cH = S+\bS$ and 
$\cJ = R_{i}(T_{i}+\bT_{i}) + E_{r} (U_{r}+\bU_{r})$. 
However, as we have seen that both the chiral superfields $U_{r}$ and $S$ have anomalous variations, these functions need to be extended to gauge invariant combinations given by
\equ{
\cH ~=~ S + \bS  \,-\, \frac 1{2\gp}\, q_{I} \cV^{I}~, 
\qquad 
\cJ ~=~  R_{i}(T_{i}+\bT_{i}) + E_{r} 
\big( U_{r}+\bU_{r} - V_{r}^{I} \cV^I \big)~, 
}
where $\cV^{I} = (\cV^{\prime I},\cV^{\prime\prime I})$ is the E$_{8}\times$E$_{8}$ vector multiplet containing the gauge fields $A^{\prime I}$ and $A^{\prime\prime I}$. The \Kh\ potential $\cK$ is thus a function of the moduli chiral multiplets and the Abelian vector multiplets. 

This fact was used in~\cite{Blumenhagen:2005ga} to obtain the loop--corrected Donaldson--Uhlenbeck--Yau theorem~\cite{Donaldson:1985,Uhlenbeck:1986}. This integrated form of the Hermitian Yang--Mills equations is derived by determining the part of the action proportional to the auxiliary fields $D^{I}$ of the Abelian vector multiplets. This amounts to expanding the \Kh\ potential to first order in $V^{I}$ after which all Grassmann components are set to zero: 
\equ{
\int_X \frac 12\,J^{2} \frac{\cF}{2\gp} ~=~ e^{2\gf}\, q_{I} H_{I} 
~=~ \frac{e^{2\gf}}{16\gp} \int_X \frac 1{(2\gp)^{3}} 
\Big( \tr {\cF'}^{2} - \half \tr \cR^{2}\Big) \cF' \,+\, ('\ra '')~.  
\label{hym}
}
In superspace the mass matrix for gauge fields is obtained by expanding the \Kh\ potential to second order in the vector multiplets, because $\int \d^4\gth\, V^2 \sim A_\gm A^\gm$ in Wess--Zumino gauge. Hence we obtain the mass matrix for the gauge fields
\equ{
M_{IJ}^2 ~=~% \frac{\der^2 \cK}{\der V^I \der V^J}\Big| ~=~ 
\frac 14 \, \Big( \frac{l_s^6}{\text{Vol}(X)} \Big)^2 \Big\{ 
e^{4\gf}\, q_I q_J \,+\, 
\int_X J^2 E_r V_r^I \, \int_X J^2 E_s V_s^J \,-\, 
\frac{\text{Vol}(X)}{l_s^6} \int_X J E_r E_s V_r^I V_s^J
\Big\}~. 
\label{gaugemass}
}
The first term arises from differentiating $-\ln \cH$ twice, the other two terms from differentiating $-\ln\int_X \cJ^3/6$. The fact that two terms arise here might be somewhat surprising, but agrees with computing the mass directly from dimensional reducing the action~\eqref{hetsugraaction} to four dimensions: The reduced $H_3 * H_3$ term contains $\cF *_6 \cF \, A_\gm A^\gm$. Using that on Calabi--Yau spaces the dual of a two--form can be expressed by~\cite{Strominger:1985ks}
\equ{
*_6 \cF = - J\cF \,-\, \frac 32\, \int_X{J^2\cF} \, \Big/ \, {\int_X\frac 16 J^3}~,
}
we confirm that two mass structures should arise. The first and the second term in~\eqref{gaugemass} are equal, when the Hermitian Yang--Mills equation is fulfilled. The mass matrix given above can be interpreted as the physical mass matrix provided that one takes into account that the kinetic terms for the gauge fields are not canonically normalized. 

However, both the Hermitian Yang--Mills equation~\eqref{hym} and the mass matrix~\eqref{gaugemass} are only valid in the supergravity regime. The reason is that the ten dimensional heterotic supergravity action~\eqref{hetsugraaction}, from which these results are derived, is lowest order in $\ga'$, up to some terms introduced for the purpose of anomaly cancelation. The full string dynamics furnish a series of $\ga'$ corrections to this effective action. Therefore, in particular the Hermitian Yang--Mills equations will receive such $\ga'$ corrections, and consequently the notion of the stability of bundles may have  to be reconsidered in this light. From the effective four dimensional perspective, this does not only result in corrections to the \Kh\ potential~\eqref{Kahler} and superpotential, but also in new higher derivative interactions in the four dimensional supergravity theory. Moreover, at a certain order in $\ga'$ also some massive string excitations become relevant. Only when the curvatures are small compared to the string scale $M_s$ the naive supergravity can be trusted. 

Precisely when we want to consider the matching of the effective supergravity description on the Calabi--Yau resolution of $T^6/\Intr_\text{6--II}$ with the orbifold theory, we run out of the regime of validity of supergravity. As we observed from~\eqref{idTwU} the orbifold limit corresponds to taking the volume parameters $b_r$ of the exceptional divisors to $-\infty$. As the classical \Kh\ cone requires that $b_r \geq 0$, this means that area and volume integrals, like $\int_X J^3$, take wrong signs, which results in sick behavior of, for example, the Hermitian Yang--Mills equation~\eqref{hym} or the mass matrix~\eqref{gaugemass}. Presumably the higher order $\ga'$ corrections will compensate for this disastrous behavior, but unfortunately we do not know these supergravity improvements explicitly. 

Approaching the matching from the orbifold side seems to become problematic as well. At the orbifold point there is an exact CFT description available which is perfectly under control (with the possible exception of the Fayet--Illiopoulos term due to the universal anomalous U(1), see~\eqref{eq:Anomaly_conditions}). However, as soon as one wants to consider a full blowup of the orbifold, one needs to allocate VEV's to at least a single twisted state $\gPs_r$ per fixed point or fixed line. As long as these VEV's are small compared to the string scale $M_s$, this corresponds to small deviations from the exactly solvable CFT. But~\eqref{idTwU} tells us that in order to match with positive values for $b_r$, where the supergravity approximation would start to make some sense, the VEV's of $\gPs_r$ are at least of the order of the string scale. Even though the convergence radius of the VEV's of say the superpotential computed from orbifold CFT's is unknown (calculations like~\cite{Choi:2007nb} attempt to get an insight into this), it would likely be not much beyond the string scale. Therefore the matching of the supergravity on smooth Calabi--Yau spaces and heterotic strings on orbifolds beyond the topological level, is very difficult: The natural place for that seems to be when $b_r \approx 0$, i.e.\ $\gPs_r \approx M_s$, but in fact there both approaches are less under control. 

\section{Resolution of a heterotic MSSM Orbifold}
\label{sc:MSSMblow}

In this Section, we want to apply the results of the previous Sections, i.e. we want to compute and solve the integrated Bianchi identities for the 3--form field strength $H$ given in (\ref{allBianchis}). For concreteness and simplicity we focus on the case where resolution i) is used exclusively to resolve all 12 $\mathbb{C}^{3}/\mathbb{Z}_{\text{6--II}}$ fixed points. Subsequently, we describe a method that can be employed for solving the Diophantine equations resulting from the Bianchi identities. Thereafter we discuss the massless spectrum of our solution. At last we discuss the identification of twisted orbifold states and line bundle vectors, which allows for another way of finding a solution for the Bianchi identites.

\subsection{The Bianchi identities with resolution i) for all fixed points}
\label{sc:ComputingBianchis}

As stated in Subsection~\ref{sc:gaugeflux}, we obtain 24 non--trivial equations from integrating (\ref{allBianchis}) over the 35 possible divisors. The Bianchi identities resulting from integrating over $E_{1,\beta\gamma}$ depend only on the local resolution of the fixed points. In addition, we obtain the two resolution independent equations coming from an integration over the inherited divisors $R_{i}$. The remaining 10 non--trivial equations result from integrating over $E_{2,1\beta}$,  $E_{4,1\beta}$, and  $E_{3,1\gamma}$. In contrast to the Bianchi identities for $E_{1,\beta\gamma}$, these equations depend on a combination of chosen resolutions. This is due to the fact that integrating over $E_{3,1\gamma}$ and $E_{2,1\beta}$,  $E_{4,1\beta}$ leaves $\beta$ and $\gamma$ unspecified, respectively. As a consequence, there remains a sum over several distinct fixed points coming from the expansion of the gauge flux and the second Chern class, which in turn leads to the fact that these 10 equations depend on the combination of resolutions at these distinct fixed points. In this example, however, we use resolution i) only, so this complication will not concern us further here.

The resolution--dependent Bianchi identities are computed as outlined in Subsection~\ref{sc:gaugeflux}. The relevant intersection numbers were given in Table \ref{indint} and Table \ref{res1int}, and the integrals over the second Chern class can be found in (\ref{c2int}). Carrying out the integration for all 24 divisors  yields the following set of non--trivial Bianchi identities:
\begin{subequations}
\label{BIeqns}
\begin{equation}
\sum \limits_{\gamma} V_{3,1 \gamma}^{2} + 3\sum \limits_{\gamma} V_{3,2 \gamma}^{2} 						= 24, 	\label{eq:BIeqnsA}\\
\end{equation}
\begin{equation}
\sum \limits_{\beta} (V_{2,1 \beta};V_{4,1 \beta}) + 2 \sum \limits_{\beta} (V_{2,3\beta};V_{4,3\beta})  					= 24, 	\label{eq:BIeqnsB}\\
\end{equation}
\begin{equation}
3 V_{1,\beta \gamma}^{2} - V_{3,1 \gamma}^{2} - (V_{2,1 \beta};V_{4,1 \beta}) 									= 0, 	\label{eq:BIeqnsC}\\
\end{equation}
\vskip0.5pt
\begin{equation}
2 V_{3,1\gamma}^{2} - V_{3,1 \gamma} \cdot \sum \limits_{\beta} V_{1,\beta \gamma} 								= 2, 	\label{eq:BIeqnsD}\\
\end{equation}
\begin{equation}
3 V_{2,1 \beta}^{2} + 4 (V_{2,1 \beta};V_{4,1 \beta}) -  3 V_{2,1 \beta} \cdot \sum \limits_{\gamma} V_{1, \beta \gamma} 	= 12,	\label{eq:BIeqnsE}\\
\end{equation}
\begin{equation}
6 V_{4,1 \beta}^{2} + 2 (V_{2,1 \beta};V_{4,1 \beta}) -  3 V_{4,1 \beta} \cdot \sum \limits_{\gamma} V_{1, \beta \gamma}	= 12.	\label{eq:BIeqnsF}
\end{equation}
\end{subequations}
Equations (\ref{eq:BIeqnsA}) and (\ref{eq:BIeqnsB}) are the resolution independent Bianchi identities coming from the integration over $R_{2}$ and $R_{3}$, i.e.\ equations (\ref{RBianchis}). The twelve equations given in (\ref{eq:BIeqnsC}) are the ones coming from integrating over $E_{1,\beta\gamma}$. The relevant data is given in the first line of Table~\ref{tb:E1Bianchis}. When integrating over $E_{3,1\gamma}$, we obtain the four equations (\ref{eq:BIeqnsD}). Finally, the two times three equations (\ref{eq:BIeqnsE}) and (\ref{eq:BIeqnsF}) are obtained from integrating over $E_{2,1\beta}$ and $E_{4,1\beta}$, respectively. Having obtained these equations, one can in principle take the bundle vector identifications given in (\ref{idV1}) -- (\ref{idV3}) together with the data from (\ref{eq:DataBM2}) and insert them into (\ref{BIeqns}). However, in general none of the Bianchi identities will be solved with this procedure. This is due to the fact that the 32 line bundle vectors are only defined up to the addition of E$_{8}\times$E$_{8}$ lattice vectors. One possibility to find a solution is to find a set of appropriate lattice vectors that is added to the 32 line bundle vectors. As already mentioned at the end of Subsection~\ref{sc:gaugeflux}, this leads to a system of 24 Diophantine equations in 512 unknowns. In order to be able to solve this, it is convenient to simplify the set of equations. How this can be done is illustrated in the next Subsection. Another approach to finding a solution to (\ref{BIeqns}) is to start with twisted orbifold states and use the identification between them and the line bundle vectors as discussed in Subsection~\ref{sc:Axions}. This method is exemplified in Subsection~\ref{sc:SolutionProperties}, after we discussed the identification in more detail.

\subsection{Solving the Bianchi identities}
\label{sc:SolvingBianchis} 

In this rather technical Subsection we give a solving procedure for the Bianchi identities. The physics of a Bianchi identity solution is discussed in the subsequent Subsections~\ref{sc:ResultingSpectrum} and \ref{sc:SolutionProperties}.

In order to obtain a solution, we start with the identifications (\ref{idV1}) -- (\ref{idV3}) for the 32 line bundle vectors. The SO(10) of the orbifold shift vector $V$ is broken down to SU(5) by the order three Wilson line $W_{3}$. The order two Wilson line $W_{2}$ further reduces the gauge group to SU(3) $\times$ SU(2). From a phenomenological point of view it is desirable to keep the SU(3) $\times$ SU(2) gauge group living in the first E$_{8}$, as this yields part of the Standard Model gauge group. Additionally, one may not want to completely break the SO(8) $\times$ SU(2) in the second E$_{8}$, as the hidden sector gauge group must not be too small in order to allow for the right gaugino condensation scale. In order to preserve the Standard Model gauge group in the first E$_{8}$, one has to think about which of the E$_{8}\times$E$_{8}$ vectors can be added to the line bundle vectors. The  E$_{8}\times$E$_{8}$ vectors must have identical entries in components $4$ and $5$ as well as in components $6$, $7$, and $8$, as this is where the SU(2) and the SU(3) live, respectively. This already reduces the number of unknowns considerably. Initially, we do not take the gauge groups of the second E$_{8}$ into consideration, as it turns out that once we find a solution, it can be quite easily changed into a solution with better features, e.g. the preservation of a big hidden sector gauge group.

A further complication arises from the fact that the Bianchi identities contain a lot of inner products between line bundle vectors, which couples many equations; this makes it hard to reduce the Bianchi identities to sets of smaller and thus easier equations. Additionally, it is much easier to solve equations containing squares of vectors than solving equations containing inner products. Hence we aim at rewriting as many equations as possible in terms of vector squares only. As can be seen from (\ref{idV24}), for each  vector in the $\theta^{2}$--sector there is a vector in the $\theta^{4}$--sector that has the same identification of orbifold shifts and Wilson lines up to a minus sign and addition of lattice vectors. By having a closer look at the Bianchi identities one realizes that exactly these pairs of vectors appear in the inner product $(a;b)=a^{2}+b^{2}-a \cdot b$. If one requires that the vectors in such pairs are exactly opposite, $V_{2,s} = -V_{4,s}$, the inner product reduces to $(V_{2,s};V_{4,s})=3 V_{2,s}^{2} = 3 V_{4,s}^{2}$ where $s$ is an appropriately chosen multi--index for ($\alpha \beta$). This allows us to replace all the inner products of the type $(\cdot\,;\cdot)$ occurring in the Bianchi identities by squares of vectors. Moreover, it further reduces the number of independent variables. For yet an additional reduction of the number of unknowns, we extend the assumption that two vectors that have the same orbifold shift vector and Wilson line identification are identical to the $\theta$-- and $\theta^{3}$--sector, meaning $V_{1,11}=V_{1,13}$, $V_{1,12}=V_{1,14}$, $V_{3,11}=V_{3,13}$, and so on. Finally, we require that all vectors coming from the same $\theta$--sector have the same absolute value squared. Making these simplifications, we can cast (\ref{BIeqns})  into the following form:
\begin{subequations}
\label{BIeqnsSimplified}
\begin{align}
V_{1,\beta\gamma}^{2} 										&= \frac{25}{18},~~ \!\!\beta \in \left\{1,2,3\right\}, ~ \!\gamma \in \left\{1,2,3,4\right\},\label{eq:BIeqnsSimplifiedA}\\[8pt]
V_{2,\alpha\beta}^{2} = V_{4,\alpha\beta}^{2} 						&= \frac{8}{9},~~ \alpha \in \left\{1,3\right\},~~~\beta \in \left\{1,2,3\right\},\label{eq:BIeqnsSimplifiedB}\\[8pt]
V_{3,\alpha\gamma}^{2} 										&= \frac{3}{2},~~ \alpha \in \left\{1,2\right\},~~~\gamma \in \left\{1,2,3,4\right\},\label{eq:BIeqnsSimplifiedC}\\[8pt]
V_{3,1\gamma} \cdot \sum\limits_{\beta = 1}^{3}V_{1,\beta\gamma} &= 1,~~ \gamma \in\left\{1,2,3,4\right\},\label{eq:BIeqnsSimplifiedD}\\
V_{2,\alpha\beta} \cdot \sum\limits_{\gamma = 1}^{4}V_{1,\beta\gamma}&= \frac{4}{9},~~ \alpha \in \left\{1,3\right\},~~~\beta \in \left\{1,2,3\right\}.\label{eq:BIeqnsSimplifiedE}
\end{align}
\end{subequations}
The two remaining inner product equations (\ref{eq:BIeqnsSimplifiedD}) and (\ref{eq:BIeqnsSimplifiedE}) come from (\ref{eq:BIeqnsD}) and (\ref{eq:BIeqnsE}), respectively. Under the simplifications, (\ref{eq:BIeqnsF}) is automatically satisfied if (\ref{eq:BIeqnsSimplifiedE}) is. It is now easy to find a set of 32 line bundle vectors that satisfy the first three conditions(\ref{eq:BIeqnsSimplifiedA}) -- (\ref{eq:BIeqnsSimplifiedC}).

However, it turns out that the assumptions made above are too restrictive, which renders it impossible to find a solution for the whole set of equations (\ref{BIeqnsSimplified}) simultaneously. Therefore, one has to abandon some of the assumptions made above. It is, however, advantageous to keep the equations decoupled, so that we do not have to give up all the line bundle vectors we just found from solving (\ref{eq:BIeqnsSimplifiedA}) to (\ref{eq:BIeqnsSimplifiedC}). This allows us to change only a small subset of equations. Relaxing the condition $V_{2,1\beta}=-V_{4,1\beta}$ leads to a violation of (\ref{eq:BIeqnsSimplifiedB}) and (\ref{eq:BIeqnsSimplifiedE}) for $\alpha = 1$. So one has to modify at most all the vectors involved in these equations. However, when modifying the solution, one has to pay attention to maintaining the vector squares as dictated by equations (\ref{eq:BIeqnsSimplifiedA}) as this guarantees that the equations are still decoupled in the sense that changing something in one equation does not influence the validity of the other equations. In our case, it was sufficient to change $V_{2,1\beta}$, $V_{4,1\beta}$, and $V_{1,\beta 4}$.\\
\indent With this procedure we can have a solution satisfying all 24 Bianchi identities. As mentioned earlier, once we find a solution, it can be easily modified. The solution presented in Table \ref{table:BISolutionVectors} was found as described above and then altered\footnote{In fact, we altered the twelve vectors $V_{2,\alpha\beta}$ and $V_{4,\alpha\beta}$ such that they additionally correspond to a twisted orbifold state, although this means that they violate some of the equations of (\ref{BIeqnsSimplified}). A detailed discussion of this issue is given in Subsection~\ref{sc:SolutionProperties}.} such that the hidden sector gauge group is SU(4), while the SU(2) is broken to U(1)'s. By construction, the SU(3) $\times$ SU(2) Standard Model gauge group in the first E$_{8}$ is conserved.

\subsection{The massless spectrum}
\label{sc:ResultingSpectrum}

For a given solution of the Bianchi identities (\ref{BIeqns}), we can compute the massless particle content of our model. In order to obtain the multiplicity of the matter representations we use equation (\ref{multiplicityOperator}).  As in the case of the Bianchi identities, the integration can again be carried out using the intersection numbers (Tables \ref{indint} and \ref{res1int}), the expansion of the gauge flux (\ref{gaugeFluxExpansion}), and the relation between the curvature and the second Chern class combined with the splitting principle (\ref{eq:spprin}). Using resolution i) for all $\mathbb{C}^{3}/\mathbb{Z}_{\text{6--II}}$ fixed points, this yields the following expression for the multiplicity operator $N$:
\begin{equation}
\begin{array}{lll}
\label{eq:multiplicityOperatorIntegratedOut}
N & = & \sum\limits_{\beta=1}^{3} \sum\limits_{\gamma=1}^{4}H_{1,\beta\gamma}\left[\left(H_{2,1\beta}\right)^{3} + \left(H_{4,1\beta}\right)^{3} - H_{2,1\beta}H_{4,1\beta} - \left(H_{1,\beta\gamma}\right)^{2} + \left(H_{3,1\gamma}\right)^{2}\right]\\
& & + \frac{1}{3} \sum\limits_{\beta=1}^{3}\left[4\left(H_{2,1\beta}\right)^{3} +4 \left(H_{4,1\beta}\right)^{3} - H_{2,1\beta} - H_{4,1\beta} -3\left(H_{2,1\beta}\right)^{2}H_{4,1\beta}\right] \\
& & + \frac{1}{3}\sum\limits_{\gamma=1}^{4}\left[4\left(H_{3,1\gamma}\right)^{3} - H_{3,1\gamma}\right].
\end{array}
\end{equation}
The expression $H_{r}=V_{r}^{I} H_{I}$ is a shorthand notation for the line bundle vectors contracted with the Cartan generators. The entries in the line bundle vectors can thus be interpreted as the expansion coefficients of the Cartan element $H_{r}$ expanded in the Cartan generators $H_{I}$. Applying the $H_{I}$ to some vector of the adjoint representation gives the corresponding weight times the vector: $H_{r} \left\vert w\right\rangle =\left( V_{r}  \cdot w \right) \left\vert w\right\rangle$ with $w$ being the eigenvector (which coincides with the root vector in case of the adjoint representation) of $\left\vert w\right\rangle$. Categorizing the 248 elements (112 vectorial roots, 128 spinorial roots, 8 Cartan generators) of each E$_{8}$ according to the representation they form under the gauge group of the Standard Model SU(3) $\times$ SU(2) $\times$ $U(1)_{Y}$ in case of the first E$_{8}$ and the gauge group of the hidden sector SU(4) in case of the second E$_{8}$, we obtain the spectrum given in Tables~\ref{table:E81SpectrumSmall} and~\ref{table:E82SpectrumSmall}.

\renewcommand{\arraystretch}{1.1}
\begin{table}[t]
\centering
\begin{tabular}{|c||c@{}cccccccc@{}c@{}cccccccc@{}|}
\hline
$V_{r}$ & \multicolumn{18}{c|}{expression for the bundle vector $V_r$}\\
\hline
\hline
$V_{1,11}, V_{1,13}$ & $($ & $-\frac{1}{6}$ & $0$ & $0$ & $-\frac{1}{2}$ & $-\frac{1}{2}$ & $-\frac{1}{2}$ & $-\frac{1}{2}$ & $-\frac{1}{2}$ & $)$  $($ & $0$ & $\frac{1}{3}$ & $0$ & $0$ & $0$ & $0$ & $0$ & $0)$\\
$V_{1,12}, V_{1,14}$ & $($ & $-\frac{5}{12}$ & $\frac{1}{4}$ & $-\frac{3}{4}$ & $-\frac{1}{4}$ & $-\frac{1}{4}$ & $\frac{1}{4}$ & $\frac{1}{4}$ & $\frac{1}{4}$ & $)$  $($ & $-\frac{1}{2}$ & $-\frac{1}{6}$ & $0$ & $0$ & $0$ & $0$ & $0$ & $0)$\\
$V_{1,21}, V_{1,23}$ & $($ & $-\frac{1}{6}$ & $0$ & $\frac{2}{3}$ & $\frac{1}{6}$ & $\frac{1}{6}$ & $\frac{1}{6}$ & $\frac{1}{6}$ & $\frac{1}{6}$ & $)$ $($ & $\frac{1}{3}$ & $-\frac{2}{3}$ & $-\frac{1}{3}$ & $-\frac{1}{3}$ & $0$ & $0$ & $0$ & $0)$\\
$V_{1,22}, V_{1,24}$ & $($ & $\frac{1}{12}$ & $-\frac{1}{4}$ & $\frac{5}{12}$ & $-\frac{1}{12}$ & $-\frac{1}{12}$ & $\frac{5}{12}$ & $\frac{5}{12}$ & $\frac{5}{12}$ & $)$ $($ & $-\frac{1}{6}$ & $-\frac{1}{6}$ & $-\frac{1}{3}$ & $\frac{2}{3}$ & $0$ & $0$ & $0$ & $0)$\\
$V_{1,31}, V_{1,33}$ &$ ($ & $-\frac{1}{6}$ & $0$ & $\frac{1}{3}$ & $-\frac{1}{6}$ & $-\frac{1}{6}$ & $-\frac{1}{6}$ & $-\frac{1}{6}$ & $-\frac{1}{6}$ & $)$ $($ & $\frac{1}{6}$ & $-\frac{1}{6}$ & $-\frac{1}{6}$ & $-\frac{1}{6}$ & $\frac{1}{2}$ & $\frac{1}{2}$ & $\frac{1}{2}$ & $\frac{1}{2})$\\
$V_{1,32}, V_{1,34}$ &$ ($ & $\frac{1}{12}$ & $-\frac{1}{4}$ & $\frac{1}{12}$ & $-\frac{5}{12}$ & $-\frac{5}{12}$ & $\frac{1}{12}$ & $\frac{1}{12}$ & $\frac{1}{12}$ & $)$ $($ & $\frac{1}{6}$ & $-\frac{1}{6}$ & $-\frac{2}{3}$ & $-\frac{2}{3}$ & $0$ & $0$ & $0$ & $0)$\\
$V_{2,11}$ & $($ & $-\frac{1}{3}$ & $0$ & $-1$ & $0$ & $0$ & $0$ & $0$ & $0$ & $)$ $($ & $0$ & $-\frac{1}{3}$ & $0$ & $-1$ & $0$ & $0$ & $0$ & $0)$\\
$V_{2,12}$ & $($ & $-\frac{1}{3}$ & $1$ & $\frac{1}{3}$ & $\frac{1}{3}$ & $\frac{1}{3}$ & $\frac{1}{3}$ & $\frac{1}{3}$ & $\frac{1}{3}$ & $)$  $($ & $-\frac{1}{3}$ & $-\frac{1}{3}$ & $\frac{1}{3}$ & $\frac{1}{3}$ & $0$ & $0$ & $0$ & $0)$\\
$V_{2,13}$ & $($ & $-\frac{5}{6}$ & $\frac{1}{2}$ & $\frac{1}{6}$ & $\frac{1}{6}$ & $\frac{1}{6}$ & $\frac{1}{6}$ & $\frac{1}{6}$ & $\frac{1}{6}$ & $)$ $($ & $\frac{1}{3}$ & $-\frac{1}{3}$ & $-\frac{1}{3}$ & $-\frac{1}{3}$ & $0$ & $0$ & $0$ & $0)$\\
$V_{2,31}$ & $($ & $\frac{2}{3}$ & $0$ & $0$ & $0$ & $0$ & $0$ & $0$ & $0$ & $)$ $($ & $0$ & $-\frac{1}{3}$ & $0$ & $-1$ & $0$ & $0$ & $0$ & $0)$\\
$V_{2,32}$ & $($ & $\frac{1}{6}$ & $-\frac{1}{2}$ & $-\frac{1}{6}$ & $-\frac{1}{6}$ & $-\frac{1}{6}$ & $-\frac{1}{6}$ & $-\frac{1}{6}$ & $-\frac{1}{6}$ & $)$ $($ & $\frac{2}{3}$ & $-\frac{1}{3}$ & $\frac{1}{3}$ & $-\frac{2}{3}$ & $0$ & $0$ & $0$ & $0)$\\
$V_{2,33}$ & $($ & $\frac{1}{6}$ & $-\frac{1}{2}$ & $\frac{1}{6}$ & $\frac{1}{6}$ & $\frac{1}{6}$ & $\frac{1}{6}$ & $\frac{1}{6}$ & $\frac{1}{6}$ & $)$ $($ & $\frac{1}{3}$ & $\frac{2}{3}$ & $\frac{2}{3}$ & $-\frac{1}{3}$ & $0$ & $0$ & $0$ & $0)$\\
$V_{3,11}, V_{3,13}$ & $($ & $0$ & $-\frac{1}{2}$ & $\frac{1}{2}$ & $0$ & $0$ & $0$ & $0$ & $0$ & $)$ $($ & $0$ & $-1$ & $0$ & $0$ & $0$ & $0$ & $0$ & $0)$\\
$V_{3,12}, V_{3,14}$ & $($ & $\frac{1}{4}$ & -$\frac{3}{4}$ & $\frac{1}{4}$ & $-\frac{1}{4}$ & $-\frac{1}{4}$ & $\frac{1}{4}$ & $\frac{1}{4}$ & $\frac{1}{4}$ & $)$ $($ & $\frac{1}{2}$ & $-\frac{1}{2}$ & $0$ & $0$ & $0$ & $0$ & $0$ & $0)$\\
$V_{3,21}, V_{3,23}$ & $($ & $0$ & $-\frac{1}{2}$ & $\frac{1}{2}$ & $0$ & $0$ & $0$ & $0$ & $0$ & $)$ $($ & $0$ & $-1$ & $0$ & $0$ & $0$ & $0$ & $0$ & $0)$\\
$V_{3,22}, V_{3,24}$ & $($ & $\frac{1}{4}$ & $-\frac{3}{4}$ & $\frac{1}{4}$ & $-\frac{1}{4}$ & $-\frac{1}{4}$ & $\frac{1}{4}$ & $\frac{1}{4}$ & $\frac{1}{4}$ & $)$ $($ & $\frac{1}{2}$ & $-\frac{1}{2}$ & $0$ & $0$ & $0$ & $0$ & $0$ & $0)$\\
$V_{4,11}, V_{4,31}$ & $($ & $\frac{1}{3}$ & $0$ & $-1$ & $0$ & $0$ & $0$ & $0$ & $0$ & $)$ $($ & $0$ & $-\frac{2}{3}$ & $0$ & $0$ & $0$ & $0$ & $0$ & $0)$\\
$V_{4,12}, V_{4,32}$ & $($ & $-\frac{1}{6}$ & $\frac{1}{2}$ & $\frac{1}{6}$ & $\frac{1}{6}$ & $\frac{1}{6}$ & $\frac{1}{6}$ & $\frac{1}{6}$ & $\frac{1}{6}$ & $)$ $($ & $\frac{1}{3}$ & $-\frac{2}{3}$ & $\frac{2}{3}$ & $-\frac{1}{3}$ & $0$ & $0$ & $0$ & $0)$\\
$V_{4,13}$ & $($ & $-\frac{1}{6}$ & $\frac{1}{2}$ & $-\frac{1}{6}$ & $-\frac{1}{6}$ & $-\frac{1}{6}$ & $-\frac{1}{6}$ & $-\frac{1}{6}$ & $-\frac{1}{6}$ & $)$ $($ & $-\frac{1}{3}$ & $-\frac{2}{3}$ & $\frac{1}{3}$ & $-\frac{2}{3}$ & $0$ & $0$ & $0$ & $0)$\\
$ V_{4,33}$ & $($ & $\frac{5}{6}$ & $-\frac{1}{2}$ & $-\frac{1}{6}$ & $-\frac{1}{6}$ & $-\frac{1}{6}$ & $-\frac{1}{6}$ & $-\frac{1}{6}$ & $-\frac{1}{6}$ & $)$ $($ & $-\frac{1}{3}$ & $\frac{1}{3}$ & $\frac{1}{3}$ & $\frac{1}{3}$ & $0$ & $0$ & $0$ & $0)$\\
\hline
\end{tabular}
\caption{Set of 32 line bundle vectors such that they solve the Bianchi identities (\ref{BIeqns}) obtained by using resolution i) for all 12 $\mathbb{C}^{3}/\mathbb{Z}_{\text{6--II}}$ fixed points.
\label{table:BISolutionVectors}}
\end{table}
\renewcommand{\arraystretch}{1.0}

\renewcommand{\arraystretch}{1.2}
\begin{table}
\begin{center}
\subfloat[Massless spectrum of the first E$_{8}$\label{table:E81SpectrumSmall}]{
\begin{tabular}[t]{|c|l||c|l|}
\hline
$\#$	& irrep 									& $\#$	& irrep 										\\
\hline
\hline
3	& $(\mathbf{3},\mathbf{2})_{1/6}$			&  3 		& $(\mathbf{\overline{3}},\mathbf{1})_{-2/3}$  		\\ 
5	& $(\mathbf{\overline{3}},\mathbf{1})_{1/3}$	&  2 		& $(\mathbf{3},\mathbf{1})_{-1/3}$  				\\ 
5	& $(\mathbf{1},\mathbf{2})_{-1/2}$			&  2 		& $(\mathbf{1},\mathbf{2})_{1/2}$		  		\\ 
6	& $(\mathbf{1},\mathbf{1})_{1}$				&  1 		& $(\mathbf{1},\mathbf{1})_{-1}$			  		\\ 
\hline
17	& $(\mathbf{1},\mathbf{1})_{0}$				&  		& 							  				\\ 
\hline
\end{tabular}
}
\subfloat[Massless spectrum of the second E$_{8}$\label{table:E82SpectrumSmall}]{
\begin{tabular}[t]{|c|l@{\hskip.2cm}||c|l@{\hskip1cm}|}
\hline
$\#$	& irrep 									& $\#$	& irrep				\\
\hline
\hline
4	& $(\rep{4})+(\crep{4})$							
& 3 	& $(\mathbf{6})$		\\
44	& $(\mathbf{1})$							&		&					\\
\hline
\end{tabular}
}
\end{center}
\caption{Chiral massless spectrum of the model. The multiplicities are calculated using (\ref{eq:multiplicityOperatorIntegratedOut}). The representations under SU(3) $\times$ SU(2) of the first E$_{8}$ and SU(4) of the second E$_{8}$ are given in boldface. The subscript denotes the hypercharge. \label{table:bothE8}}
\end{table}
\renewcommand{\arraystretch}{1.0}

\begin{table}[t!]
\begin{tabular}{|c||c|c|c|c|c|c|c|c|}
\hline
E$_{8}\times$E$_{8}$  root vector & $\text{irrep}$ & $Q_{Y}$ & $Q_{2}$ & $Q_{3}$ & $Q_{4}$ & $Q_{5}$ & $Q_\text{anom}$ & $N$ \\
\hline
\hline
$(0,0,0,1,0,1,0,0)(0^{8})$ & $(\mathbf{3},\mathbf{2})$ & $\frac{1}{6}$ & $0$ & $0$ & $0$ & $2$ & $-\frac{2}{3}$ & $-2$ \\
$(\frac{1}{2},-\frac{1}{2},\frac{1}{2},-\frac{1}{2},\frac{1}{2},-\frac{1}{2},-\frac{1}{2},\frac{1}{2})(0^{8})$ & $(\mathbf{3},\mathbf{2})$ & $\frac{1}{6}$ & $\frac{1}{2}$ & $-\frac{1}{2}$ & $\frac{1}{2}$ & $-\frac{1}{2}$ & $-\frac{2}{3}$ & $-1$ \\
\hline
$(0,0,0,0,0,1,1,0)(0^{8})$ & $(\mathbf{3},\mathbf{1})$ & $-\frac{2}{3}$ & $0$ & $0$ & $0$ & $2$ & $-\frac{2}{3}$ & $-2$ \\
$(-\frac{1}{2},\frac{1}{2},\frac{1}{2},-\frac{1}{2},-\frac{1}{2},-\frac{1}{2},\frac{1}{2},\frac{1}{2})(0^{8})$ & $(\mathbf{3},\mathbf{1})$ & $-\frac{2}{3}$ & $-\frac{1}{2}$ & $\frac{1}{2}$ & $\frac{1}{2}$ & $-\frac{1}{2}$ & $\frac{8}{3}$ & $-1$ \\
\hline
$(0,-1,0,0,0,-1,0,0)(0^{8})$ & $(\mathbf{3},\mathbf{1})$ & $\frac{1}{3}$ & $0$ & $-1$ & $0$ & $-1$ & $-\frac{2}{3}$ & $-1$  \\
$(0,0,-1,0,0,-1,0,0)(0^{8})$ & $(\mathbf{3},\mathbf{1})$ & $\frac{1}{3}$ & $0$ & $0$ & $-1$ & $-1$ & $-\frac{4}{3}$ & $-1$ \\
$(0,0,1,0,0,-1,0,0)(0^{8})$ & $(\mathbf{3},\mathbf{1})$ & $\frac{1}{3}$ & $0$ & $0$ & $1$ & $-1$ & $2$ & $-2$ \\
$(1,0,0,0,0,-1,0,0)(0^{8})$ & $(\mathbf{3},\mathbf{1})$ & $\frac{1}{3}$ & $1$ & $0$ & $0$ & $-1$ & $-2$ & $-1$ \\
$(-\frac{1}{2},\frac{1}{2},\frac{1}{2},\frac{1}{2},\frac{1}{2},-\frac{1}{2},\frac{1}{2},\frac{1}{2})(0^{8})$ & $(\mathbf{3},\mathbf{1})$ & $\frac{1}{3}$ & $-\frac{1}{2}$ & $\frac{1}{2}$ & $\frac{1}{2}$ & $\frac{3}{2}$ & $2$ & $2$ \\
\hline
$(0,-1,0,-1,0,0,0,0)(0^{8})$ & $(\mathbf{1},\mathbf{2})$ & $-\frac{1}{2}$ & $0$ & $-1$ & $0$ & $-1$ & $-\frac{2}{3}$ & $-2$ \\
$(0,0,1,-1,0,0,0,0)(0^{8})$ & $(\mathbf{1},\mathbf{2})$ & $-\frac{1}{2}$ & $0$ & $0$ & $1$ & $-1$ & $2$ & $-1$ \\
$(1,0,0,-1,0,0,0,0)(0^{8})$ & $(\mathbf{1},\mathbf{2})$ & $-\frac{1}{2}$ & $1$ & $0$ & $0$ & $-1$ & $-2$ & $-2$ \\
$(\frac{1}{2},-\frac{1}{2},\frac{1}{2},-\frac{1}{2},\frac{1}{2},\frac{1}{2},\frac{1}{2},\frac{1}{2})(0^{8})$ & $(\mathbf{1},\mathbf{2})$ & $-\frac{1}{2}$ & $\frac{1}{2}$ & $-\frac{1}{2}$ & $\frac{1}{2}$ & $\frac{3}{2}$ & $-\frac{4}{3}$ & $2$ \\
\hline
$(0,0,0,1,1,0,0,0)(0^{8})$ & $(\mathbf{1},\mathbf{1})$ & $1$ & $0$ & $0$ & $0$ & $2$ & $-\frac{2}{3}$ & $-2$ \\
$(-\frac{1}{2},\frac{1}{2},\frac{1}{2},\frac{1}{2},\frac{1}{2},-\frac{1}{2},-\frac{1}{2},-\frac{1}{2})(0^{8})$ & $(\mathbf{1},\mathbf{1})$ & $1$ & $-\frac{1}{2}$ & $\frac{1}{2}$ & $\frac{1}{2}$ & $-\frac{1}{2}$ & $\frac{8}{3}$ & $1$ \\
$(\frac{1}{2},-\frac{1}{2},\frac{1}{2},\frac{1}{2},\frac{1}{2},-\frac{1}{2},-\frac{1}{2},-\frac{1}{2})(0^{8})$ & $(\mathbf{1},\mathbf{1})$ & $1$ & $\frac{1}{2}$ & $-\frac{1}{2}$ & $\frac{1}{2}$ & $-\frac{1}{2}$ & $-\frac{2}{3}$ & $-4$ \\
\hline
$(-1,-1,0,0,0,0,0,0)(0^{8})$ & $(\mathbf{1},\mathbf{1})$ & $0$ & $-1$ & $-1$ & $0$ & $0$ & $\frac{4}{3}$ & $-1$ \\
$(-1,0,-1,0,0,0,0,0)(0^{8})$ & $(\mathbf{1},\mathbf{1})$ & $0$ & $-1$ & $0$ & $-1$ & $0$ & $\frac{2}{3}$ & $1$ \\
$(-1,0,1,0,0,0,0,0)(0^{8})$ & $(\mathbf{1},\mathbf{1})$ & $0$ & $-1$ & $0$ & $1$ & $0$ & $4$ & $-2$ \\
$(-1,1,0,0,0,0,0,0)(0^{8})$ & $(\mathbf{1},\mathbf{1})$ & $0$ & $-1$ & $1$ & $0$ & $0$ & $\frac{10}{3}$ & $2$ \\
$(0,-1,1,0,0,0,0,0)(0^{8})$ & $(\mathbf{1},\mathbf{1})$ & $0$ & $0$ & $-1$ & $1$ & $0$ & $\frac{2}{3}$ & $1$ \\
$(-\frac{1}{2},-\frac{1}{2},-\frac{1}{2},-\frac{1}{2},-\frac{1}{2},-\frac{1}{2},-\frac{1}{2},-\frac{1}{2})(0^{8})$ & $(\mathbf{1},\mathbf{1})$ & $0$ & $-\frac{1}{2}$ & $-\frac{1}{2}$ & $-\frac{1}{2}$ & $-\frac{5}{2}$ & $\frac{2}{3}$ & $4$ \\
$(-\frac{1}{2},\frac{1}{2},\frac{1}{2},-\frac{1}{2},-\frac{1}{2},-\frac{1}{2},-\frac{1}{2},-\frac{1}{2})(0^{8})$ & $(\mathbf{1},\mathbf{1})$ & $0$ & $-\frac{1}{2}$ & $\frac{1}{2}$ & $\frac{1}{2}$ & $-\frac{5}{2}$ & $\frac{10}{3}$ & $4$ \\
$(\frac{1}{2},-\frac{1}{2},\frac{1}{2},-\frac{1}{2},-\frac{1}{2},-\frac{1}{2},-\frac{1}{2},-\frac{1}{2})(0^{8})$ & $(\mathbf{1},\mathbf{1})$ & $0$ & $\frac{1}{2}$ & $-\frac{1}{2}$ & $\frac{1}{2}$ & $-\frac{5}{2}$ & $0$ & $-2$ \\
\hline
\end{tabular}
\caption{Detailed massless chiral spectrum of the first E$_{8}$ computed from the solution given in Table~\ref{table:BISolutionVectors}. The $Q$'s are the eight U(1) charges computed using the operators defined in \cite{Lebedev:2007hv}. The vectors are grouped by their value $Q_{Y}$ under the hypercharge operator (\ref{eq:Hypercharge}). $Q_\text{anom}$ is the anomalous U(1) charge (\ref{eq:anomalousCharge}) on the orbifold. As we are looking at the first E$_{8}$ only, $Q_{6}$ to $Q_{8}$ are zero. $N$ is the eigenvalue of the corresponding E$_{8}\times$E$_{8}$ root vector under the multiplicity operator.
\label{table:E81DetailedSpectrum}}
\end{table}

A few technical remarks concerning the spectrum are in order. First of all, each of the eigenvalues of $N$ are integer--valued on any of the $240+240$ states, see Table~\ref{table:E81DetailedSpectrum} for the eigenvalues of $N$ in the observable E$_8$. Looking at the complicated structure of $N$ in (\ref{eq:multiplicityOperatorIntegratedOut}) and at the rational entries that appear in the solution of the Bianchi identities (Table \ref{table:BISolutionVectors}), this is a highly non--trivial observation, as the sum in $N$ contains terms like $\frac{7}{6} \cdot \frac{1}{3}^{3}=\frac{7}{162}$. We take this feature as a strong check that the methods we employ in Section~\ref{sc:resZsix} to calculate the intersection numbers are consistent. From the first row block in Table~\ref{table:E81DetailedSpectrum} we conclude that the multiplicity operator on the quark-doublets $q_i$ takes the value $-3$, hence there are 3 quark--doublets. From the second row block we read off that there are also 3 $\bar u_i$. Consequently, from the third row block we infer that there are 5 $\bar d_i$ and 2 charge conjugates $d_i$, because there are two states with the opposite (positive) eigenvalue. Using this analysis the Tables \ref{table:E81SpectrumSmall} and \ref{table:E82SpectrumSmall} are composed.
For the singlet states that do not carry hypercharge, we are not able to make a distinction between states or their charge conjugates w.r.t. the Standard Model gauge group, hence we simply add the absolute value of the eigenvalues of the multiplicity operator.

After these more technical comments we conclude this Subsection with some more physical remarks concerning the spectrum in Tables~\ref{table:E81SpectrumSmall} and~\ref{table:E82SpectrumSmall}. First of all the spectrum is free of any non--Abelian anomalies. (The $\rep{6}$ of SU(4) is self--conjugate hence does not contribute to non--Abelian anomalies.) From the spectrum we read off that there are vector--like exotics for the right--handed down--quarks, the left--handed lepton doublet, and the right--handed electron singlets. 
Disregarding some of these vector--like pairs the spectrum is identical to that of the Standard Model, except for two additional right--handed electrons $(\rep{1},\rep{1})_1$. This means that the spectrum has anomalous U(1)s and in particular the hypercharge is anomalous. This is not a computational error, but rather confirms the general analysis presented in Subsections~\ref{sc:anom} and~\ref{sc:Axions}: The hypercharge is necessarily broken as it is part of the structure group of the bundle~\cite{Distler:1987ee}. By explicitly computing the inner product it is immediately apparent that exactly those line bundle vectors, whose identification contains  the Wilson line $W_{2}$ have non--vanishing inner product with the hypercharge operator, and hence it is anomalous. This Wilson line is responsible for breaking the SU(5) to the SU(3) $\times$ SU(2) Standard Model gauge group. 
In the conclusions we discuss a couple of possibilities how one can avoid that this implies that the hypercharge is broken.

\subsection{Identification of line bundle vectors with twisted states}
\label{sc:SolutionProperties}

Another interesting observation is that the conditions on the squares of the vectors that we obtained from the simplified Bianchi identities (\ref{BIeqnsSimplified}) closely resemble the mass--shell condition (\ref{eq:massless}). In the $\theta$--sector, one finds from the condition in the massless case $M_L=0$ with $(p+V_g)=V_{1,\beta\gamma}$ and $\delta c = \frac{11}{36}$: 
\begin{equation*}
V_{1,\beta\gamma}^{2} = \frac{25}{18} - 2 \tilde{N}~, \qquad \beta \in \left\{1,2,3\right\},~\gamma \in \left\{1,2,3,4\right\}.
\end{equation*}
\noindent In the case of vanishing oscillator number, this is exactly the condition found in (\ref{eq:BIeqnsSimplifiedA}). A similar observation is made when considering the $\theta^{3}$--sector. Here the massless equation reads with $\delta c = \frac{1}{4}$:
\begin{equation*}
V_{3,\alpha\gamma}^{2} = \frac{3}{2} - 2 \tilde{N}~, \qquad \alpha \in \left\{1,2\right\}
\end{equation*}
\noindent which is again the same condition as we obtained in (\ref{eq:BIeqnsSimplifiedC}) with the oscillator number set to zero. However, for the $\theta^{2}$-- and $\theta^{4}$--sector, things look a bit different. Here, $\delta = \frac{2}{9}$, and one finds
\begin{eqnarray*}
V_{2,\alpha\beta}^{2} & = & \frac{14}{9} - 2 \tilde{N}~, \qquad \alpha \in \left\{1,3\right\}\nonumber\\
V_{4,\alpha\beta}^{2} & = & \frac{14}{9} - 2 \tilde{N}~, \qquad \alpha \in \left\{1,3\right\}.\nonumber
\end{eqnarray*}
In this sector the simplified Bianchi identities (\ref{eq:BIeqnsSimplifiedB}) dictate a non--vanishing $\tilde{N}$ for the massless spectrum. The solution we are giving in Table \ref{table:BISolutionVectors} was modified such that $V_{2,\alpha\beta}^{2} = V_{4,\alpha\beta}^{2} = \frac{14}{9}$ holds for as many of these vectors as  considered possible (i.e. for all but $V_{2,11}$ and $V_{2,12}$). Note that it was also condition (\ref{eq:BIeqnsSimplifiedB}) that was relaxed in order to find a solution. This shows that it is also possible to demand $V_{2,\alpha\beta}^{2} =  V_{4,\alpha\beta}^{2} = \frac{14}{9}$ instead of (\ref{eq:BIeqnsSimplifiedB}). However, this does not change the fact that $(V_{2,\alpha\beta},V_{4,\alpha\beta})\stackrel{!}{=}\frac{8}{3}$ which is needed to solve (\ref{eq:BIeqnsB}), but it does not allow for the identification $V_{2,\alpha\beta}=-V_{4,\alpha\beta}$.\\
Interestingly, in our solution $V_{2,11}^{2}=V_{2,12}^{2}=\frac{20}{9}$, which cannot be satisfied with a non--negative oscillation number. So these two states have a non--zero mass,
\begin{equation*}
M_{L}^{2}=\frac{8}{3}+8 \tilde{N},
\end{equation*}
and hence correspond to massive twisted states. The level--matching condition $M_{L}^{2}=M_{R}^{2}$ given below (\ref{eq:massless}) can still be satisfied by choosing an appropriate SO(8) vector $q$. The fact that these states have a non--zero mass seems to imply that the fixed points $E_{2,11}$ and $E_{2,12}$ do not acquire a vacuum expectation value and hence are not blown up. However, these fixed points are nevertheless resolved in the sense that the singularities have been cut out and a resolution has been glued in as discussed in Section~\ref{sc:resZsix}. The resolution simply does not have a finite volume, i.e. its K\"ahler modulus vanishes.

As mentioned above, the solution for the Bianchi identities is by far not unique. The non--uniqueness is two--fold: On the one hand, given a combination of resolutions for the twelve fixed points (in our example resolution i) twelve times), it is possible to find different combinations of line bundle vectors that satisfy the associated Bianchi identities. Different solutions exhibit different behavior with respect to the unbroken gauge groups and the number of scalars and vector--like exotics in the model. For example, if one adds the E$_{8}\times$E$_{8}$ lattice vector $(0^{3},1^{2},0^{3})(0^{8})$ to $V_{1,11}$ and/or to $V_{1,13}$, the particle content of the model is changed with respect to the exotics, yet the new set of vectors still satisfies the Bianchi identities. On the other hand, given a set of 32 line bundle vectors, there exist different combinations of local resolutions such that the resulting Bianchi identities are satisfied by this set of vectors. In our case, one could for example use any combination of the five possible triangulations at the fixed points $E_{1,31}$ and $E_{1,33}$.

\begin{table}
\centering
$
\begin{array}{|l|l|l|l|}
\hline
\theta \text{--sector} 						& \theta^{2} \text{--sector} 					& \theta^{3} \text{--sector} 						& \theta^{4} \text{--sector}			\\
\hline
\hline
V_{1,11} \leftrightarrow \overline{n}_{1} 		& V_{2,11} \leftrightarrow  \text{massive}		& V_{3,11} \leftrightarrow  \text{projected out} 	& V_{4,11} \leftrightarrow s_{26}^{0}	\\ 
V_{1,12} \leftrightarrow s_{2}^{-}				& V_{2,12} \leftrightarrow  \text{massive} 	& V_{3,12} \leftrightarrow s_{14}^{+}				& V_{4,12} \leftrightarrow h_{7}		\\ 
V_{1,13} \leftrightarrow \overline{n}_{4} 		& V_{2,13} \leftrightarrow \overline{n}_{15}	& V_{3,13} \leftrightarrow  \text{projected out} 	& V_{4,13} \leftrightarrow h_{12}	\\ 
V_{1,14} \leftrightarrow s_{5}^{-}				& V_{2,31} \leftrightarrow h_{20} 			& V_{3,14} \leftrightarrow s_{18}^{+}				& V_{4,31} \leftrightarrow s_{28}^{0}	\\
V_{1,21} \leftrightarrow n_{4}				& V_{2,32} \leftrightarrow h_{21} 			& V_{3,21} \leftrightarrow  \text{projected out}		& V_{4,32} \leftrightarrow h_{9}		\\
V_{1,22} \leftrightarrow x_{1}^{-}				& V_{2,33} \leftrightarrow h_{25}			& V_{3,22} \leftrightarrow s_{15}^{+}				& V_{4,33} \leftrightarrow n_{8}		\\
V_{1,23} \leftrightarrow n_{6}				&										& V_{3,23} \leftrightarrow  \text{projected out}		& 								\\
V_{1,24} \leftrightarrow x_{2}^{-}				& 										& V_{3,24} \leftrightarrow s_{19}^{+}				& 								\\
V_{1,31} \leftrightarrow w_{1}				&										& 											& 								\\
V_{1,32} \leftrightarrow s_{7}^{-}				&										& 											& 								\\
V_{1,33} \leftrightarrow w_{2}				&										& 											& 								\\
V_{1,34} \leftrightarrow s_{10}^{-}			&										&			 								& 								\\
\hline
\end{array}
$
\caption{Identification between the orbifold states and the line bundle vectors. The nomenclature of the twisted states is summarized in Table~\ref{bm2spectrum} and taken from \cite{Lebedev:2007hv}. Here ``massive'' means that the vector corresponds to a massive orbifold state. The vectors tagged with ``projected out'' are present in the six dimensional theory but are projected out in four dimensions.}
\label{table:vectorIdentifications}
\end{table} 

As discussed in Subsection~\ref{sc:anom} below (\ref{multipletScalarAxionsEquations}), the axionic states can be identified with twisted states from the orbifold. The spectrum of the orbifold for the chosen model is given in the appendix of \cite{Lebedev:2007hv}. The identifications can be made by comparing the bundle vectors with the weights of the corresponding twisted states (or equivalently by comparing the charges and non--Abelian representations given in the table of \cite{Lebedev:2007hv}). Since the complex scalars in chiral multiplets are composed of two real scalars, that are each other's charge conjugates, these identifications are made up to overall signs of the weight vectors. Table \ref{table:vectorIdentifications} gives a list of the line bundle vectors and the corresponding orbifold states. In the $\theta^{2}$--sector, the two vectors $V_{2,11}$ and $V_{2,12}$ that acquire a mass do not have a matching orbifold state. All other states from this sector can be identified with line bundle vectors. In the $\theta^{3}$ case this is similar. For $V_{3,2\gamma}$, also each line bundle vector can be found in the third twisted sector. The four vectors $V_{3,1\gamma}$ are present in the six dimensional spectrum, but are projected out in four dimensions. In the $\theta$-- and $\theta^{4}$--sector, there is an orbifold state for each line bundle vector.\\
\indent The possibility of identifying orbifold states and line bundle vectors suggests a different approach to solving the Bianchi identities. Namely one starts with a set of 32 line bundle vectors that satisfy the resolution independent Bianchi identities (\ref{RBianchis}) and scans over possible combinations of triangulations. At first sight this seems hopeless due to the vast amount of physically inequivalent models that can be obtained by combining the five different resolutions (cf. Section~\ref{sc:counting} in the appendix). However, as stated at the beginning of Subsection~\ref{sc:ComputingBianchis}, the twelve Bianchi identities obtained from integrating over $E_{1,\beta\gamma}$ depend on the local resolution only. This makes it possible to check for each of the twelve $\mathbb{C}^{3}/\mathbb{Z}_{\text{6--II}}$ fixed points which of the five resolutions are allowed. One can hope that this leaves a subset small enough to compute the Bianchi identities for all possible combinations of the subset and check whether there exists a combination of resolutions such that the associated Bianchi identities are solved by the initially chosen set of line bundle vectors.

\section{Conclusions}
\label{sc:concl}

One of the objectives of string phenomenology is to construct string realizations of MSSM--like models. Heterotic orbifolds have been  successful in achieving this goal, especially those which build on the $T^6/\Intr_\text{6--II}$ orbifold. Heterotic orbifolds can be exactly described using CFT techniques, while Calabi--Yau compactifications are mostly described in the supergravity regime. Orbifolds as geometrical spaces are often considered as singular limits of smooth Calabi--Yau manifolds; it is therefore interesting to investigate what happens to these orbifold models in blowup, i.e.\ when all singularities are smoothed out. In this paper we explained how  $T^6/{\mathbb Z}_\text{6--II}$ orbifold models can be recovered in a supergravity language. We achieved this as follows: 

To set the stage we began by reviewing the construction of heterotic MSSM $T^6/\Intr_\text{6--II}$ orbifolds. These orbifold models are characterized by a geometrical shift, and a gauge shift and Wilson lines that act on the gauge degrees of freedom, which are severely constrained by modular invariance. The twisted sectors live on codimension six singularities $\Cplx^3/\Intr_\text{6--II}$ and two types of codimension four singularities, $\Cplx^2/\Intr_3$ and $\Cplx^2/\Intr_2$. To allow for the interpretation of some of the heterotic $T^6/\Intr_\text{6--II}$ orbifolds as MSSM--like models it was crucial to identify a non--anomalous hypercharge. 

To determine a supergravity realization of an  $T^6/\Intr_\text{6--II}$ orbifold model, the first step was to characterize the topological properties of the smooth Calabi--Yau geometry corresponding to the blowup of this orbifold. The isolated conical singularities of this orbifold were resolved using toric geometry techniques that identify exceptional divisors hidden inside them. On these exceptional cycles the twisted states can be thought of being localized in the resolved picture, but in the blow down regime. These local resolutions were glued together according to the procedure described in detail in~\cite{Lust:2006zh}. This gluing process adds three more divisors inherited from the covering torus $T^6$, giving 35 divisors in total. We described their intersection numbers, the characteristic classes and the K\"ahler cone of the global resolution. The $T^6/{\mathbb Z}_\text{6--II}$ orbifold can be resolved in many topologically distinct ways: Since each of the 12 $\Cplx^3/{\mathbb Z}_\text{6--II}$ singularities has five possible resolutions, there are almost two millions ways to do so.

In the second step we singled out the analog of the gauge shift vector and the Wilson lines from the orbifold theory in blowup. We considered wrapped Abelian gauge fluxes on the exceptional divisors as line bundles for two reasons: This choice automatically fulfills the requirement that they must be (1,1)--forms, and for such gauge backgrounds we could formulate a precise identification with these orbifold inputs (gauge flux on the inherited divisors would correspond to orbifolds with magnetized tori). The 24 resolution dependent consistency conditions are obtained by integrating the Bianchi identity  of the anti--symmetric tensor field over 35 divisors of the resolution of $T^6/{\mathbb Z}_\text{6--II}$. These conditions play a similar role as the modular invariance conditions on the orbifold: They severely constrain the possible choices for the 32 bundle vectors that characterize the line bundles embedded in E$_{8}\times$E$_{8}$. 

After this we studied the basic physical properties of the resulting blowup models that can be analyzed by topological means only. We computed the spectrum of the resolved models by integrating the ten dimensional gaugino anomaly polynomial  on the resolved $T^6/\Intr_\text{6--II}$ orbifold in the presence of the Abelian gauge fluxes. The multiplicities of the states, that appear in the decomposition of the E$_{8}\times$E$_{8}$ w.r.t.\ the unbroken gauge group, can be determined from the resulting four dimensional anomaly polynomial (this approach is more sensitive than standard index theorems, because any state, even if it is not charged under the Standard Model, is chiral with respect to some of the U(1) factors singled out by the gauge flux). Moreover, this analysis shows that the non--Abelian anomalies are absent, provided that the 24 Bianchi identities are satisfied, but in accordance with~\cite{Blumenhagen:2005ga,Blumenhagen:2005pm} multiple anomalous U(1)'s are possible. These anomalous U(1)'s are canceled by anomalous variations of axions $\gb_r$ that appear in the expansion of the anti--symmetric tensor $B_2$ on the resolution. Their shift transformations reveal that they are the reincarnation of the twisted states $\gPs_r$, that generated the blowup from the orbifold point of view by taking non--vanishing VEV's (one per fixed point or fixed line). Their identification $\Psi_r\sim e^{2\pi U_r}$, where the chiral superfields $U_r$ contain the axions $\gb_r$ (see~\cite{Groot Nibbelink:2007ew}) and the \Kh\ moduli $b_r$ that measure the volumes of the exceptional divisors, has some far reaching consequences.

As the identification of a non--anomalous hypercharge was a crucial ingredient in the orbifold construction of MSSM--like models, it was important to study the fate of the hypercharge in blowup. The anomaly analysis on the resolution showed that the hypercharge is always among the anomalous U(1)'s: The identification of the line bundle vectors with the orbifold shift vector and Wilson lines implies that the hypercharge is never perpendicular to all of them, because one of the Wilson lines is responsible for the breaking of a GUT group down to the Standard Model. The identification between the blowup axions and the twisted states with VEV's lead to the same conclusion: If the orbifold is completely blown up, for each of the 32 exceptional divisors a twisted state takes a non--vanishing VEV. Since any of the heterotic MSSM models in the
``mini--landscape''~\cite{Lebedev:2006kn}  has fixed points where all the twisted fields are charged under the hypercharge (or under some other MSSM gauge interactions), the Standard Model gauge group can only be preserved if not all singularities are blown up. 

Such a  ``partial'' blowup does not render the construction of the orbifold resolution inconsistent, because the amount of blowup is irrelevant as a resolution is defined by topological information only, but signals the loss of control over the supergravity construction: 
Defining a (partial) blow down by having vanishing VEV's for (some of) the twisted states implies via the identification $\Psi_r\sim e^{2\pi U_r}$ that (some of) the \Kh\ moduli $b_r \ra -\infty$: The volumes (of some) exceptional cycles do not tend to zero but become minus infinite using the standard ``classical'' geometrical notion of volumes. According to~\cite{Aspinwall:1993xz} this regime can only be described as a ``string'' geometry, where string corrections to the supergravity description become dominant. This is in particular true for the orbifold point in the moduli space where the orbifold CFT provides the correct description. 

Only purely topological quantities are unaffected by the break down of the supergravity description in (a partial) blow down. At the level of gauge symmetry and spectrum the only minor caveat is  that in the blow down regime the blowup moduli $\Psi_r$ have vanishing VEV's. This might result in gauge symmetry enhancement, axions reappearing as twisted states, and some vector--like states might become massless via vanishing Yukawa couplings that involve these blowup moduli. These effects are completely under control in the orbifold description, but cannot be followed in the supergravity description as it needs to receive string correction which are (mostly) unknown. As these stringy effects lead to corrections to the Hermitian Yang--Mills equations, in particular the stability of the bundle also becomes a subtle issue in (a partial) blow down. However, at least for line bundles, stability is just a constraint on the K\"ahler moduli of the space, but not on the existence of these Abelian bundles itself. This means that our identification of the bundle vectors and the orbifold gauge shift and Wilson lines is self-consistent. 

In order to check and to illustrate our general findings we specified our study to the construction of a blowup of a heterotic orbifold  having the MSSM spectrum, the so--called ``benchmark model~2''~\cite{Lebedev:2006kn,Buchmuller:2007zd,Lebedev:2007hv}. For the explicit study of a blowup it is crucial to solve the 24 resolution dependent Bianchi identities. This is a highly non--trivial task in general, but we found two procedures to address this problem. In the first approach one chooses a particular resolution of the 12 $\Cplx^3/\Intr_\text{6--II}$ fixed points; in our case we took resolution i) for all of them. By making some simplifying assumptions on the bundle vectors, the system of equation reduces to a set from which a solution can be guessed much easier. Once such a solution is found, it can be easily modified to satisfy additional requirements. In this way we found a resolution which preserves the SU(2) and SU(3) of the Standard Model, and an SU(4) in the hidden sector. The spectrum in blowup is essentially that of the MSSM with some exotics and two additional right-handed electrons. But as the general analysis indicated the hypercharge is broken in full blowup. 

The identification between the blowup axions and the twisted states whose VEV's generate the blowup suggests a second method to construct blowup models: The twisted states are identified by weight vectors (or shifted momenta) on the gauge E$_{8}\times$E$_{8}$ lattice. The identification implies that these weight vectors can be interpreted as bundle vectors that define the blowup using Abelian fluxes. This allows one to chose 32 bundle vectors corresponding to twisted states at each of the fixed points and lines, and then search for a resolution of the $T^6/\Intr_\text{6--II}$ for which the 24 Bianchi identities are solved by this choice. As our explicit example using resolution i) everywhere showed, these twisted states can be those that exist as massless twisted states in an effective four dimensional orbifold model, as massless twisted states in six dimensions that are projected out, and even as massive twisted states. Because there are almost two million possible resolutions of the $T^6/\Intr_\text{6--II}$, determining solutions in this way can be a time consuming process. 

Finally, let us comment on possibilities to overcome the problem that 
none of the models studied in~\cite{Lebedev:2006kn} will have an unbroken U(1)$_{Y}$ in complete blowup. First of all, from the string moduli space perspective it is not impossible that some cycles remain so small that the hypercharge is effectively unbroken. This would imply that the Standard Model is only realized close to special points in moduli space with enhanced symmetry with singular geometry. The only price one pays is a large fine tuning between the volumes of different cycles that may be considered unnatural. Secondly one could imagine that the VEV of twisted states (i.e.\ Standard Model Higgses) is such that electroweak symmetry breaking corresponds to blowing up some fixed points. Since the electroweak scale is much smaller than the Planck scale this reintroduces the fine tuning problem. 

Another option could be to consider a model where the GUT group is not broken in any of the orbifold singularities, i.e.  by any of the orbifold actions {\it having fixed points}
in the internal space (see e.g. \cite{Hebecker:2004ce}). Unfortunately we are not aware of any model built on such geometries having just the MSSM spectrum. Alternatively, one could think about models where the hypercharge is not arising from a GUT breaking like SU(5)$\rightarrow$ SU(3)$\times $SU(2)$\times $U(1), but it includes a mixing
with other U(1)'s present in heterotic models. Such a non--GUT embedding of the hypercharge in the heterotic orbifold has been considered in~\cite{Raby:2007yc}. However, for the only model described explicitly there, the hypercharge is not perpendicular to all the bundle vectors that arise from the chosen gauge shift and Wilson lines, hence still the hypercharge is broken in complete blowup in that particular model. Whether this is a general feature of all models with non--GUT embedded hypercharge would require further study.

\subsection*{Acknowledgements}

We would like to thank Tae-Won Ha, Babak Haghighat, Arthur Hebecker, Albrecht Klemm, Hans-Peter Nilles, Stuart Raby, Michael Ratz, Emanuel Scheidegger and Timo Weigand for discussions, and Dieter L\"ust for helpful correspondence. The work of MT is supported by the European Community through the contract N 041273 (Marie Curie Intra-European Fellowships). He is also  partially supported by the ANR grant  ANR-05-BLAN-0079-02, the RTN contracts MRTN-CT-2004-005104 and MRTN-CT-2004-503369, the CNRS PICS \#~2530, 3059 and 3747, and by the European Union Excellence Grant  MEXT-CT-2003-509661. The work of P.V. is supported by LMU Excellent, he would like to thank the Institute for Theoretical Physics of Heidelberg University for hospitality and support.

\appendix 
\def\theequation{\thesection.\arabic{equation}} 
\setcounter{equation}{0}

\section{Counting of different triangulations}
\label{sc:counting}
\begin{table}
\centering
\begin{tabular}{p{3.5cm}p{3.5cm}p{3.5cm}p{3.5cm}}
$ R_1R_2R_3=6 $&$ R_{2} E_{3,1\,\gamma}^2=-2 $&$ R_{2} E_{3,2\,\gamma}^2=-6 $&$ R_{3} E_{2,1\,\beta}^2=-2 $\\
$ R_{3} E_{2,3\,\beta}^2=-4 $&$ R_{3} E_{4,1\,\beta}^2=-2 $&$ R_{3} E_{4,3\,\beta}^2=-4 $&$ R_{3} E_{2,1\,\beta} E_{4,1\,\beta}=1 $\\
$ R_{3} E_{2,3\,\beta} E_{4,3\,\beta}=2$
\end{tabular}
\caption{The triangulation independent intersection numbers of Res$\left(T^6/\mathbb{Z}_{\text{6--II}}\right)$.}
\label{tab:indinttable}
%\end{table}
%\begin{table}
\centering
\subfloat{
\begin{tabular}{p{3.5cm}p{3.5cm}p{3.5cm}p{3.5cm}}
\multicolumn{4}{c}{Triangulation i)}\\
$E_{1,\beta\gamma}^3=6 $&$ E_{2,1\,\beta}^3=8 $&$ E_{3,1\,\gamma}^3=8 $&$E_{4,1\,\beta}^3=8 $\\
$ E_{1,\beta\gamma}E_{2,1\,\beta}^2=-2 $&$ E_{1,\beta\gamma}E_{3,1\,\gamma}^2=-2 $&$E_{1,\beta\gamma}E_{4,1\,\beta}^2=-2 $&$ E_{1,\beta\gamma}E_{2,1\,\beta}E_{4,1\,\beta}=1 $\\
$ E_{2,1\,\beta}^2E_{4,1\,\beta}=-2$
\end{tabular}}
\\
\subfloat{
\begin{tabular}{p{3.5cm}p{3.5cm}p{3.5cm}p{3.5cm}}
\multicolumn{4}{c}{Triangulation ii)}\\
$ E_{1,\beta\gamma}^3=7 $&$ E_{2,1\,\beta}^3=8 $&$ E_{3,1\,\gamma}^3=5 $&$ E_{4,1\,\beta}^3=4 $\\
$ E_{1,\beta\gamma} E_{2,1\,\beta}^2=-2 $&$ E_{1,\beta\gamma} E_{3,1\,\gamma}^2=-1 $&$ E_{1,\beta\gamma}^2 E_{3,1\,\gamma}=-1 $&$ E_{1,\beta\gamma} E_{4,1\,\beta}^2=-1 $\\
$ E_{3,1\,\gamma} E_{4,1\,\beta}^2=-1 $&$ E_{1,\beta\gamma}^2 E_{4,1\,\beta}=-1 $&$ E_{2,1\,\beta}^2 E_{4,1\,\beta}=-2 $&$ E_{3,1\,\gamma}^2 E_{4,1\,\beta}=-1 $\\
$E_{1,\beta\gamma} E_{2,1\,\beta} E_{4,1\,\beta}=1 $&$ E_{1,\beta\gamma} E_{3,1\,\gamma} E_{4,1\,\beta}=1$
\end{tabular}}
\\
\subfloat{
\begin{tabular}{p{3.5cm}p{3.5cm}p{3.5cm}p{3.5cm}}
\multicolumn{4}{c}{Triangulation iii)}\\
$ E_{1,\beta\gamma}^3=8 $&$ E_{2,1\,\beta}^3=4 $&$ E_{3,1\,\gamma}^3=2 $&$ E_{4,1\,\beta}^3=8 $\\
$ E_{1,\beta\gamma} E_{2,1\,\beta}^2=-1 $&$ E_{1,\beta\gamma}^2 E_{2,1\,\beta}=-1 $&$  E_{1,\beta\gamma}^2 E_{3,1\,\gamma}=-2 $&$ E_{2,1\,\beta} E_{3,1\,\gamma}^2=-1 $\\
$ E_{2,1\,\beta}^2 E_{3,1\,\gamma}=-1 $&$ E_{2,1\,\beta} E_{4,1\,\beta}^2=-4 $&$ E_{2,1\,\beta}^2 E_{4,1\,\beta}=2 $&$ E_{3,1\,\gamma} E_{4,1\,\beta}^2=-2 $\\
$ E_{1,\beta\gamma}E_{2,1\,\beta}E_{3,1\,\gamma}=1$&$E_{2,1\,\beta}E_{3,1\,\gamma}E_{4,1\,\beta}=1$
\end{tabular}}
\\
\subfloat{
\begin{tabular}{p{3.5cm}p{3.5cm}p{3.5cm}p{3.5cm}}
\multicolumn{4}{c}{Triangulation iv)}\\
$ E_{1,\beta\gamma}^3=9 $&$ E_{2,1\,\beta}^3=8 $&$ E_{3,1\,\gamma}^3=-1 $&$ E_{4,1\,\beta}^3=8 $\\
$ E_{1,\beta\gamma} E_{3,1\,\gamma}^2=1 $&$ E_{1,\beta\gamma}^2 E_{3,1\,\gamma}=-3 $&$ E_{2,1\,\beta}^2 E_{3,1\,\gamma}=-2 $&$ E_{2,1\,\beta} E_{4,1\,\beta}^2=-4 $\\
$ E_{2,1\,\beta}^2 E_{4,1\,\beta}=2 $&$ E_{3,1\,\gamma} E_{4,1\,\beta}^2=-2 $&$ E_{2,1\,\beta} E_{3,1\,\gamma} E_{4,1\,\beta}=1$
\end{tabular}}
\\
\subfloat{
\begin{tabular}{p{3.5cm}p{3.5cm}p{3.5cm}p{3.5cm}}
\multicolumn{4}{c}{Triangulation v)}\\
$ E_{1,\beta\gamma}^3=8 $&$ E_{2,1\,\beta}^3=8 $&$ E_{3,1\,\gamma}^3=8 $&$ E_{1,\beta\gamma} E_{2,1\,\beta}^2=-2 $\\
$ E_{1,\beta\gamma}^2 E_{4,1\,\beta}=-2 $&$ E_{2,1\,\beta}^2 E_{4,1\,\beta}=-2 $&
$E_{1,\beta\gamma} E_{2,1\,\beta} E_{4,\beta\gamma}=1$
\end{tabular}}
\caption{Triangulation dependent intersection numbers of Res$\left(T^6/\mathbb{Z}_{\text{6--II}}\right)$ for the five cases in which all twelve $\mathbb{C}^3/\mathbb{Z}_{\text{6--II}}$ fixed points are resolved with the same triangulation.}
\label{tab:inttable}
\end{table}

Here we want to describe the manner in which we counted different triangulation possibilities. In order to do so, we remind the reader that the self--intersection numbers are obtained by multiplying the linear equivalence relations (\ref{eq:leq}) with all curves and solving the system of linear equations. From this it follows that self--intersections containing only a certain $\beta$ depend on the triangulations of all fixed points labeled by this $\beta$. Consider for example
\begin{gather}
R_1 ~\sim~ 6D_{1,1}+\sum\limits_{\beta^{'}=1}^3\sum\limits_{\gamma=1}^4{E_{1,\beta^{'}\gamma}}+\sum\limits_{\beta^{'}=1}^3{\left(2E_{2,1\,\beta^{'}}+4E_{4,1\,\beta^{'}}\right)}+3\sum\limits_{\gamma=1}^4{E_{3,1\,\gamma}}\quad |\cdot E_{2,1\,\beta}^2\\
\Rightarrow\quad0~\sim~\sum\limits_{\gamma=1}^4{E_{1,\beta\gamma}E_{2,1\,\beta}^2}+2E_{2,1\,\beta}^3+4E_{2,1\,\beta}^2E_{4,1\,\beta}+3\sum\limits_{\gamma=1}^4{E_{2,1\,\beta}^2E_{3,1\,\gamma}}~.\nonumber
\end{gather}
The intersection numbers that do not contain $\gamma$ depend via those containing $\gamma$ on the triangulation of all fixed points over which is summed. These are all the fixed points that are labeled by $\beta$. The sum over $\gamma$ implies that only the total number of occurring triangulations is important, but not which triangulation is chosen for which $\gamma$ in particular.

The same argument holds if one considers self--intersections that contain only a certain $\gamma$. Those numbers depend on the triangulations of all fixed points labeled by this $\gamma$, but again only the total number of occurring triangulations is important.

To visualize this we introduce a $3\times4$ matrix
\begin{equation}
M=\left(\begin{array}{cccc} 
t_{1,1}&t_{1,2}&t_{1,3}&t_{1,4}\\
t_{2,1}&t_{2,2}&t_{2,3}&t_{2,4}\\
t_{3,1}&t_{3,2}&t_{3,3}&t_{3,4}\\
\end{array}\right)~,\nonumber
\end{equation}
where each entry $t_{\beta,\gamma}$ can take a value from $1$ to $5$. Therefore $M$ represents a specific triangulation of the resolved orbifold. The triangulations of the fixed points with a fixed $\beta$ are given by a row of $M$, while the columns represent the triangulations of the fixed points with a fixed $\gamma$. The above considerations mean, that two triangulations are equivalent if the corresponding matrices can be transferred into each other by permutations of whole columns and whole rows.

An estimate of the inequivalent triangulations is obtained by taking one column as one index which runs from $1$ to $5^3$, symmetrize in the four indices and divide the result by $3!$ (the permutation symmetry factor of a column). This gives
\begin{equation}
p_1=\frac{1}{3!}\left(\begin{array}{c} 5^3+4-1\\4\end{array}\right)\approx1.78\times10^6~.
\end{equation}
Taking one row as one index (running from $1$ to $5^4$), symmetrizing the resulting three indices and dividing by $4!$ gives
\begin{equation}
p_2=\frac{1}{4!}\left(\begin{array}{c} 5^4+3-1\\3\end{array}\right)\approx1.70\times10^6~.
\end{equation}
These two estimates are not equal. This is due to the fact that we divided by the full symmetry factor of columns and rows, respectively. By doing this we underestimate the total number, since for example the case $t_{\beta,\gamma}=1$ (for all $\beta,\gamma$) is invariant under permutation and should not be divided by the symmetry factor. By using a computer to check how many matrices there are that cannot be converted into each other by interchanging rows and columns we found that the number of physically different triangulations is
\begin{equation}
\#(\text{triangulations})=1.797.090~,
\end{equation}
which is quite close to the two estimates made above.

\begin{table}
\centering
\begin{tabular}{|l||l|}
\hline
divisor $S$ & Vol$(S)=\frac{1}{2}J^2S$\\
\hline
\hline
$E_{1,\beta\gamma}$ & $3 b_{1,\beta\gamma}^2- b_{3,1\,\gamma}^2 -(b_{2,1\,\beta};b_{4,1\,\beta})$\\
$E_{2,1\,\beta}$ & $a_3(2b_{2,1\,\beta}b_{4,1\,\beta})+4b^2_{2,1\,\beta}-2b_{2,1\,\beta}b_{4,1\,\beta}+\sum\limits_{\gamma=1}^4{b_{1,\beta\gamma}}(-2b_{2,1\,\beta}+b_{4,1\,\beta})$\\
$E_{4,1\,\beta}$ & $-a_3(b_{2,1\,\beta}-2b_{4,1\,\beta})-b^2_{2,1\,\beta}+4b^2_{4,1\,\beta}+\sum\limits_{\gamma=1}^4{b_{1,\beta\gamma}}(b_{2,1\,\beta}-2b_{4,1\,\beta})$\\
$E_{2,3\,\beta}$ & $2a_3(2b_{2,3\,\beta}-b_{4,3\,\beta})$\\
$E_{4,3\,\beta}$ & $-2a_3(b_{2,3\,\beta}-2b_{4,3\,\beta})$\\
$E_{3,1\,\gamma}$ & $2a_2b_{3,1\,\gamma}+4b^2_{3,1\,\gamma}-2b_{3,1\,\gamma}\sum\limits_{\beta=1}^3{b_{1,\beta\gamma}}$\\
$E_{3,2\,\gamma}$ & $6a_2b_{3,2\,\gamma}$\\
$R_1$ & $6a_2a_3$\\
$R_2$ & $6a_1a_3-\sum\limits_{\gamma=1}^4{b^2_{3,1\,\gamma}}-3\sum\limits_{\gamma=1}^4{b^2_{3,2\,\gamma}}$\\$R_3$ & $6a_1a_2-\sum\limits_{\beta=1}^3{(b_{2,1\,\beta};b_{4,1\,\beta})}-2\sum\limits_{\beta=1}^3{(b_{2,3\,\beta};b_{4,3\,\beta})}$\\
\hline
\end{tabular}
\caption{The volumes of the divisors of Res$\left(T^6/\mathbb{Z}_\text{6--II}\right)$. }
\label{tab:DivVol}
\end{table}

\section{Details of $\boldsymbol{T^6/\Intr_\text{6--II}}$ resolutions}
\label{sc:detailsres}

In this appendix we give some details of resolutions of ${T^6/\Intr_\text{6--II}}$ that are triangulation dependent. In Table~\ref{tab:indinttable} and Table~\ref{tab:inttable} we give the triangulation independent intersection numbers  and the triangulation dependent intersection numbers for the cases in which all twelve $\mathbb{C}^3/\mathbb{Z}_{\text{6--II}}$ fixed points have the same triangulation.

Table~\ref{tab:DivVol} gives the volumes of the exceptional and inherited divisors. The volumes of the inherited divisors have to be larger than zero. Exceptional divisors of blown up fixed points also obtain positive volumes.

The volumes of compact curves are given in Table~\ref{tab:DepCurvVol} and Table~\ref{tab:IndCurvVol}. To get positive volumes, $a_i$ and $b_i$ have to be positive. All volumes depend on the chosen triangulation. The results given in the tables are obtained if all fixed points are resolved with triangulation i). Furthermore, the curves in Table~\ref{tab:DepCurvVol} exist only in this case, while those of Table~\ref{tab:IndCurvVol} exist for all triangulations.

\begin{table}
\centering
\begin{tabular}{|p{2.5cm}|p{4.5cm}||p{2.5cm}|p{4.5cm}|}
\hline
curve $C$&Vol$(C)=J C$&curve $C$ &Vol$(C)=J C$\\
\hline
\hline
$D_{1,1}E_{1,\beta\gamma}$ & $b_{1,\beta\gamma}-b_{3,1\,\gamma}-b_{4,1\,\beta}$&
$E_{1,\beta\gamma}E_{4,1\,\beta}$ & $-b_{2,1\,\beta}+2 b_{4,1\,\beta}$\\
$E_{1,\beta\gamma}E_{2,1\,\beta}$ & $2 b_{2,1\,\beta}-b_{4,1\,\beta}$&
$D_{2,\beta}E_{1,\beta\gamma}$ & $2 b_{1,\beta\gamma}-b_{2,1\,\beta}$\\
$D_{3,\gamma}E_{1,\beta\gamma}$ & $3 b_{1,\beta\gamma}-b_{3,1\,\gamma}$&
$E_{1,\beta\gamma}E_{3,1\,\gamma}$ & $b_{3,1\,\gamma}$\\
\hline
\end{tabular}
\caption{Volumes of the curves that exist if all $\mathbb{Z}_{\text{6--II}}$ fixed points are resolved with triangulation~i).}
\label{tab:DepCurvVol}
%\end{table}
%\begin{table}
\centering
\begin{tabular}{|p{2.5cm}|p{4.5cm}||p{2.5cm}|p{4.5cm}|}
\hline
curve $C$&Vol$(C)=J C$&curve $C$&Vol$(C)=J C$\\
\hline
\hline
$R_1R_{2}$ & $6a_{3}$&$R_3E_{2,1\,\beta}$ & $2 b_{2,1\,\beta}-b_{4,1\,\beta}$\\
$R_1R_{3}$ & $6a_{2}$&$R_3E_{2,3\,\beta}$ & $2 b_{2,3\,\beta}-b_{4,3\,\beta}$\\
$R_2R_{3}$ & $6a_{1}$&$R_3E_{4,1\,\beta}$ & $-b_{2,1\,\beta}+2 b_{4,1\,\beta}$\\
$R_1D_{2,\beta}$ & $2a_{3}$&$R_{3}E_{4,3\,\beta}$ & $-b_{2,3\,\beta}+2 b_{4,3\,\beta}$\\
$R_1D_{3,\gamma}$ & $3a_{2}$&$R_2E_{3,1\,\gamma}$ & $2b_{3,1\,\gamma}$\\
$R_2D_{1,1}$ & $a_{3}-\sum\limits_{\gamma=1}^4{b_{3,1\,\gamma}}$&$R_{2}E_{3,2\,\gamma}$ & $b_{3,2\,\gamma}$\\
$R_2D_{1,2}$ & $3a_{3}-3\sum\limits_{\gamma=1}^4{b_{3,2\,\gamma}}$&$D_{1,1}E_{3,1\,\gamma}$ & $a_{2}-\sum\limits_{\beta=1}^3{b_{1,\beta\gamma}}+3 b_{3,1\,\gamma}$\\
$R_2D_{1,3}$ & $2a_{3}$&$D_{1,2}E_{3,2\,\gamma}$ & $3a_{2}$\\
$R_2D_{3,\gamma}$ & $3 a_{1}-b_{3,1\,\gamma}-3 b_{3,2\,\gamma}$&$D_{1,1}E_{4,1\,\beta}$& $a_{3}-\sum\limits_{\gamma=1}^4{b_{1,\beta\gamma}}+4 b_{4,1\,\beta}$\\
$R_3D_{1,1}$ & $a_{2}-\sum\limits_{\beta=1}^3{b_{4,1\,\beta}}$&$D_{1,3}E_{4,3\,\beta}$ & $2a_{3}$\\
$R_3D_{1,2}$ & $3a_{2}$&$D_{2,\beta}E_{2,1\,\beta}$ & $a_{3}-\sum\limits_{\gamma=1}^4{b_{1,\beta\gamma}}+2 b_{2,1\,\beta}$\\
$R_3D_{1,3}$ & $a_{2}-\sum\limits_{\beta=1}^3{b_{4,3\,\beta}}$&$D_{2,\beta}E_{2,3\,\beta}$ & $2a_{3}$\\
$R_3D_{2,\beta}$ & $2 a_{1}-b_{2,1\,\beta}-2 b_{2,3\,\beta}$&$D_{2,\beta}E_{3,2\,\gamma}$ & $2b_{3,2\,\gamma}$\\
$D_{1,2}D_{2,\beta}$ & $a_{3}-\sum\limits_{\gamma=1}^4{b_{3,2\,\gamma}}$&$D_{3,\gamma}E_{3,1\,\gamma}$ & $a_{2}-\sum\limits_{\beta=1}^3{b_{1,\beta\gamma}}+b_{3,1\,\gamma}$\\
$D_{2,\beta}D_{3,\gamma}$ & $a_{1}-b_{1,\beta\gamma}-b_{2,3\,\beta}+b_{3,2\,\gamma}$&$D_{3,\gamma}E_{3,2\,\gamma}$ & $3a_{2}$\\
$D_{1,3}E_{4,3\,\beta}$ & $2a_{3}$&$D_{3,\gamma}E_{2,3\,\beta}$ & $2 b_{2,3\,\beta}-b_{4,3\,\beta}$\\
&&$D_{3,\gamma}E_{4,3\,\beta}$ & $-b_{2,3\,\beta}+2 b_{4,3\,\beta}$\\
&&$E_{2,1\,\beta}E_{4,1\,\beta}$ & $a_{3}-\sum\limits_{\gamma=1}^4{b_{1,\beta\gamma}}+2 b_{2,1\,\beta}$\\
&&$E_{2,3\,\beta}E_{4,3\,\beta}$ & $2a_{3}$\\
\hline
\end{tabular}
\caption{The volumes of compact curves existing independently of the triangulation of Res$\left(T^6/\mathbb{Z}_\text{6--II}\right)$ for the case in which all $\mathbb{Z}_{\text{6--II}}$ fixed points are resolved with triangulation i). }
\label{tab:IndCurvVol}
\end{table}

\newpage
\bibliographystyle{paper}
{\small
\providecommand{\href}[2]{#2}\begingroup\raggedright\endgroup

}
\end{document}